%% file: main.tex
\documentclass[acmtog]{acmart}
\acmSubmissionID{508}

\usepackage{booktabs} 
\usepackage{amsmath,bm,mathtools}
\usepackage{wrapfig}
\usepackage{floatflt}
\usepackage{subfigure}
\usepackage{cases, enumerate}

\usepackage{threeparttable}
\usepackage[linesnumbered,ruled,vlined]{algorithm2e} 
\SetKwInput{kwPartI}{Part I}
\SetKwInput{kwPartII}{Part II}
\SetKwInput{kwState}{State Variables}
\usepackage{algpseudocode}

\SetAlFnt{\small}
\SetAlCapFnt{\small}
\SetAlCapNameFnt{\small}
\SetAlCapHSkip{0pt}

\acmJournal{TOG}

\input{bo_macro}
\usepackage{comment}

\setcopyright{acmcopyright}\acmJournal{TOG}\acmYear{2021}\acmVolume{40}\acmNumber{4}\acmArticle{120}\acmMonth{8}\acmDOI{10.1145/3450626.3459862}

\begin{document}
\title{Solid-Fluid Interaction with Surface-Tension-Dominant Contact}
\author{Liangwang Ruan}
\affiliation{\institution{CFCS, Peking University \& AICFVE, Beijing Film Academy}}
\email{ruanliangwang@pku.edu.cn}
\authornote{Joint first authors}

\author{Jinyuan Liu}
\affiliation{\institution{Dartmouth College}}
\email{jinyuan.liu.gr@dartmouth.edu}
\authornotemark[1]

\author{Bo Zhu}
\affiliation{\institution{Dartmouth College}}
\email{bo.zhu@dartmouth.edu}

\author{Shinjiro Sueda}
\affiliation{\institution{Texas A\&M University}}
\email{sueda@tamu.edu}

\author{Bin Wang}
\affiliation{\institution{AICFVE, Beijing Film Academy}}
\email{binwangbuaa@gmail.com}
\authornote{Corresponding authors}

\author{Baoquan Chen}
\affiliation{\institution{CFCS, Peking University \& AICFVE, Beijing Film Academy}}
\email{baoquan.chen@gmail.com}
\authornotemark[2]

\begin{abstract}
We propose a novel three-way coupling method to model the contact interaction between solid and fluid driven by strong surface tension.
At the heart of our physical model is a thin liquid membrane that simultaneously couples to both the liquid volume and the rigid objects, facilitating accurate momentum transfer, collision processing, and surface tension calculation.
This model is implemented numerically under a hybrid Eulerian-Lagrangian framework where the membrane is modelled as a simplicial mesh and the liquid volume is simulated on a background Cartesian grid.
We devise a monolithic solver to solve the interactions among the three systems of liquid, solid, and membrane.
We demonstrate the efficacy of our method through an array of rigid-fluid contact simulations dominated by strong surface tension, which enables the faithful modeling of a host of new surface-tension-dominant phenomena including: objects with higher density than water that remains afloat; ‘Cheerios effect’ where floating objects attract one another; and surface tension weakening effect caused by surface-active constituents.


\end{abstract}

%
%


%
%


\begin{teaserfigure}
  \centering
  \includegraphics[width=\textwidth]{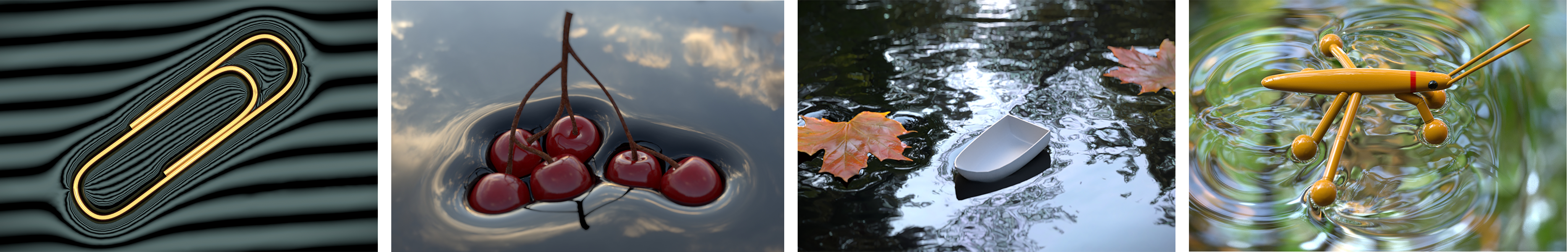}
  \caption{Our novel three-way coupling method can simulate the dynamics of surface-tension-dominant contact between solid and liquid, including static contact between a steel paperclip, cherries sitting on top of water, autumn leaves floating and rotating in a brook, and a water strider robot actuated by its joints.}  
 \label{fig:teaser}
\end{teaserfigure}


\maketitle
\input{Sections/introduction}
\input{Sections/related_work}
\input{Sections/physics_model}
\input{Sections/discretization}
\input{Sections/threeway_coupling}
\input{Sections/time_scheme}

\input{Sections/result}
\input{Sections/discussion_and_conclusion}

\begin{acks}
We thank the anonymous reviewers for their constructive comments. This work was supported in part by National Key R\&D Program of China (2018YFB1403900).
Dartmouth authors acknowledge the funding support from the Burke Research Initiation Award and Toyota TEMA North America Inc.
TAMU author acknowledges the funding support from the National Science Foundation (CAREER-1846368).
We credit the Houdini Education licenses for the video generations.
\end{acks}

\bibliographystyle{ACM-Reference-Format}
\bibliography{main}
\end{document}

%% file: bo_macro.tex
\usepackage{xcolor}

\newif\ifsingle

\ifsingle

\else

\fi

\newcommand{\jy}[1]{\textcolor{orange}{[jyuan: #1]}}

\newif\ifverbose
\verbosetrue

\ifverbose

\newcommand{\vb}[1]{\textcolor{red}{#1}}
\else

\newcommand{\vb}[1]{}
\fi


%% file: Sections/introduction.tex
\section{Introduction}

Interactions between solids and fluids driven by the strong interfacial capillary forces are seen ubiquitously, especially in a miniature environment.
Examples include small insects walking on a pond surface, autumn leaves floating and swirling in a brook stream, breakfast cereal clumping together in a bowl of milk, and many other small-scale interactions between creatures, plants, and their living aquatic environment (see Fig.~\ref{fig:teaser}).
These phenomena' most appealing visual aspects are their largely curved and bent liquid surface, the floating and swaying objects, and the intricate and often unstable balance between the object and the liquid surface.
Even a small wrinkle on the water surface might cause the floating object to sink.
These surface-tension-driven phenomena have been drawing attention from both theoretical and experimental fluid mechanics \cite{janssens_chaurasia_fried_2017,10.1088/978-0-7503-1554-8ch3,doi:10.1146/annurev-fluid-122316-045034,Navascues_1979}, motivating new design of various miniature biomimetic robots \cite{Hu2010,Ozcan2014STRIDEIA,Song2007SurfaceTensionDrivenBI,Suhr2005BiologicallyIM}, and inspiring artistic creations of many beautiful photographs and slow-motion videos. 

\begin{figure}[t]
    \includegraphics[trim=1.7cm 1.5cm 1.7cm 1cm,clip,width=.98\linewidth]{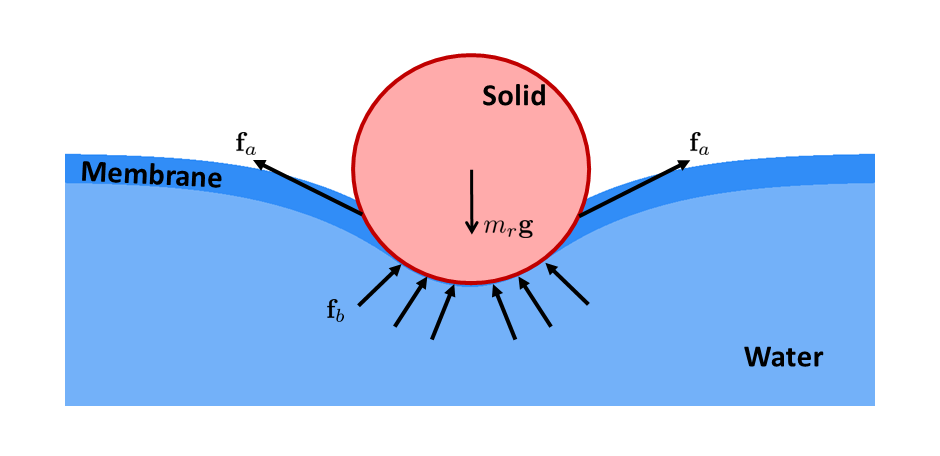}
\caption{Solid and fluid interaction. The solid circle stays afloat on the water under the equilibrium among gravity force $mg$, buoyancy force $\mathbf{f}_b$, and capillary force $\mathbf{f}_a$.} 
	\label{fig:ThreewayPrinciple}
\end{figure}

The most salient feature of such a strongly coupled solid-fluid system is the feasibility of holding an object on a liquid surface whose density is significantly higher than the fluid underneath.
For example, the density of a steel paperclip can be as high as $8$ times as that of water. 
The underpinning governing physics lies in the additional curvature force that contributes to balancing the body's gravity in the vertical direction.  
As illustrated in Figure~\ref{fig:ThreewayPrinciple}, for a solid object sitting on a surface-tensioned water surface, its force balance can be understood as an equilibrium among three forces: $m_r\mathbf{g}=\mathbf{f}_a +\mathbf{f}_b$, with the body's gravity force $m_r\mathbf{g}$, the water's buoyancy force $\mathbf{f}_b$, and the water surface's capillary force $\mathbf{f}_a$.
The buoyancy force's effect is deduced by integrating the fluid pressure over the body in contact with water, and the capillary force is calculated by integrating the surface tension along the contact perimeter of the body.

From a computational perspective, accurately modeling the interaction among these three forces requires proper treatment of three subsystems --- the liquid body, the solid body, and the strongly tensioned liquid interface between them.
However, in both the computational physics and the computer graphics communities, the problem of simulating a strongly coupled surface-tension-dominant contact process remains largely unexplored, due to the lack of effective computational tools to precisely simulate the interactions among the three subsystems. 
The mainstream numerical paradigms of using an implicit level-set method to model the free-surface flow and its interaction with a Lagrangian solid systems \cite{robinson2008two,Carlson2004RigidFA,batty2007fast} suffer from limitations in the following three aspects:
First, capturing the liquid-solid contact perimeter is difficult for an implicit geometry representation. 
Second, using an implicit surface to calculate the curvature as well as the fine-scale capillary wave propagation on the interface is less accurate compared with an explicit mesh representation.
Third, this further adds difficulty in building a monolithic system to accurately couple the two systems, especially when the interface dominates the dynamics.
In a conventional two-way coupled system, the solid and fluid systems interact with each other via an (implicitly) transferred momentum term between the liquid volume and the solid volume (e.g., see \cite{robinson2008two}). 
The surface tension force on the interface contributes to the coupling mechanics by enforcing a pressure jump on the liquid's boundary, which (indirectly) affects the solid's dynamics by changing the liquid pressure.
There is no direct pathway to bridge the liquid interface and the solid, which makes it infeasible to model the term of $\mathbf{f}_a$ that is essential for a surface-tension-dominant contact process.

To tackle these computational challenges, we propose a novel ``three-way'' coupling mechanism to model a solid-fluid coupling system driven by strong surface tension.  
Our key innovation is to treat the surface-tension-dominant interface as a Lagrangian thin membrane that is simultaneously coupled with both the liquid volume and the solid object.
Under this new Lagrangian perspective, the interface is no longer an infinitely thin numerical carrier to transfer the surface tension force from the boundary to the interior; instead, it is a ``virtual'' liquid membrane with a finite, small thickness that can directly impose both buoyancy and surface tension forces onto its contacting objects.
In this sense, the two-way coupled system now becomes a three-way coupled one, with an additional thin liquid layer separating the liquid bulk and the solid.
Thanks to the Lagrangian nature and its explicit mesh representation of the additional liquid membrane, various physical forces can be effectively discretized and enforced in the coupled evolution of both solid and fluid volume.
We develop a full set of numerical infrastructures centered around this ``three-way'' coupling idea to comprehensively accommodate the treatments of incompressibility, buoyancy, surface tension, rigid articulation, and their various intricate interactions.
One significant characteristic of our numerical solution is its ability to handle the coupling between liquid and high-density-ratio solid systems, which was infeasible for all the previous methods.
In particular, our simulator can accurately reproduce the surface-tension-dominant floating phenomena of various thin objects, such as paperclip, thin shell boat, pushpin etc., all implemented with their physical densities (with the maximum density ratio up to $8$---the density ratio of iron to water).

We summarize our technical contributions as follows:
\begin{itemize}
    \item A novel Lagrangian thin membrane representation to accurately capture the contacting interactions between solid and fluid driven by strong surface tension.
    \item A monolithic coupling framework satisfying all velocity and position constraints while conserving momentum.
    \item A prediction-correction contact handling scheme to accurately process fluid-solid contact with realistic physical parameters.
\end{itemize}

%% file: Sections/related_work.tex
\section{Related works}
\input{Figures/ball-3d}

\paragraph{Solid-Fluid Coupling}
State-of-the-art solvers typically use the Eulerian method for fluid and the Lagrangian method for solid, and the coupling between them is usually done by using the solid velocity as a fluid boundary condition and integrating the pressure on the interface. \citet{genevaux2003simulating} use a simple scheme to integrate fluid and solid separately with a coherent-behavior-guaranteed interface force as interaction. \citet{guendelman2005coupling} follow a similar idea but use the predicted displacement of solid as the velocity boundary constraints in the projection step. These kinds of weak-coupling methods usually lead to limited stability and accuracy, which can be largely solved by monolithic strong coupling methods, i.e., integrating fluid and solid in one system. 
The first fully implicit stable fluid-rigid two-way coupling solver is presented by \citet{chentanez2006simultaneous}, who combine the fluid pressure projection equation with velocity update equation of deformable bodies into one asymmetric linear equation. \citet{batty2007fast} introduce an SPD (symmetric positive definite) equation to solve the coupling between fluid and rigid body by considering kinetic energy minimization. \citet{robinson2008two} further make the coupling system to conserve momentum by using a momentum conservation equation for the fluid-body-mixed cells instead of directly interpolating the pressure on the grid to compute forces and torques on the interface. \citet{robinson2011symmetric} further make the equation in \cite{robinson2008two} SPD by using algebraic transformations.
\citet{robinson2009accurate} use Lagrangian multipliers method to make the boundary condition be free-slipping, which means only the normal component of the velocity is constrained. 
\citet{zarifi2017positive} generalize the cut-cell method into the coupling system, managing to achieve free-slipping boundary condition without Lagrangian multipliers by rewriting the incompressible condition in the fluid-body-mixed cells, and also make it SPD. \citet{hyde2019unified} generalize the method in fluid-solid coupling to fluid coupling with sub-grid solids. \citet{aanjaneya2018efficient} develops an efficient solver to accelerate the fluid-rigid coupling.

\paragraph{Surface Tension}
Surface tension is important for the simulation of free-surface fluid. It can be represented in different ways depending on the discretization methods. From an Eulerian point of view, surface tension can be described as pressure discontinuity across the interface between different phases. From a Lagrangian point of view, surface tension can be accommodated as forces between surface particles. Surface tension can be treated either explicitly or implicitly. Implicit treatment often provides better numerical stability. 
\citet{Kang2000ABC} extend the ghost fluid method (GFM) to treat multi-phase incompressible flow including the effects of surface tension.
\citet{Wang2005WaterDrop} use surface tension as an explicit boundary condition in the projection step, and use a virtual level set surface penetrating solid to simulate the contact angles of water drops. However, these explicit surface tension treatments will cause instability when surface tension is dominating the scene, such as for the simulation of bubbles. 
\citet{zheng2009simulation} propose a semi-implicit surface tension method based on the level set method, by considering the advection of the surface curvature using velocity spatial divergence on the surface, to produce realistic bubble dynamics. 
\citet{sussman2009stable} use a notably different scheme by solving a volume-preserving equation based on mean curvature instead of deriving surface tension formulation. 
\citet{chen2020cutcell} propose an extended cut-cell method for handling liquid structures with surface tension that are smaller than a grid cell.
Compared to the grid-based methods, Lagrangian surface tension exhibits its computational merits for its explicit geometry discretization. There are mainly two categories of Lagrangian methods. Mesh-based methods with connectivity information define the differential operators on a mesh node and its incident triangles. \citet{Zhang-2011-DMS} simulate droplets on a surface mesh using deformation operators. \citet{DVS2012} model thin viscous sheets with the single-layer mesh with thickness. \citet{dbwg15} model complex bubble structures with a Lagrangian non-manifold triangle mesh and integrate the surface tension into vertex-based circulations. \citet{dhbwg16} develop a surface-only simulation framework that adapts a general 3D fluid solver onto the surface mesh.  \citet{10.1145/3130800.3130835} replace the curvature with the gradient of the surface area function, to handle the non-manifold junctions of films and bubbles robustly. Lagrangian representations of surface tension are also easier to be treated implicitly. \citet{schroeder2012semi} developed a hybrid framework that formulates implicit surface tension force on a Lagrangian mesh and carries out the pressure solve on background grid. \citet{zheng2015new} introduce an implicit surface tension scheme based on the particle-constructed surface tracking method. \citet{Zhu2014CodimensionalST,Zhu2016CodimensionalNF} propose a unified framework that simulates codimensional phenomena using codimensional simplicial complexes. Surface differential operators can also be approximated by pure particles. The traditional smoothed particle hydrodynamics (SPH) method introduced by \cite{Gingold1977SmoothedPH} uses radial symmetrical smoothing kernels to approximate the differential operators. Other approximation methods include graph Laplacian \cite{BELKIN20081289}, local triangular mesh \cite{Belkin2009ConstructingLO,Lai2013ALM},  closest point method \cite{Cheung2015ALM} and moving least squares \cite{Lancaster1981SurfacesGB,Wang2020CodimensionalST}.

%% file: Figures/ball-3d.tex
\begin{figure}
    \centering
    \includegraphics[width=0.98\linewidth]{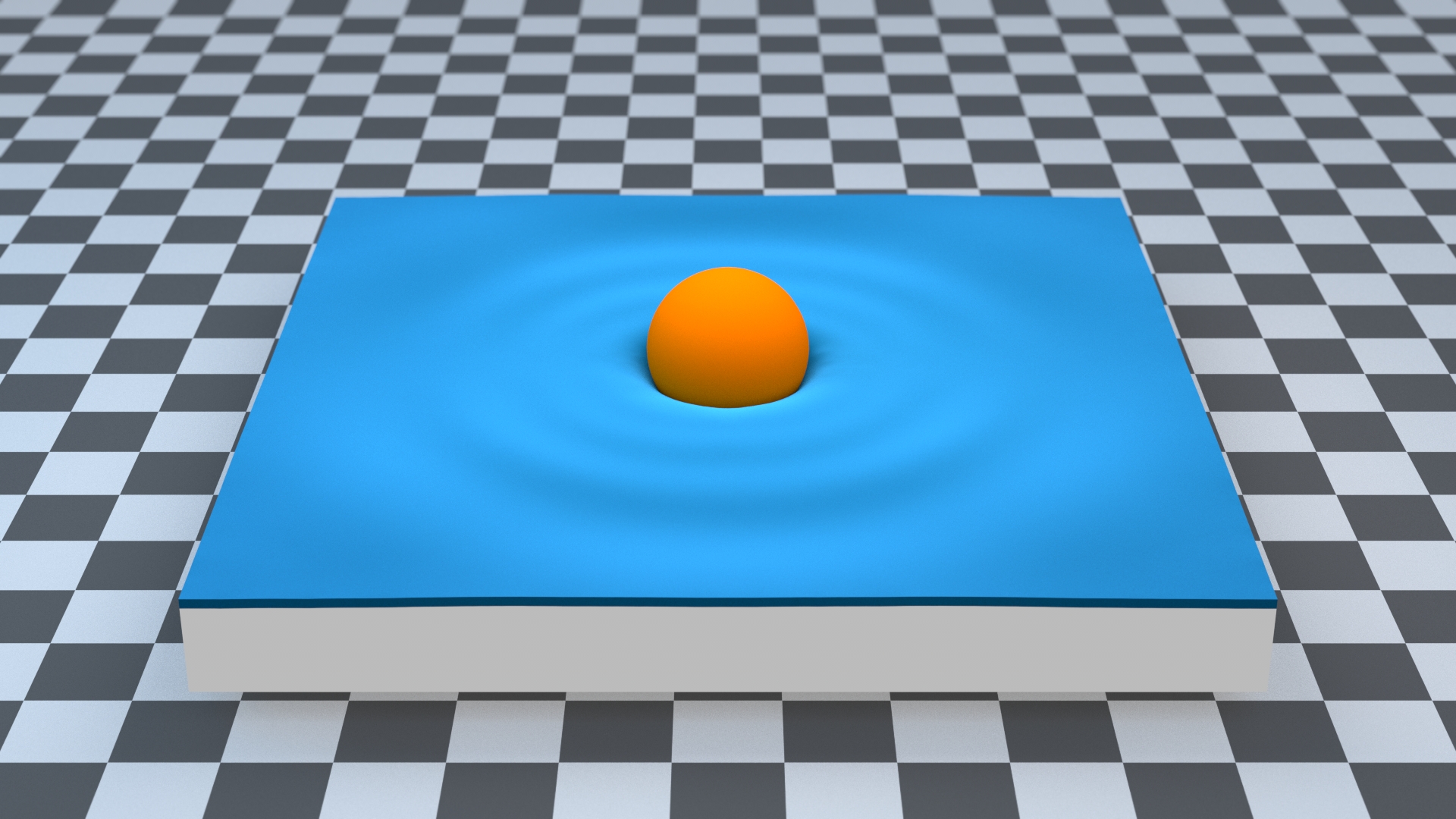}
    \caption{Sphere Falling into Water. Due to the mesh representation of the thin liquid membrane, we obtain fine-scale wave propagation stimulated by the solid’s motion.}
    \label{fig:ball-3d}
\end{figure}

%% file: Sections/physics_model.tex
\section{Physics Model}

In this section, we will first go over the notation convention that we use throughout this paper.
Then we will describe the three physical systems: volumetric fluid, surface membrane, and the rigid body.

\input{Figures/paperclip}

\paragraph{Notation convention} We use plain text to denote scalar quantities (e.g., $\rho$), and boldface lowercase and uppercase letters to denote vectors (e.g., $\mathbf{u}$) and matrices (e.g., $\mathbf{M}$) respectively. 
Subscripts $r$ and $s$ are used to distinguish quantities between the rigid body and the surface membrane, respectively.
Quantities in their continuous or discretized form are not explicitly distinguished.

\paragraph{Volumetric Fluid}
The motion of the fluid is governed by the standard incompressible Navier-Stokes equations:
\begin{equation}
    \rho(\frac{\partial \mathbf{u}}{\partial t} + \mathbf{u}\cdot \nabla \mathbf{u}) = -\nabla p + \mu \nabla^2 \mathbf{u} + \mathbf{f},
    \label{eq:fluid:ns}
\end{equation}
\begin{equation}
    \nabla \cdot \mathbf{u} = 0,
\end{equation}
where $\mathbf{u}$ is fluid's velocity; $p$ is the pressure; $\rho$ and $\mu$ are the fluid's density and dynamic viscosity. We ignore viscosity in our algorithm for simplicity. $\mathbf{f}$ denotes all body forces applied to the fluid.
We enforce the Neumann boundary constraint $\mathbf{u}\cdot \mathbf{n} = 0$ on wall boundaries and the Dirichlet boundary constraint  $p=0$ on the free surface. 
At the regions in contact with rigid bodies, we enforce $\mathbf{u}=\dot{\mathbf{x}}$, where $\dot{\mathbf{x}}$ is the velocity of the contact position on the rigid body.

\paragraph{Surface membrane}
We define the geometry of the surface membrane as the thin region around the fluid's free surface with a constant thickness $h$. 
The governing equations of the fluid membrane are the same as the fluid volume, except that a surface tension force is applied on the membrane due to the unbalanced cohesion force between the fluid molecules near the interface.
For a given area surrounded by boundary line $\Gamma$ on the membrane, its capillary force is calculated as 
\begin{equation}
    \mathbf{f}_{c}=\oint_{\Gamma} \sigma \mathrm{d}\mathbf{l}\times \mathbf{n},
\end{equation}
where $\sigma$ is the surface tension coefficient, and $\mathbf{n}$ is the surface normal at $d\mathbf{l}$ on $\Gamma$; here $\mathbf{f}_{c}$ is used to distinguish the capillary force acting on the fluid membrane from the capillary force $\mathbf{f}_{a}$ acting on the solid (Eq.~\ref{eq:attraction_force}). From the energy perspective, we can also treat the capillary force as the outcome of surface energy
\begin{equation}
    E_c = \sigma A,
\end{equation}
where $A$ is the total area of the surface. This energy perspective is what we use in the later analysis. At the same time, the surface membrane exchanges momentum with both the fluid volume underneath and the contacting rigid body.

\paragraph{Rigid body}
\label{par:pm:rb}
We assume that the (articulated) rigid body is expressed by generalized position, $\mathbf{q}_r$, and generalized velocity, $\mathbf{v}_r$. For a single rigid body, these generalized coordinates are 6D vectors; for each additional degree of freedom (e.g., joint angles), we add one dimension to these vectors.
The equations of motion of an articulated rigid body can be written as:
\begin{equation}
\label{eq:rigidEOM}
    \mathbf{M}_r(\mathbf{q}_r) \frac{\mathrm{d}\mathbf{v}_r}{\mathrm{d}t} = \mathbf{f}_r(\mathbf{q}_r,\mathbf{v}_r),
\end{equation}
where $\mathbf{M}_r$ is the generalized inertia, and $\mathbf{f}_r$ is the generalized force, which includes gravity, contact forces, Coriolis force, and all other quadratic velocity vectors that may result from the reduced coordinates of the articulated system \cite{Murray1994,Wang2019}.
We also keep track of the Jacobian matrix, $\mathbf{J}_r(\mathbf{q}_r)$, which maps the rigid body's generalized velocity to the 3D world velocity of points that are defined with respect to the rigid body. The transpose of this Jacobian, $\mathbf{J}_r^\mathrm{T}$, maps the 3D world forces applied at these points to the generalized force of the rigid body. These mappings will be used in later sections to couple the rigid body to the rest of the system.

%% file: Figures/paperclip.tex
\begin{figure*}[t]
    \newcommand{\formattedgraphics}[1]{\includegraphics[width=0.32\textwidth,trim= 0cm 0 0cm 0,clip]{#1}}
    \centering
    \formattedgraphics{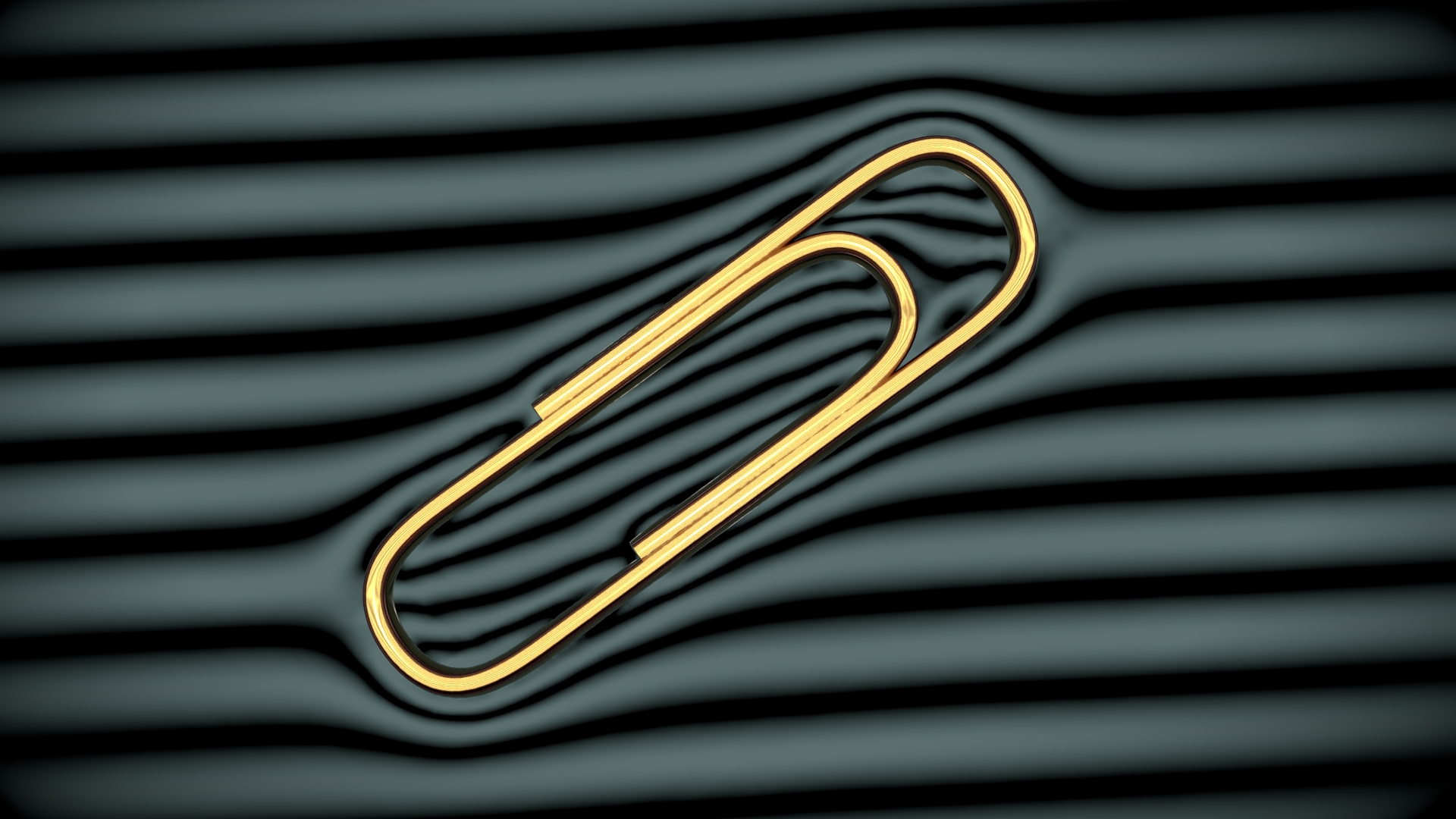}
    \formattedgraphics{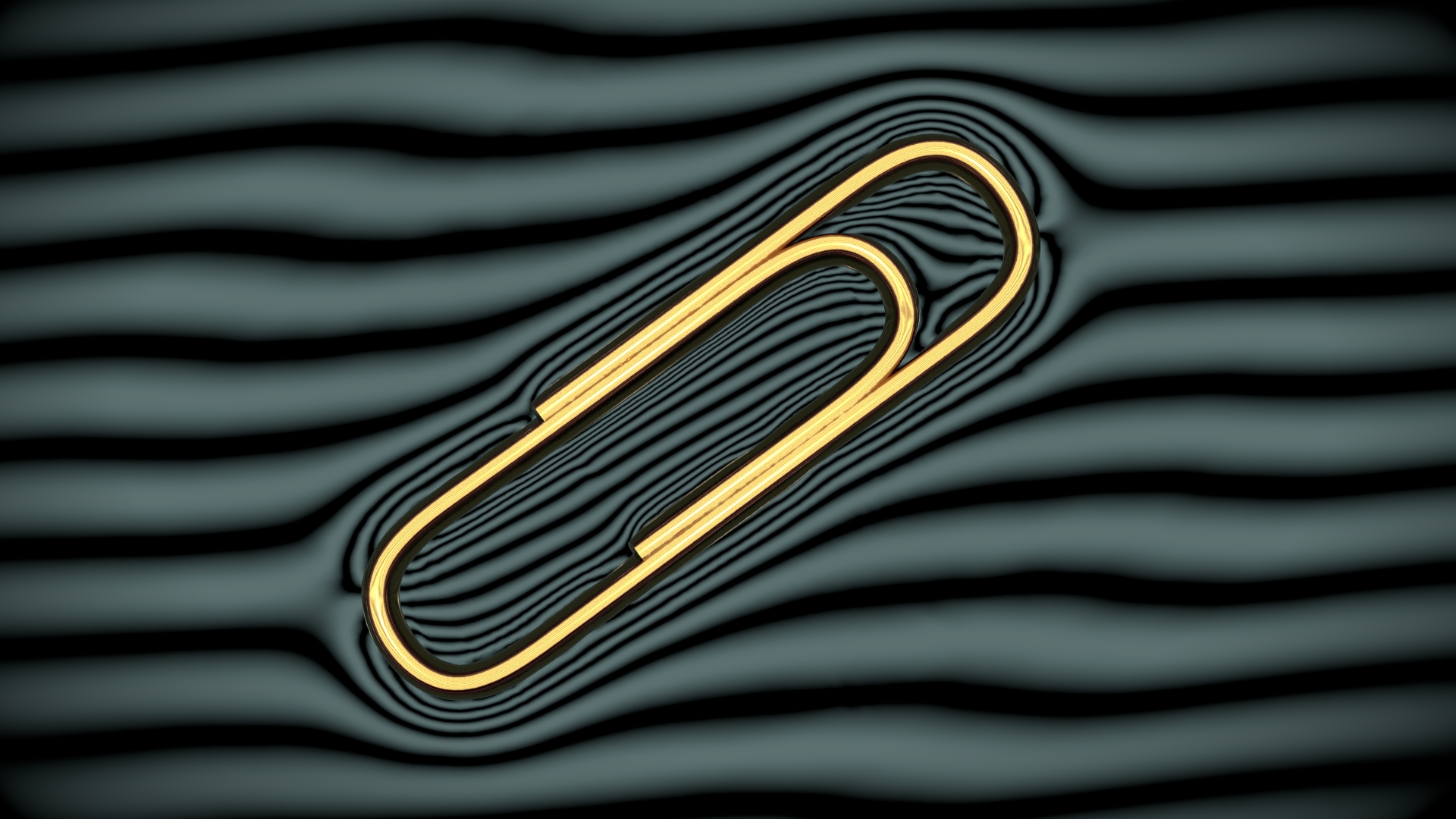}
    \formattedgraphics{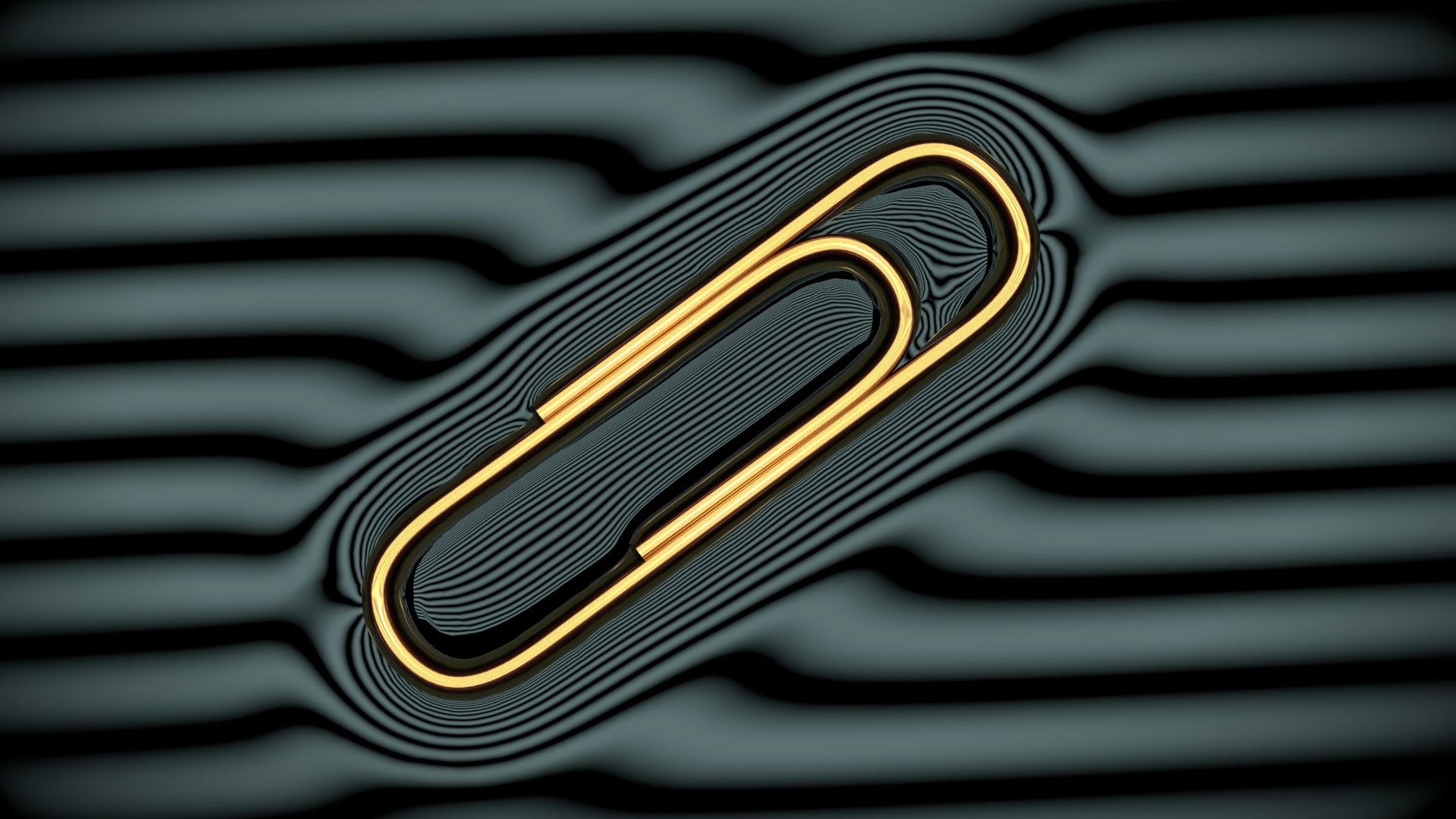}
    \caption{Paperclips with increasing density stay afloat on the surface of the water. From left to right, the density are $2.0$, $5.0$, and $7.9$ $g/cm^3$, respectively. Heavier paperclip deforms the water surface in a larger range with sharper curvature, which can be observed from the degree of distortion of water refraction. }
    \label{fig:paperclip}
\end{figure*}

%% file: Sections/discretization.tex
\section{Discretization}

In this section we introduce our discretization for different parts of our system. We use an Eulerian description for the volumetric fluid, a single-layer surface mesh for the surface membrane, and a Lagrangian description for the rigid body.

\subsection{Fluid}
A standard marker-and-cell (MAC) grid is used to discretize the fluid volume by storing the velocity on faces and pressure on cell centers. We take a classical operator-splitting approach to solve the fluid equations \cite{Jos1999}, including semi-Lagrangian advection, applying external forces, and pressure projection.
In particular, the projection step calculates the velocity $\mathbf{u}^{n+1}$ for the next time step based on the intermediate velocity $\mathbf{u}^*$:
\begin{equation}
    -\mathbf{G}^T\mathbf{u}^{n+1} = 0,
    \label{eq:fluid:inc}
\end{equation}
\begin{equation}
    \mathbf{u}^{n+1} = \mathbf{u}^* - \frac{\Delta t}{\rho }\mathbf{G}  \mathbf{p},
    \label{eq:fluid:pp}
\end{equation}
where $\mathbf{G}$ is the discretized gradient operator and $-\mathbf{G}^T$ is the discretized divergence operator.
Substituting $\mathbf{u}^{n+1}$ from Eq.~\ref{eq:fluid:pp} into Eq.~\ref{eq:fluid:inc} leads to a Poisson equation for pressure:
\begin{equation}
    \frac{1}{\rho}\mathbf{G}^T\mathbf{G}\hat{\mathbf{p}} = \mathbf{G}^T\mathbf{u}^*,
\end{equation}
where $\hat{\mathbf{p}}=\mathbf{p}\cdot \Delta t$ for brevity. 

\subsection{Membrane}

We discretize the surface membrane using a triangle mesh $\mathcal{S}$ in 3D, and a line segment mesh in 2D. We will only focus on the 3D case in this section.
The mesh resolution depends on the background grid resolution. In all our experiments, we set the mesh resolution at the initial state to be two times higher than the resolution of the background grid. Other resolutions such as 1.5x and 3x should also work without producing significantly different behaviors. 
 
\setlength{\columnsep}{9pt}
\begin{wrapfigure}[10]{r}{0.4\linewidth}
	\vspace{-1em}
	\centering
	\includegraphics[width=\linewidth,trim=40 50 40 50,clip,]{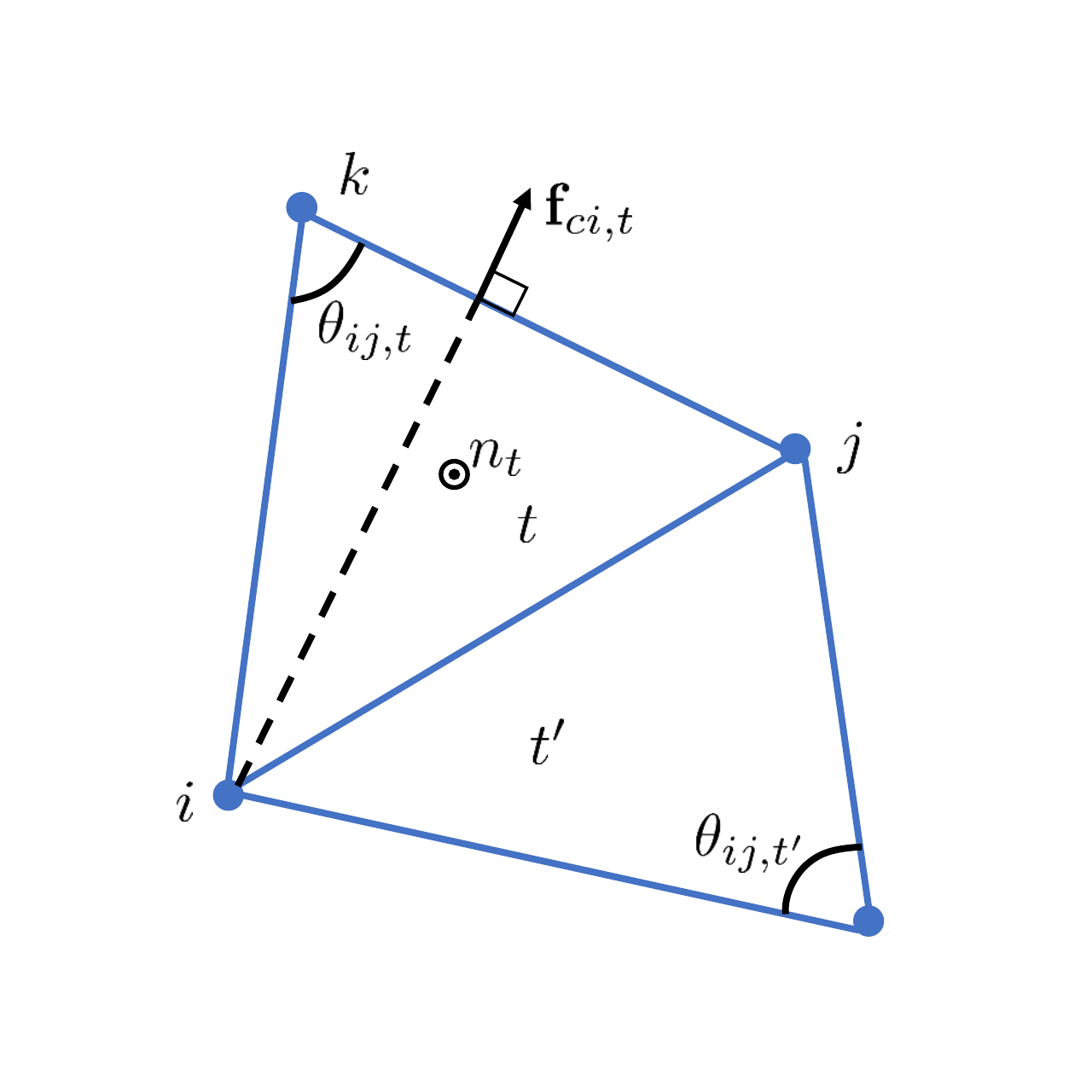}
\end{wrapfigure}
 Each mesh node is interpreted as a particle with mass. The mass of an individual particle is proportional to its surrounding area, and can be calculated as $m_{i} = \rho h\sum_{t\in \mathcal{N}_i}A_t/3$, where $\mathcal{N}_i$ is the 1-ring neighborhood of particle $i$, and $A_t$ is the area of its incident triangle $t$.
 With this discretization, we model the surface tension as capillary forces similarly as \cite{zheng2015new}:
  \begin{equation}
 \mathbf{f}_{ci} = \sum_{t\in \mathcal{N}_i} \mathbf{f}_{ci,t} =  -\sum_{t\in \mathcal{N}_i}\sigma \frac{\partial A_t}{\partial \mathbf{x}_{i}}, 
 \end{equation}
where $\mathbf{f}_{ci,t}$ is the capillary force on particle $i$ from triangle $t$. The exact form of $\mathbf{f}_{ci,t}$ is
\begin{equation}
    \mathbf{f}_{ci,t} = \frac{\sigma}{2}\mathbf{l}_{jk}\times \mathbf{n}_t,
\end{equation}
where $\mathbf{l}_{jk}=\mathbf{x}_{k}-\mathbf{x}_{j}$, and $\mathbf{n}_t$ is the normal of triangle $t$ pointing out of the surface. 
To compute $\mathbf{f}_{ci}$ implicitly, we need the derivative of $\mathbf{f}_{ci}$ with respect to the particle position. Since this derivative is indefinite, we follow \cite{zheng2015new} using an all-directional implicit model instead:
 \begin{equation}
     \mathbf{K}_{cij,t} = -\frac{\partial(\mathbf{f}_{ci,t}, \mathbf{f}_{cj,t})}{\partial(\mathbf{x}_{i},\mathbf{x}_{j})} = \frac{\sigma \cot{\theta_{ij,t}}}{2}\begin{pmatrix}
     \mathbf{I} & -\mathbf{I}\\
     -\mathbf{I} & \mathbf{I}
     \end{pmatrix}.
     \label{eq:mem:Kc}
 \end{equation}
Assembling $\mathbf{K}_{cij,t}$ for all pairs of $i$, $j$ and triangle $t$, we get $\mathbf{K}_c$. Since the surface membrane is affected by both the capillary force and fluid's pressure force, we delay the discussion about the membrane's equation of motion to Sec.~\ref{sec:fluid-membrane}.

\subsection{Rigid Body}
\label{sec:dis:rigid}
To match the time-stepping formulation of the fluid, we discretize the equations of motion of the (articulated) rigid body as follows.
At each time step, we first update the rigid body's generalized position as $\mathbf{q}_r^{n+1}=\mathbf{q}_r^n+\mathbf{v}_r^n\Delta t$.
Then we discretize Eq.~\ref{eq:rigidEOM} semi-implicitly as
\begin{equation}
    \mathbf{M}_r\frac{\mathbf{v}_r^{n+1}-\mathbf{v}_r^n}{\Delta t} = \mathbf{f}_r(\mathbf{q}_r^{n+1}+(\mathbf{v}_r^{n+1}-\mathbf{v}_r^n)\Delta t, \mathbf{v}_r^{n+1}).
\end{equation}
This equation can be further rewritten as
\begin{equation}
    \begin{aligned}
    &(\mathbf{M}_r+\mathbf{D}_r\Delta t+\mathbf{K}_r\Delta t^2)\mathbf{v}_r^{n+1}\\
    & \quad = \mathbf{M}_r\mathbf{v}_r^{n} + \mathbf{f}_r(\mathbf{q}_r^{n+1},\mathbf{v}_r^n)\Delta t + \mathbf{D}_r\mathbf{v}_r^n\Delta t+\mathbf{K}_r\mathbf{v}_r^{n}\Delta t^2,
    \end{aligned}
    \label{eq:rigid:full}
\end{equation}
where $\mathbf{D}_r = -\frac{\partial \mathbf{f}_r}{\partial \mathbf{v}_r}|_{\mathbf{q}_r^{n+1},\mathbf{v}_r^n}$, $\mathbf{K}_r = -\frac{\partial \mathbf{f}_r}{\partial \mathbf{q}_r}|_{\mathbf{q}_r^{n+1},\mathbf{v}_r^n}$. 
Finally, we obtain:
\begin{equation}
    \hat{\mathbf{M}}_r\mathbf{v}_r^{n+1} = \mathbf{M}_r\mathbf{v}_r^n + \hat{\mathbf{f}}_r(\mathbf{q}_r^{n+1}, \mathbf{v}_r^n),
\end{equation}
where $\hat{\mathbf{M}}_r=\mathbf{M}_r+\mathbf{D}_r\Delta t+\mathbf{K}_r\Delta t^2$, and $\hat{\mathbf{f}}_r=\mathbf{f}_r+\mathbf{D}_r\mathbf{v}_r^n + \mathbf{K}_r\mathbf{v}_r^n\Delta t$. 
This semi-implicit linear system uses the updated generalized positions, $\mathbf{q}_r^{n+1}$, and the previous generalized velocities, $\mathbf{v}_r^n$, to solve for the next generalized velocities, $\mathbf{v}_r^{n+1}$.

%% file: Sections/threeway_coupling.tex
\section{Three-way Coupling}

\begin{figure}
    \centering
    \includegraphics[width=0.98\linewidth]{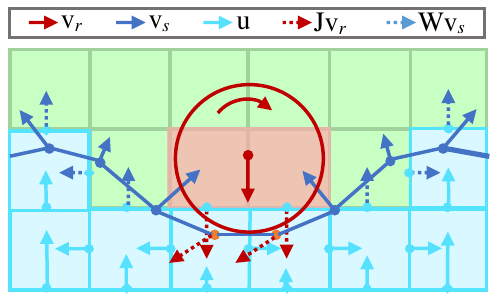}
    \caption{System DOFs. Solid red arrows show the rigid body's generalized velocities. Solid blue arrows show the surface membrane's particle velocities. Solid cyan arrows show the volumetric fluid's face velocities. Dotted red arrows show the velocities of membrane particles attached to the rigid body and the projected velocities on fluid faces. Dotted blue arrows show the velocities of fluid faces projected from the membrane.}
    \label{fig:dof}
\end{figure}
In this section, we introduce the derivation of our monolithic three-way coupling system for fluid, membrane, and rigid body. Firstly, the membrane-rigid (M-R) and fluid-membrane (F-M) coupling equations are described individually. Then, we explain how we assemble our final three-way coupling equation. 

\subsection{Fluid-Membrane Coupling}
\label{sec:fluid-membrane}

\setlength{\columnsep}{9pt}
\begin{wrapfigure}[6]{r}{0.4\linewidth}
\vspace{-2em}
\centering
\includegraphics[width=\linewidth,trim=4 18 2 18,clip]{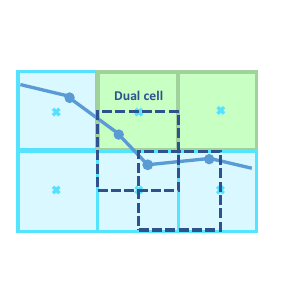}
\end{wrapfigure}
We employ the scheme proposed by \citet{robinson2008two} to model the F-M coupling, in which fluid uses an Eulerian representation, while surface membrane uses a Lagrangian representation.  
We involve a new type of cell, named ``dual cell.'' As illustrated in the inset figure, the dual cell is centered around the face center, with the same size as the MAC grid. In the entire simulation domain, there are three different kinds of dual cells: fluid-rigid mixed, fluid-air mixed, and all-fluid dual cell. The type of each dual cell is determined by the type of its two incident MAC grid cells without considering the membrane's thickness. 

The velocity of the fluid-air dual cell is interpolated from the surface membrane using the interpolation matrix $\mathbf{W}$, with row size equal to the number of dual cells, and column size equal to the number of membrane particles. This matrix is used to map the velocity of the membrane particles to the velocity of the fluid's grid. For each dual cell, we traverse each interpenetrated mesh triangle and compute its intersection area. This area is then accumulated to the corresponding entries in $\mathbf{W}$. Then, $\mathbf{W}$ are normalized along row dimension. The transpose, $\mathbf{W}^T$, maps the pressure difference at each dual cell to the corresponding particles to provide impulse. It also lumps the mass of fluid at the free surface to the particles. Using $\mathbf{W}$ we can rewrite the incompressibility equation for fluid as
\begin{equation}
-\mathbf{G}^T(\mathbf{u}^*-\frac{1}{\rho}\mathbf{G}\hat{\mathbf{p}} + \mathbf{W}\mathbf{v}^{n+1}_s) = 0.
\label{eq:mem:inc}
\end{equation}

To analyze the dynamics of the membrane, we need to compute the pressure force from the fluid grid. The pressure impulse applied on a dual cell is $-V\mathbf{G}\hat{\mathbf{p}}$, where $V$ is the dual cell's volume. The fluid momentum change is $\mathbf{M}(\mathbf{W}\mathbf{v}_s^{n+1}-\mathbf{u}^*)$, where $\mathbf{M}$ is a diagonal matrix constructed with the fluid mass in each dual cell. Then, the pressure impulse transferred from dual cell to surface membrane is
\begin{equation}
\mathbf{I} = \mathbf{W}^T(-\mathbf{G}V\hat{\mathbf{p}}-\mathbf{M}(\mathbf{W}\mathbf{v}_s^{n+1}-\mathbf{u}^*)).
\label{eq:mem:Idc}
\end{equation}

\input{Figures/Cherrio}

Consequently, the momentum of membrane changes due to the collective effect of pressure impulse and capillary force: 
\begin{equation}
\mathbf{M}_s(\mathbf{v}_s^{n+1}-\mathbf{v}_s^n) = \mathbf{f}_c(\mathbf{x}_{s}^{n+1}+(\mathbf{v}_s^{n+1}-\mathbf{v}_s^n)\Delta t)\Delta t + \mathbf{I}.
\label{eq:mem:mom}
\end{equation}
For the same reason as we explained in Sec.~\ref{sec:dis:rigid}, we use the corrected mesh position in $\mathbf{f}_c(\cdot)$. 
By substituting $\mathbf{I}$ in Eq.~\ref{eq:mem:mom} into Eq.~\ref{eq:mem:Idc}, we obtain the motion equation for membrane:
\begin{equation}
\hat{\mathbf{M}}_s\mathbf{v}_s^{n+1} + \mathbf{W}^T\mathbf{G}V\hat{\mathbf{p}} = \mathbf{M}_s \mathbf{v}_s^n + \hat{\mathbf{f}}_c(\mathbf{x}_s^{n+1})\Delta t + \mathbf{W}^T\mathbf{M}\mathbf{u}^*,
\label{eq:mem:mot}
\end{equation}
where $\hat{\mathbf{M}}_s=\mathbf{M}_s + \mathbf{K}_c\Delta t^2 + \mathbf{W}^T\mathbf{M}\mathbf{W}$, and $\hat{\mathbf{f}}_c=\mathbf{f}_c+\mathbf{K}_c\mathbf{v}_s^n\Delta t$. 

Finally, combining  Eq.~\ref{eq:mem:inc} and Eq.~\ref{eq:mem:mot}, the two-way F-M coupling system becomes:
\begin{equation}
\label{fluid_shell}
\begin{bmatrix}
\frac{V}{\rho}\mathbf{G}^T\mathbf{G} & -V\mathbf{G}^T\mathbf{W}\\
-\mathbf{W}^T\mathbf{G}V & -\hat{\mathbf{M}}_s 
\end{bmatrix}
\begin{bmatrix}
\hat{\mathbf{p}}\\
\mathbf{v}^{n+1}_s
\end{bmatrix}
=
\begin{bmatrix}
V\mathbf{G}^T\mathbf{u}^* \\
-\mathbf{M}_s\mathbf{v}_s^n - \mathbf{W}^T\mathbf{M}\mathbf{u}^* - \hat{\mathbf{f}}_c\Delta t
\end{bmatrix}.
\end{equation}

\subsection{Membrane-Rigid Coupling}
\label{sec:tc:sr}
The critical point for M-R coupling resides on how to correctly evaluate adhesion forces between the membrane and the rigid body.
We first divide the mesh particles into two sets: free particles $\mathcal{F}$ and contact particles $\mathcal{C}$. We set up $\mathcal{C}$ using these three steps:
\begin{enumerate}[(1)]
    \item If vertex $i$ and none of its neighboring vertices are in a fluid-air dual cell, put $i$ in $\mathcal{C}$.
    \item If not (1), but the distance from $\mathbf{x}_i$ to the rigid body's surface is less than a threshold $\epsilon$ (we use one tenth of grid size), put $i$ in $\mathcal{C}$.
    \item If vertex $i\in \mathcal{C}$ but none of its neighboring vertices are in $\mathbf{C}$, remove $i$ from $\mathcal{C}$.
\end{enumerate}
Since contact particles’ velocities are constrained by the rigid body, they are not involved in the coupling system. We use the set $\mathcal{M}$ to represent the triangle meshes that contain two types of particles, which are usually located around contact boundary regions. We further use $\mathcal{C}_\mathcal{M}$ to indicate contact particles belonging to triangle meshes $\mathcal{M}$.
The attraction force on particles $\in \mathcal{C}_\mathcal{M}$ can indirectly affect the rigid body through contact. We formulate the capillary force on each of these particles as the summation of the attraction force from all of its incident neighborhood triangles $\in \mathcal{M}$. The equivalent adhesion force on the rigid body is $\sum_{i \in \mathcal{C}_\mathcal{M}} \sum_{t\in \mathcal{N}_i\cap \mathcal{M}}\mathbf{f}_{ci,t}$. We can also get the generalized force acting on the rigid body using the Jacobian transpose:
\begin{equation}
\label{eq:attraction_force}
    \mathbf{f}_a = \sum_{i \in \mathcal{C}_\mathcal{M}} \sum_{t\in \mathcal{N}_i\cap \mathcal{M}}\mathbf{J}_{ri}^T\mathbf{f}_{ci,t}.
\end{equation}
To compute $\mathbf{f}_a$ implicitly, we require the derivative of $\mathbf{f}_a$ with respect to the free particles' position $\mathbf{x}_s$ and the rigid body's generalized position $\mathbf{q}_r$. 
The derivative with respect to $\mathbf{x}_s$ can be obtained using Eq.~\ref{eq:mem:Kc}:
\begin{equation}
    \mathbf{K}_{a,sj}=-\frac{\partial \mathbf{f}_a}{\partial \mathbf{x}_j} = \sum_{i\in \mathcal{C}_\mathcal{M}}\sum_{t\in \mathcal{N}_i\cap \mathcal{M}}\mathbf{J}^T_{ri}(-\frac{\sigma \cot{ \theta_{ij,t}}}{2}).
\end{equation}
The derivative with respect to $\mathbf{q}_r$ can be similarly derived using the chain rule:
\begin{equation}
    \mathbf{K}_{a,r}=-\frac{\partial \mathbf{f}_a}{\partial \mathbf{q}_r}
    =-\sum_{i \in \mathcal{C}_\mathcal{M}}\frac{\partial \mathbf{f}_a}{\partial \mathbf{x}_i}\frac{\partial \mathbf{x}_i}{\partial \mathbf{q}_r}
    =-\sum_{i \in \mathcal{C}_\mathcal{M}}\frac{\partial \mathbf{f}_a}{\partial \mathbf{x}_i}\frac{\partial \mathbf{v}_i}{\partial \mathbf{v}_r}
    =-\sum_{i \in \mathcal{C}_\mathcal{M}}\frac{\partial \mathbf{f}_a}{\partial \mathbf{x}_i}\mathbf{J}_{ri}.
\end{equation}
According to Newton's third law, we can easily get the adhesion force applied to corresponding free particles. $\mathbf{f}_c$ turns out to be a function of both $\mathbf{x}_s$ and $\mathbf{q}_r$, where the derivative of $\mathbf{f}_c$ with respect to $\mathbf{q}_r$ is $\mathbf{K}_{c,r}=\mathbf{K}_{a,s}^T$.

\subsection{Three-way Coupling}
\label{cpl:threeway}

The complete description of our system's DOFs can be seen in Fig.~\ref{fig:dof}.
\paragraph{Fluid}
The fluid velocity on each dual cell can be summarized as:
\begin{equation}
    \label{eq:fluid:velocity}
    \mathbf{u} = \begin{cases}
        \mathbf{u}^* - \frac{1}{\rho}\mathbf{G\mathbf{\hat{p}}}, &\mbox{all-fluid dual cell}\\
        \mathbf{W}\mathbf{v}_s, &\mbox{fluid-air dual cell}\\
        \mathbf{J}_{r}\mathbf{v}_r, &\mbox{fluid-solid dual cell}
    \end{cases}
\end{equation}
where $\mathbf{J}_r$ is the rigid body Jacobian introduced in Sec.~\ref{par:pm:rb}. 
Correspondingly, the fluid incompressibility equation under three-way coupling is:
\begin{equation}
    \label{eq:fluid:coupling}
    -\mathbf{G}^T(\mathbf{u}^* - \frac{1}{\rho}\mathbf{G}\hat{\mathbf{p}} + \mathbf{W}\mathbf{v}_s + \mathbf{J}_r\mathbf{v}_r) = 0.
\end{equation}

\paragraph{Membrane}
By complementing Eq.~\ref{eq:mem:mot} with adhesion force, we have the membrane's equation of motion under three-way coupling:
\begin{equation}
        \tilde{\mathbf{M}}_s \mathbf{v}_s^{n+1} + \mathbf{K}_{c,r}\Delta t^2\mathbf{v}_r^{n+1} + \mathbf{W}^T\mathbf{G}V\hat{\mathbf{p}}
        = \mathbf{M}_s \mathbf{v}_s^n + \tilde{\mathbf{f}}_c\Delta t + \mathbf{W}^T\mathbf{M}\mathbf{u}^*,
    \label{eq:shell:coupling}
\end{equation}
where $\tilde{\mathbf{M}}_s=\mathbf{M}_s+\mathbf{K}_c\Delta t^2+\mathbf{W}^T\mathbf{M}\mathbf{W}$, and $\tilde{\mathbf{f}}_c=\mathbf{f}_c(\mathbf{x}_s^{n+1},\mathbf{q}_r^{n+1})+\mathbf{K}_c\mathbf{v}_s^n\Delta t + \mathbf{K}_{c,r}\mathbf{v}_r^n\Delta t$. As we discussed in Sec.~\ref{sec:tc:sr}, only free particles are involved in this solve.

\paragraph{Rigid Body}
The buoyancy force $\mathbf{f}_b$ on the rigid body is computed using the impulse from the dual cell similar to Eq.~\ref{eq:mem:Idc}:
\begin{equation}
    \mathbf{f}_b\Delta t = \mathbf{J}_r^T(-\mathbf{G}V\hat{\mathbf{p}}-\mathbf{M}(\mathbf{J}_r\mathbf{v}_r^{n+1}-\mathbf{u}^*)).
\end{equation}
By simultaneously applying buoyancy force $\mathbf{f}_b$, adhesion force $\mathbf{f}_a$, and $\mathbf{f}_r$, we get its motion equation under three-way coupling as:
\begin{equation}
        \tilde{\mathbf{M}}_r\mathbf{v}_r^{n+1} + \mathbf{K}_{a,s}\Delta t^2 \mathbf{v}_s^{n+1} + \mathbf{J}_r^T\mathbf{G}V\hat{\mathbf{p}}
        = \mathbf{M}_r\mathbf{v}_r^n + \tilde{\mathbf{f}}_{a}\Delta t + \tilde{\mathbf{f}}_r\Delta t + \mathbf{J}_r^T\mathbf{M}\mathbf{u}^*,
    \label{eq:rigid:coupling}
\end{equation} 
where $\tilde{\mathbf{M}}_r = \mathbf{M}_r + \mathbf{D}_r\Delta t + \mathbf{K}_r\Delta t^2 + \mathbf{K}_{a,r}\Delta t^2 + \mathbf{J}_r^T\mathbf{M}\mathbf{J}_r$, $\tilde{\mathbf{f}}_r=\mathbf{f}_r+\mathbf{D}_r\mathbf{v}_r^n+\mathbf{K}_r\mathbf{v}_r^n\Delta t$, and $\tilde{\mathbf{f}}_a=\mathbf{f}_a+\mathbf{K}_{a,r}\mathbf{v}_r^n\Delta t+\mathbf{K}_{a,s}\mathbf{v}_s^n\Delta t$.

Finally we can assemble Eq.~\ref{eq:fluid:coupling}, Eq.~ \ref{eq:shell:coupling}, and Eq.~\ref{eq:rigid:coupling} into one symmetric three-way coupling system:
\begin{equation}
    \begin{aligned}
        &\begin{bmatrix}
            \frac{V}{\rho}\mathbf{G}^T\mathbf{G} & -V\mathbf{G}^T\mathbf{W} & -V\mathbf{G}^T\mathbf{J}_r\\
            -\mathbf{W}^T\mathbf{G}V & -\tilde{\mathbf{M}}_s & -\mathbf{K}_{c,r}\Delta t^2\\
            -\mathbf{J}_r^T\mathbf{G}V & -\mathbf{K}_{a,s}\Delta t^2 & -\tilde{\mathbf{M}}_r
        \end{bmatrix}
        \begin{bmatrix}
            \hat{\mathbf{p}}\\
            \mathbf{v}_s^{n+1}\\
            \mathbf{v}_r^{n+1}
        \end{bmatrix}\\
        &= 
        \begin{bmatrix}
            V\mathbf{G}^T\mathbf{u}^* \\
           -\mathbf{M}_s\mathbf{v}_s^n -\tilde{\mathbf{f}}_c\Delta t 
           - \mathbf{W}^T\mathbf{M}\mathbf{u}^*\\
            -\mathbf{M}_r\mathbf{v}_r^n -\tilde{\mathbf{f}}_{a}\Delta t - \tilde{\mathbf{f}}_r \Delta t - \mathbf{J}_r^T\mathbf{M}\mathbf{u}^*
        \end{bmatrix}.
    \end{aligned}
    \label{eq:coupling}
\end{equation}
After we compute $\hat{\mathbf{p}}$, we compute $\mathbf{u}^{n+1}$ using Eq.~\ref{eq:fluid:pp}.

%% file: Figures/Cherrio.tex
\begin{figure*}[t]
    \newcommand{\formattedgraphics}[1]{\includegraphics[width=0.19\textwidth,trim= 0 0 0 0,clip]{#1}}
    \centering
    \formattedgraphics{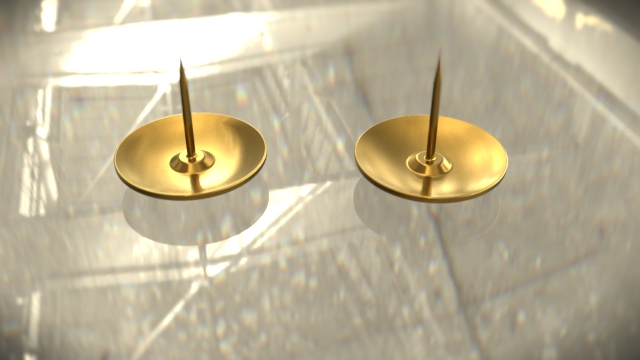}
    \formattedgraphics{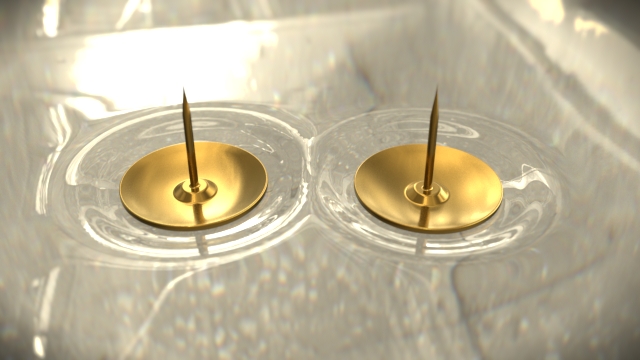}
    \formattedgraphics{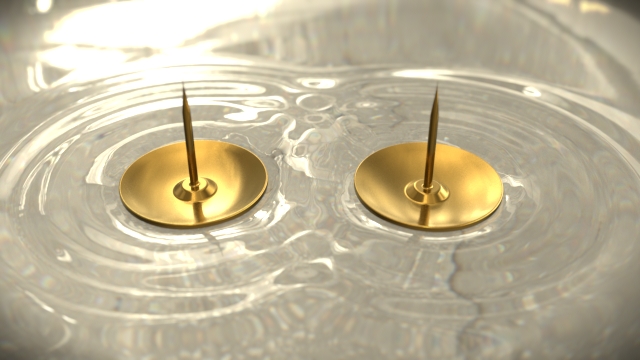}
    \formattedgraphics{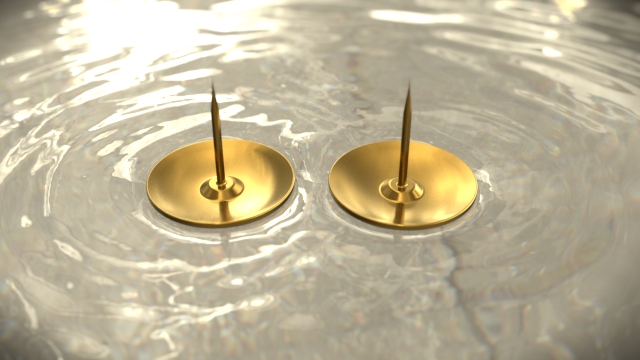}
    \formattedgraphics{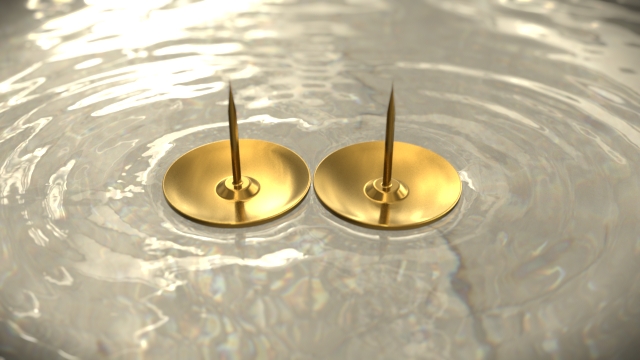}
    \caption{Attraction between Two Pushpins. Water surface between hydrophobic objects is always slightly lower compared to the rest of the surface. Thus, the capillary force along the perimeter points inward, which forces the pushpins to move slowly toward each other.}
    \label{fig:pushpin-attraction}
\end{figure*}

%% file: Sections/time_scheme.tex
\section{Time Integration Scheme}
\label{sec:ts}

\begin{figure*}[htp]
    \centering
    \includegraphics[width=0.98\textwidth]{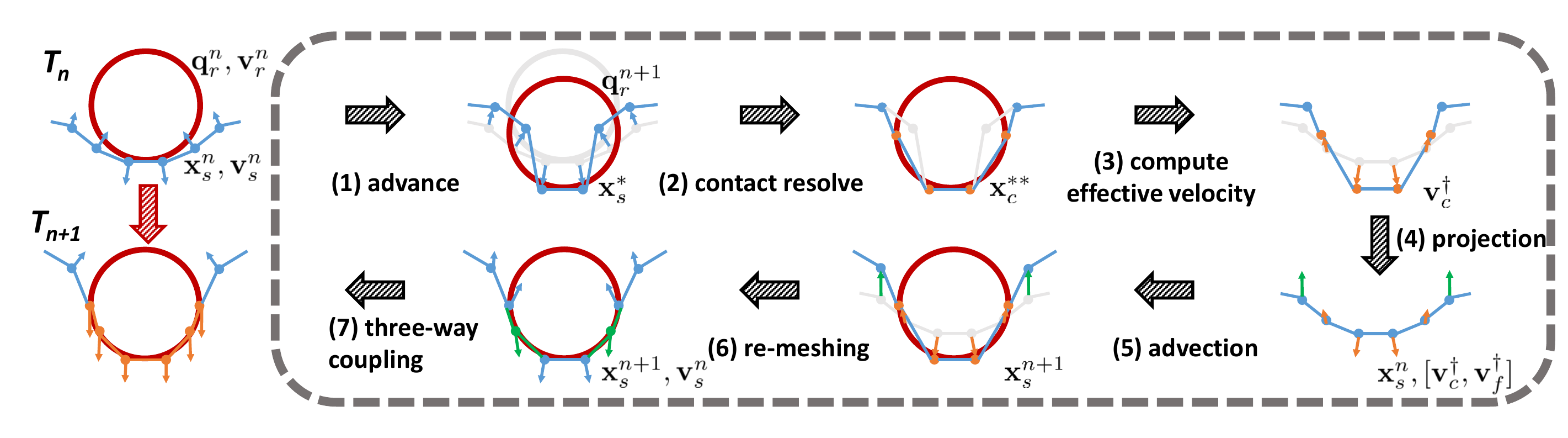}
    \caption{ Our prediction-correction time scheme. In each timestep: (1) Advance the rigid body's and the membrane's position; (2) Resolve contact between the membrane and the rigid body by projecting the contact particles' position; (3) Compute contact particles' effective velocities to resolve interpenetration; (4) Solve the F-M coupling system to get membrane and fluid's advection velocities that guarantee volume preservation of the fluid; (5) Advect the fluid velocity and the membrane's position; (6) Remesh and re-categorize the mesh particles' and grid cells' types; (7) Construct and solve the three-way coupling system. After step (4), we reach a coherent advection velocity that leads to a collision free, leakage free position without any fluid volume loss; after step (7), the three-way coupling system is solved to guarantee momentum preservation, volume preservation, and interpenetration free velocities. }
    \label{fig:time_scheme}
\end{figure*}

We first give an overview of our time integration scheme.
We devise a prediction-correction scheme to resolve the interaction between the fluid, membrane, and rigid body in each time step. In the prediction step, we update the membrane and the rigid body's positions separately without taking their interaction into consideration. In the correction stage, we first resolve the potential penetration that may have occurred between the membrane and the rigid body by correcting the membrane's contact particles' positions. 
Then we enforce fluid incompressibility by solving the F-M coupling system. After this two-way coupling solve, we obtain a coherent advection velocity that leads to a collision-free, leakage-free positions without any fluid volume loss. Lastly, the three-way coupling system is solved to guarantee momentum preservation, volume preservation, and interpenetration free velocities.

In each timestep, the algorithm updates the states of the system using the following steps, as depicted in Fig.~\ref{fig:time_scheme}:
\begin{enumerate}[(1)]
	\item \label{stp:advance} Advance the positions of the rigid body and the membrane particles for one time step. 
	\begin{equation}
	    \begin{dcases}
	    \mathbf{q}_r^{n+1} = \mathbf{q}_r^n + \mathbf{v}_r^n\Delta t\text{,}\\
	    \mathbf{x}_s^* = \mathbf{x}_s^n + \mathbf{v}_s^n\Delta t\text{.}
	    \end{dcases}
	\end{equation}
	
	\item \label{stp:contactsolve} Perform collision detection and resolve contact by projecting the membrane's contact particles $\mathcal{C}$'s positions onto the nearest rigid body surface as $\mathbf{x}_c^{**}$.  
	
	\item \label{stp:effective} Compute the effective advection velocities $\mathbf{v}_c^\dagger$ of the contact particles $\mathcal{C}$ using
	\begin{equation}
	\mathbf{v}_c^\dagger = \frac{(\mathbf{x}_c^{**} - \mathbf{x}_c^{n})}{\Delta t}.
	\end{equation}
	
	\item \label{stp:imcompressibility} Revert the membrane's position back to $\mathbf{x}_s^n$. Then we solve the F-M coupling system Eq.~\ref{fluid_shell}, but remove the surface tension part (i.e., remove all the $\mathbf{f}_c$ and $\mathbf{K}_c$) and use $\mathbf{v}_c^\dagger$ as the Neumann boundary condition, to get $\hat{\mathbf{p}}^\dagger$ in all fluid cells and $\mathbf{v}_f^\dagger$ for all free particles. By solving this coupling system, we get the fluid's advection velocity
	\begin{equation}
	    \mathbf{u}^\dagger = \mathbf{u}^n-\frac{1}{\rho}\hat{\mathbf{p}}^\dagger,
	\end{equation}
	and also the membrane's advection velocity as $\mathbf{v}_s^\dagger=[\mathbf{v}_f^\dagger, \mathbf{v}_c^\dagger]$.
	$\mathbf{u}^\dagger$ and $\mathbf{v}_s^\dagger$ ensures that the fluid preserves volume and that the particles do not collide with the rigid body.
	
	\item \label{stp:advection} Advect the membrane's position $\mathbf{x}^n_s$ and the fluid velocity $\mathbf{u}^n$ respectively by the advection velocity $\mathbf{v}_s^\dagger$ and $\mathbf{u}^\dagger$ solved above:
	\begin{equation}
	\begin{dcases}
	\mathbf{x}_s^{n+1}=\mathbf{x}_s^n+\mathbf{v}_s^\dagger\Delta t,\\
	\frac{\mathbf{u}^*-\mathbf{u}^n}{\Delta t} + \mathbf{u}^\dagger\cdot \nabla \mathbf{u}^n = 0.
	\end{dcases}
	\end{equation} \\
    Here we follow the standard semi-Lagrangian advection scheme for the fluid.
	
	\item \label{stp:remeshing} Remesh the membrane, and then update the grid cell types and mesh particle types. Our meshing algorithm relies on a set of local operations for topological repair on a triangle mesh. The main idea follows the series of previous works such as EI Topo \cite{brochu09}. Then we revert the membrane velocity back to the beginning of the current timestep as $\mathbf{v}_s^n$, because $\mathbf{v}_s^\dagger$ is only used to update the membrane's position. We also perform velocity interpolation on particles newly created by the mesher.
	
	\item \label{stp:3way} Apply external force $\mathbf{f}$ (gravity) to the fluid velocity $\mathbf{u}^*$ and the rigid body's velocity $\mathbf{v}_r^n$:
	\begin{equation}
	\begin{dcases}
	\mathbf{u}^{**} = \mathbf{u}^*+
	\textbf{f}\Delta t \\
	\mathbf{v}_r^* = \mathbf{v}_r^n + \textbf{f}_r\Delta t.\\
	\end{dcases}
	\end{equation}
	Then we construct and solve the three-way coupling system. Finally, the velocities of contact particles $\mathcal{C}$ are enforced by the corresponding rigid body's velocity. 
	
\end{enumerate}

%% file: Sections/result.tex
\section{Results}
\input{Figures/validation}
\input{Figures/ellipsoid}
\input{Figures/2D}
\input{Figures/boat}
\input{Figures/Sink}
\input{Figures/cherry}
\input{Figures/waterstrider}
In this section, we first describe experiments that help to validate the correctness of our algorithm. Then, we demonstrate the efficacy of our method through an array of rigid-fluid contact simulations dominated by strong surface tension.

\subsection{Validation}
\label{sec:validation}
\paragraph{Membrane Thickness}
In our system, the membrane is not massless but has a virtual thickness $h$. In our experiments, we find that $h$ around $0.5\Delta x$ is a reasonable choice, where $\Delta x$ is the grid cell size. As $h$ increases, the mass and the inertia of the membrane grow accordingly. This will slow down the membrane's local deformation propagation, and consequently, make the surface tension effects less obvious. This can be observed from the comparison result in Fig.~\ref{fig:validation}, in which $h = 0.5\Delta x$ and $h=10\Delta x$ are used respectively.
\paragraph{Comparison with Level Set}
We compare our membrane's representation of surface tension with the level set's pressure jump method through a bouncing-droplets test shown in Fig.~\ref{fig:ellipsoid}. It shows if we choose $h=0.5\Delta x$ or $h=\Delta x$ the droplet simulated using our method can bounce almost synchronously with the one with the level set, but if we set $h=10\Delta x$ the oscillation is significantly slower.  

\paragraph{One-time Solve v.s.\ Two-time Solve}
Our algorithm needs to solve two systems within every single timestep: (1) solving the F-M two-way coupling system to find the effective advection velocity; and (2) solving the three-way coupling system to satisfy all velocity constraints. This is necessary to accommodate our special treatment on contact particles $\mathcal{C}$. As we enforce both the position and the velocity constraints of the contact particles, solving only the three-way coupling system will make the contact particles unable to separate from the rigid body in contact by themselves. A comparison between one-time solving (2 only) and two-time solving (1 and 2) is shown in Fig.~\ref{fig:validation}. Without two solves, there are obvious sticking artifacts.
\paragraph{Surface Tension Coefficient}
We validate the correctness of our surface tension model by setting up a series of simple 2D tests, see Fig.~\ref{fig:2d}. The first row is to verify a classic phenomenon called the ``cheerios effect.'' The water surface between two hydrophobic objects will be slightly lower compared to its surrounding, causing an unbalanced force that drives both objects toward each other. The second and third rows aim at comparing the sliding performance of 2D articulated robots between low and high surface tension fluid. Many aquatic creatures rely on surface tension to move and navigate. Our experiments show that high surface tension fluid can help our robot slide more easily. Sliding on a low surface tension fluid is more challenging, making the robot struggle to move forward. A similar phenomenon occurs when the robot tries to jump up from the fluid surface. High surface tension fluid can produce more counter-impulse, thus helping the robot jump higher.

\subsection{Examples}
\label{sec:examples}
The physical parameters and performance numbers for all following examples are
listed in Tab.~\ref{tab:experiments}. We rendered the liquid surface and rigid bodies with triangle meshes in Houdini [\citeyear{houdini}].
\paragraph{Sphere Falling into Water} Fig.~\ref{fig:ball-3d} shows a sphere falling into a tank of water from the air. The image sequence shows the transition of the sphere from dynamic to static. The surface tension effect is demonstrated by the smoothly curved water surface around the sphere. By modeling the membrane, small-scale wavelets can be naturally generated and diffused even at a low grid resolution.
\paragraph{Cherries Falling into Water and Milk}
Fig.~\ref{fig:cherry} shows the comparison of a complex geometry falling into two types of fluid with different surface tensions. When falling into a high surface tension fluid (water), the cherries will drain away more fluid during the impact and thus slow down significantly. The cherries can be supported at the interface by buoyancy and capillary forces. For milk, however, the surface tension is not strong enough to counter the impact and support the cherries. Eventually, the cherries will sink to the bottom.
\paragraph{Paperclips with Different Densities}
Previous demos only show objects with densities similar to the fluid. The power of surface tension, however, is the capability of supporting materials with a much larger density. Our solver can simulate the stable contact between water and objects with up to $8$ times higher density. Fig.~\ref{fig:paperclip} shows the water surface deformed by a set of paperclips with densities ranging from $2$ to $7.9$. The real density for a paperclip made of steel is about $8\,\text{g/cm}^3$. We can see from the refraction in the rendering that heavier paperclips deform the water surface with a longer range and with sharper curvature.
\paragraph{Breaking Surface and Sinking} When putting an overweight object on the fluid surface, it will first break the interface then sink. Our solver naturally handles the sinking of such objects and can simulate the progress of water gradually ``climbing'' onto the object until it completely submerges the object. Fig.~\ref{fig:sink} shows such progress.

\paragraph{Boat and Leaves} 
Our solver cannot directly solve the contact between thin shells and fluid surface, but with a small modification in the second step of our time scheme---forcing the contact particle to be projected to the bottom side of the thin shell when the particle penetrates it---we can make a thin shell such as a boat and leaves float on top of the water.
\paragraph{Attraction between Two Pushpins} Surface tension effect helps generate many interesting natural phenomena. Experiments show that two hydrophobic objects lying on the free surface tend to attract each other. Our simulator can produce this attraction effect, as shown in Fig.~\ref{fig:pushpin-attraction}. The water surface between hydrophobic objects is always slightly lower compared to the rest of the surface. Thus, the capillary force along the perimeter is pointing inward, which forces the objects to move slowly toward each other.
\paragraph{Water Strider Robot} Water striders live their lives on the water surface. All of their motions (stand, slide, jump, etc.) rely on surface tension. As shown in Fig.~\ref{fig:waterstrider}, we model a water strider as an articulated rigid body connected by appropriate joints. We kinematically move the joints, causing the insect to move forward. With their middle legs, they push down slightly on the surface of the water, and then push back on the slight ridge resulting from surface tension. When the middle legs finish pushing the water, they are lifted up and brought forward to continue the motion. As can be seen from the accompanying video, the contact points separate properly, allowing the leg to detach from the water without any hindrance.

\subsection{Performance}
We parallelize most of the steps in our method using OpenMP \cite{openmp}, and implement a GPU version of the Diagonal Preconditioned Conjugate Gradient solver using CUDA \cite{cuda}. Our 2D examples are performed on a PC with a 16-core 3.40 GHz
CPU and an NVIDIA GeForce RTX 2070 graphics card, which also perform the attraction and sinkage of push pin in 3D. Other 3D examples are performed on a workstation with a Xeon W-3175X CPU and an NVIDIA Quadro RTX 8000 graphics card.
We summarize our simulation setups, parameters, scale, and time consumption details in Tab.~\ref{tab:experiments}. The main bottlenecks are from remeshing and the two solves. Even though our system is indefinite, we use the Diagonal Preconditioned Conjugate Gradient (DPCG) method, with tolerance of 1e-4. We cannot conclude that a DPCG solver can facilitate positive indefinite systems in all cases, but it works well in practice for our examples.
\begin{table*}[t]
	\centering
	\vspace{-5pt}
	\caption{Simulation parameters for the examples.}
	\vspace{-8pt}
	\label{tab:experiments}
	\begin{threeparttable}
		\begin{tabular}{c|c|c|c|c|c|c|c}
			\hline
			Figure & 2D Scenes\tnote{\dag} & $\rho_r$ ($\text{g} / \text{cm}^2$) & Resolution & Cell Size ($\text{cm}$) & First Solve (s) & Second Solve (s) & Time/Step (s)\\ 
			\hline
			\ref{fig:2d}(a) & Sphere Attraction & 1.1 & $160\times160$ & 1.25e-2 & 0.044 & 0.047 & 0.17\\
			\ref{fig:2d}(b,c) & Robot Swimming\tnote{\ddag} & 1.1 & $320\times320$ & 6.25e-3 & 0.13 & 0.19& 0.69\\
			\ref{fig:2d}(d,e) & Robot Jumping\tnote{\ddag} & 1.1 & $320\times320$ & 6.25e-3 & 0.22 & 0.21& 0.75\\
			\hline
			 Figure & 3D Scenes & $\rho_r$ ($\text{g} / \text{cm}^3$) & Resolution & Cell Size ($\text{cm}$) & First Solve (s) & Second Solve (s)&Time/Step (s)\\
			\hline
			\ref{fig:ball-3d} & Sphere Falling into water & 0.4 & $80\times 80\times 40$ & 2.5e-1 & 0.05 & 0.15 &2.1\\
			\ref{fig:cherry} & Cherry Water/Milk\tnote{\S} & 1.1 & $160\times 160\times 80$ & 1.25e-2 & 0.32 & 0.71 &10\\ 
			\ref{fig:paperclip} & Paperclip & 2.0/5.0/7.9 & $256\times 256\times 64$ & 7.8125e-2 & 0.30 & 6.3 & 31\\
			\ref{fig:sink} & Push Pin Sinkage & 6.8 & $256\times 256\times 64$ & 6.25e-2 &  1.8 & 4.3& 22\\
			\ref{fig:pushpin-attraction} & Push Pin Attraction & 3.8 & $256\times 256\times 64$ & 6.25e-2 &  1.9 & 5.8& 25.8\\
			\ref{fig:boat} & Boat and Leaves & -\tnote{$\ast$} & $80\times 80\times 40$ & 2.5e-1 & 0.14 & 0.32 &1.5\\
			\ref{fig:waterstrider} & Water Strider Robot & 0.4 & $160\times 200 \times 40$ & 2.5e-1 & 0.31 & 3.5 & 13\\
			\hline
		\end{tabular}
		\begin{tablenotes}
			\footnotesize
			\item[\dag] For 2D scenes, the default surface tension coefficient is $\sigma=0.9\text{ dyn}$, fluid density is $\rho=1.0\text{ g/cm}^2$, gravity acceleration is $g=9.8 \text{ cm/s}^2$. For 3D scenes, default surface tension coefficient is $\sigma=72.8\text{ dyn/cm}$, fluid density is $\rho=1.0\text{ g/cm}^3$, gravity acceleration is $g=980\text{ cm/s}^2$.
			\item[\ddag] 2D robot motions are compared with small surface tension coefficient $\sigma = 0.3\text{ dyn}$.
			\item[\S] The surface tension coefficient for milk is $\sigma=46\text{ dyn/cm}$.
			\item[$\ast$] The surface density for thin shell is $0.1 \text{ g/cm}^2$.
		\end{tablenotes}
	\end{threeparttable}
\end{table*}

%% file: Figures/validation.tex
\begin{figure}[t]
    \newcommand{\formattedgraphics}[1]{\includegraphics[width=0.45\linewidth,trim= 10cm 6cm 10cm 19cm,clip]{#1}}
    \centering
    \subfigure[$h=0.5\Delta x; \text{Solve Twice}$]{
        \formattedgraphics{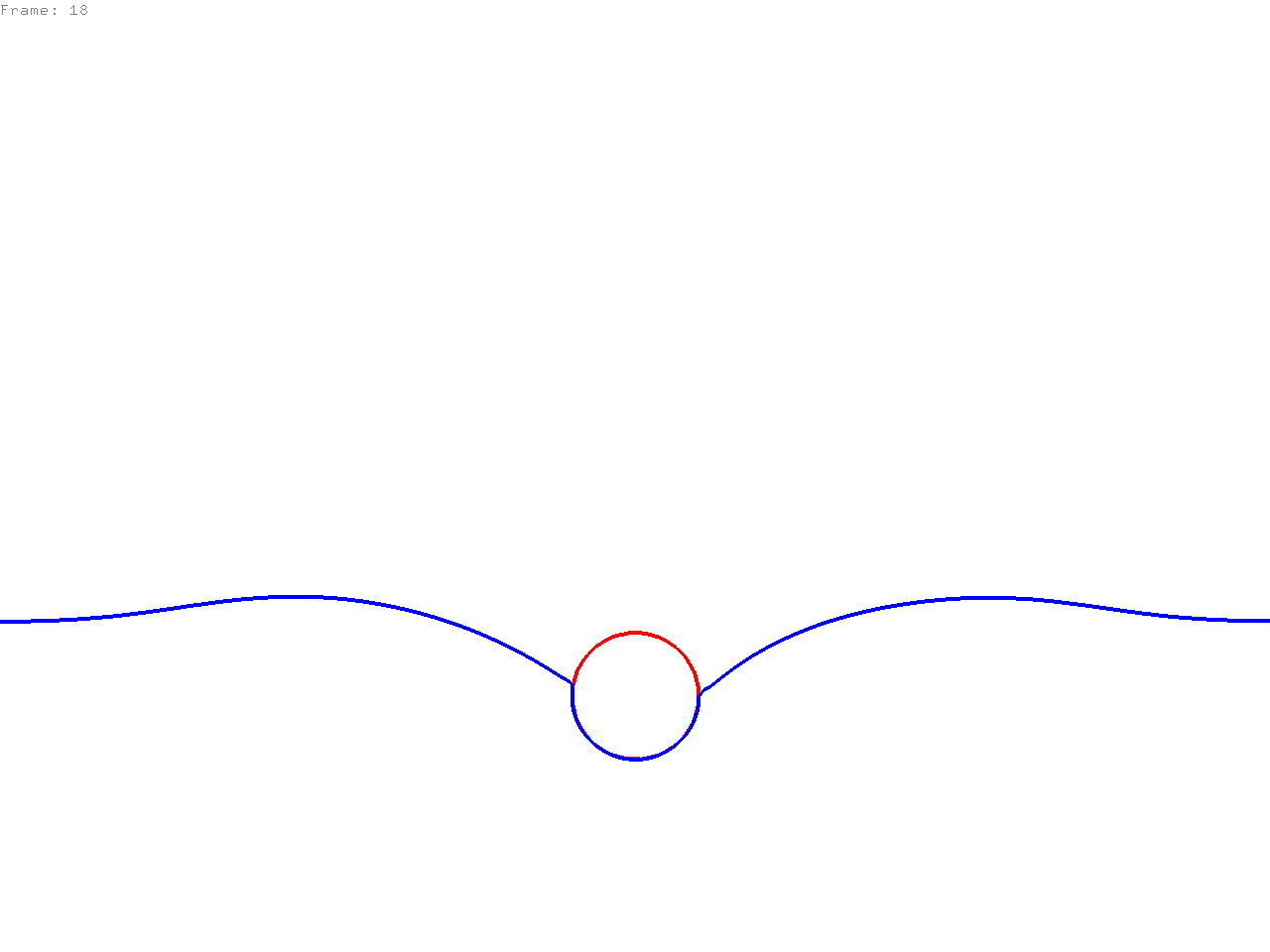}\hspace{0.3cm}
        \formattedgraphics{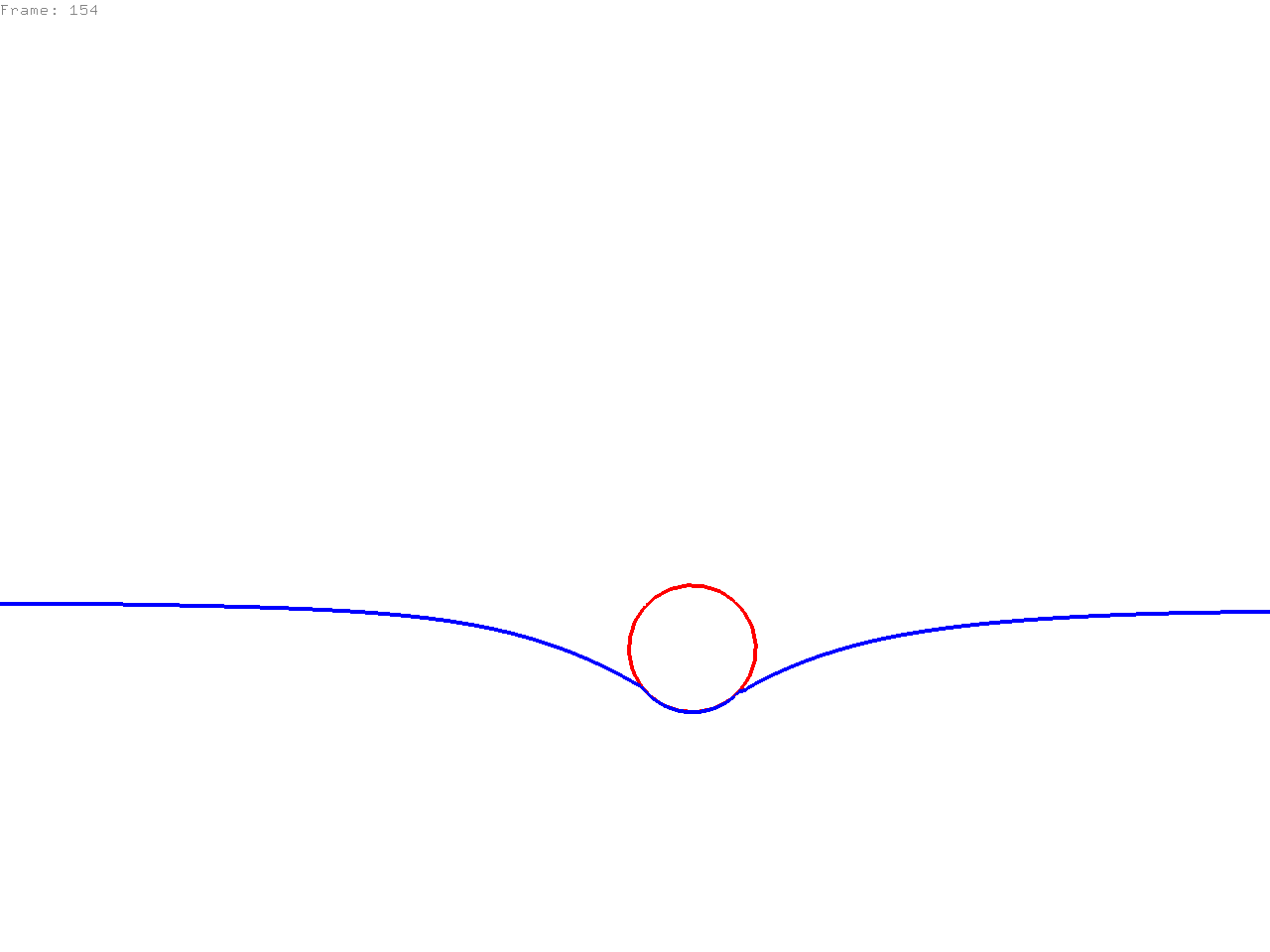}
    }
    \subfigure[$h=10\Delta x; \text{Solve Twice}$]{
        \formattedgraphics{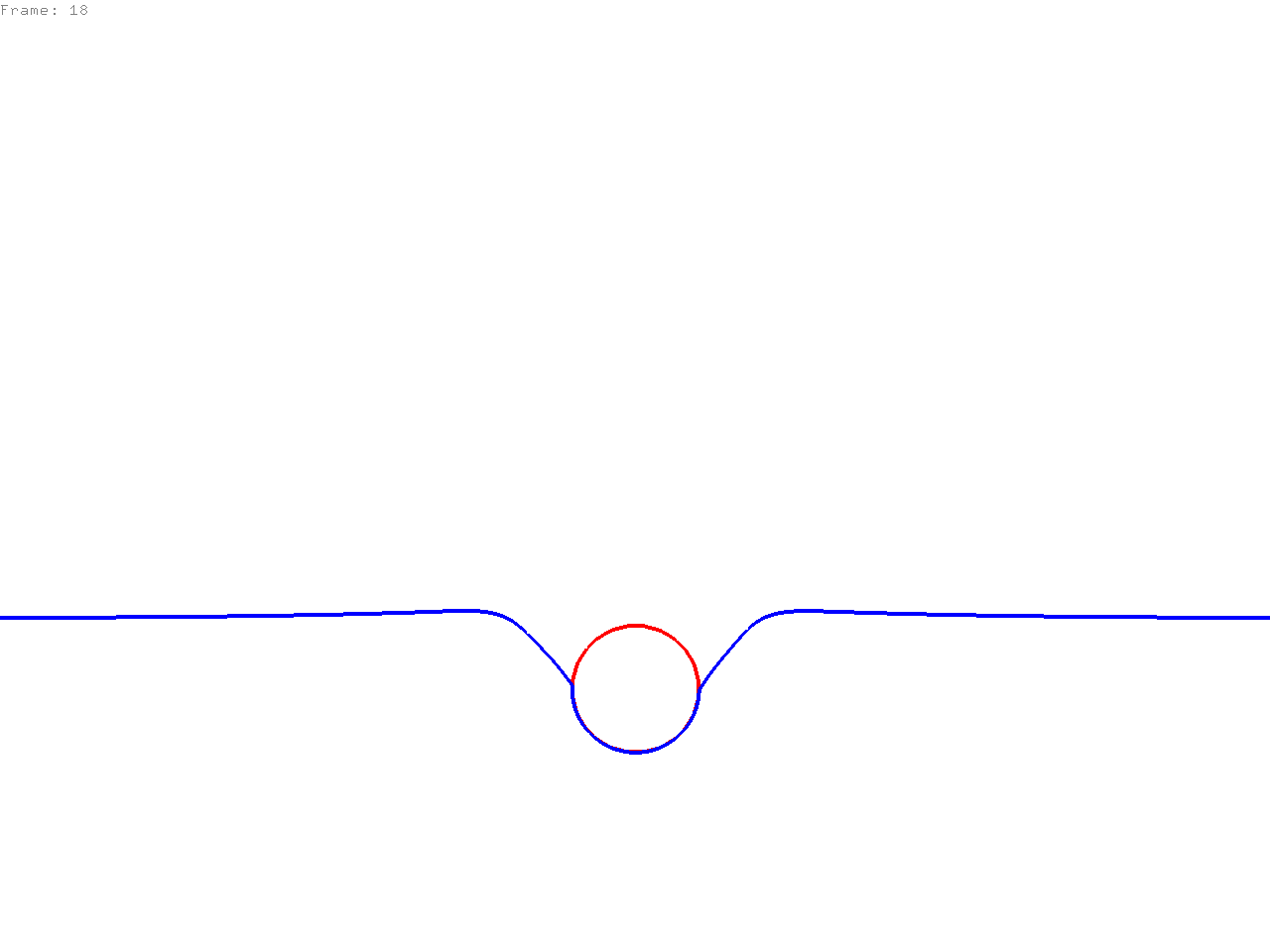}\hspace{0.3cm}
        \formattedgraphics{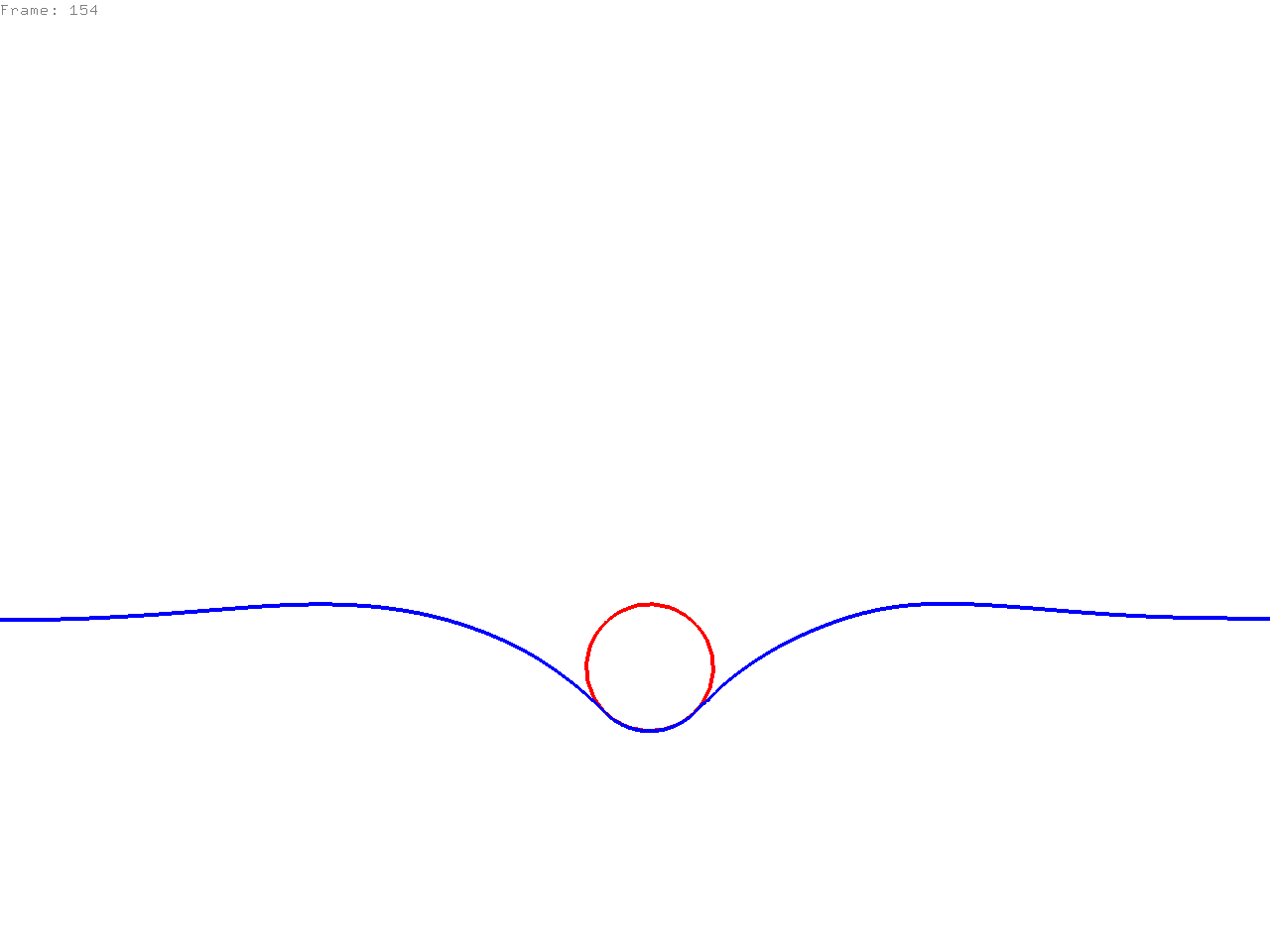}
    }
    \subfigure[$h=0.5\Delta x; \text{Solve Once}$]{
        \formattedgraphics{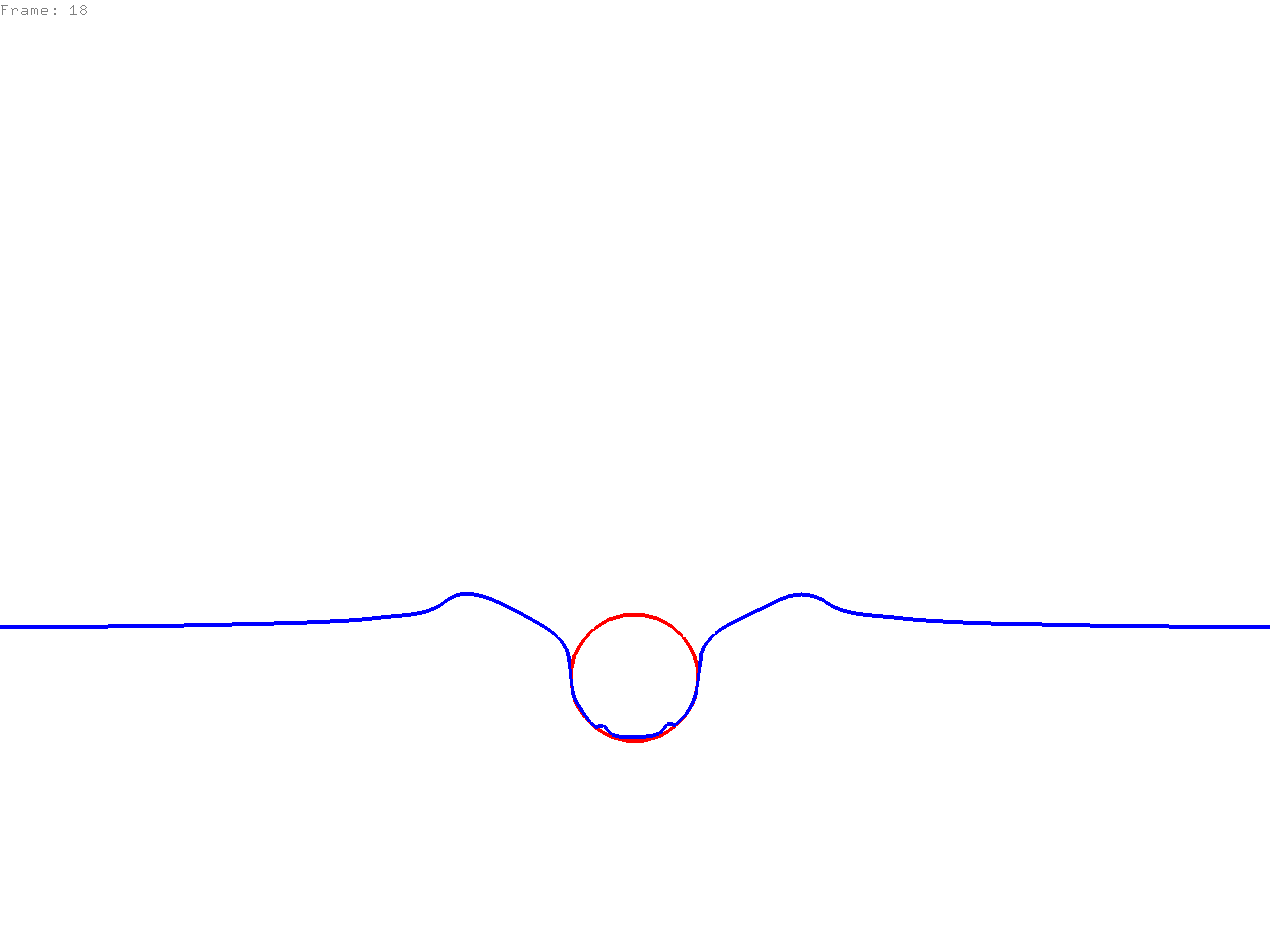}\hspace{0.3cm}
        \formattedgraphics{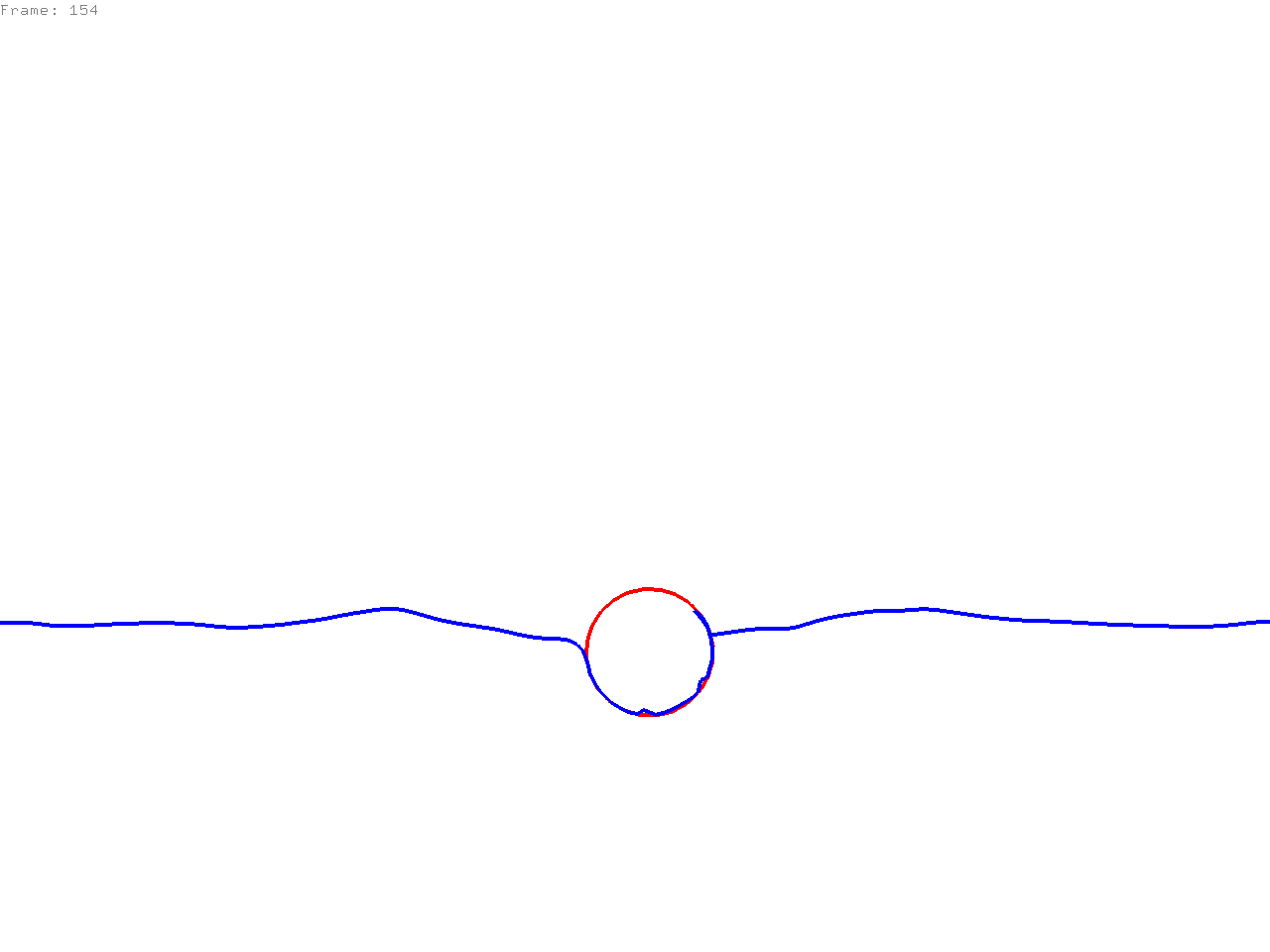}
    }
    \caption{Validation test. Each row shows two snapshots from our validation tests using different membrane thickness and solving schemes. The configuratio of (a) is the normal setup for all our examples; the membrane thickness of (b) is $20$ times larger, which makes the membrane surface deformation not propagate far enough; without two-time solve, the membrane surface in (c) has severe sticky artifacts.}
    \label{fig:validation}
\end{figure}

%% file: Figures/ellipsoid.tex
\begin{figure}
    \centering
    \includegraphics[width=0.98\linewidth]{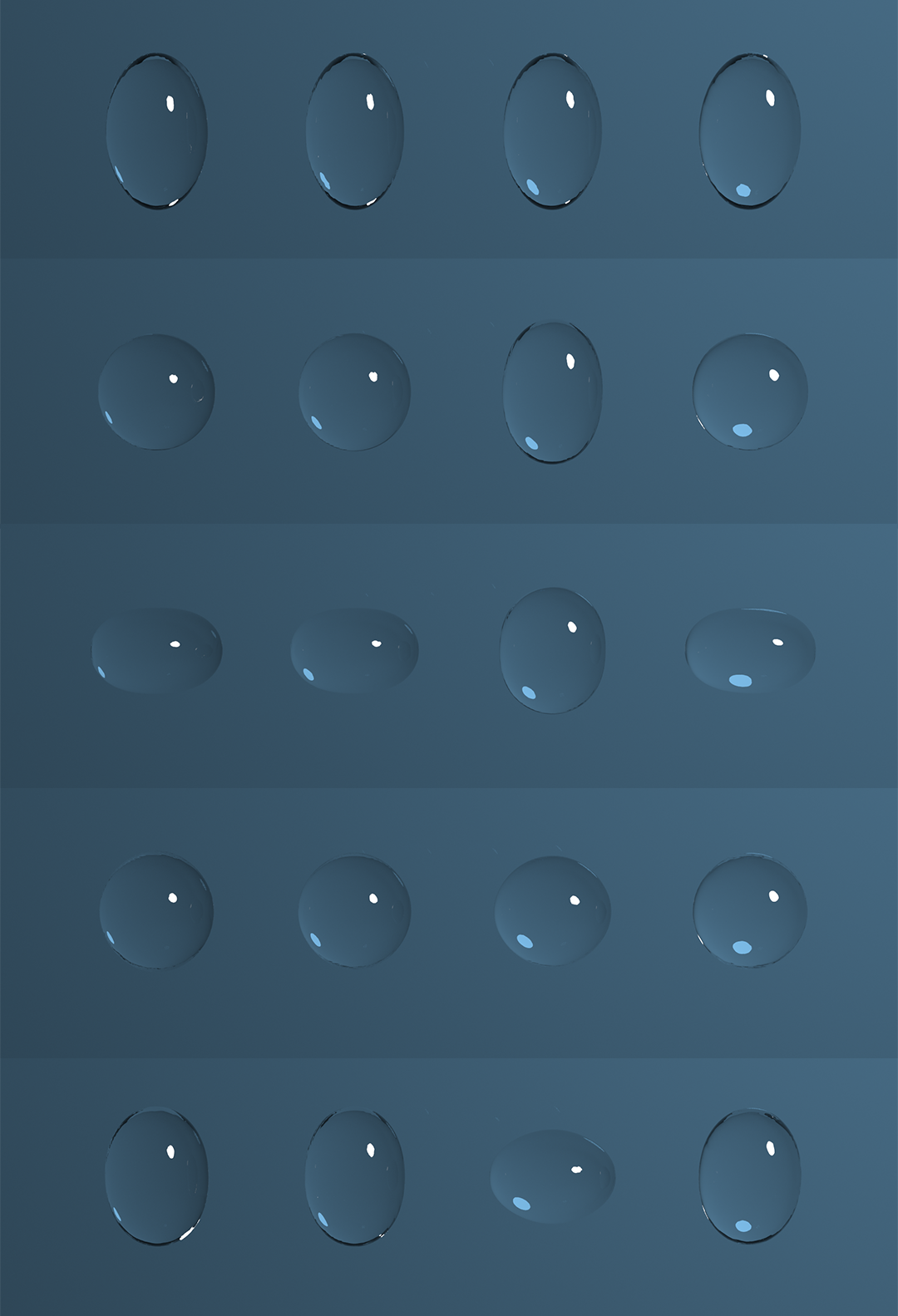}
    \caption{Bouncing Droplets. Here we show the 1st, 25th, 47th, 67th, and 88th frames of our simulation. The droplets are initialized to be in the same shape and size with $\sigma=6 \text{ dyn/cm}$. From left to right the four droplets are simulated with $h=0.5\Delta x$, $h=\Delta x$, $h=10\Delta x$ and the level set.   }
    \label{fig:ellipsoid}
\end{figure}

%% file: Figures/2D.tex
\begin{figure*}[t]
    \newcommand{\formattedgraphics}[1]{\includegraphics[width=0.24\textwidth,trim=8cm 5cm 8cm 5cm,clip]{#1}}
    \centering
    \subfigure[Sphere Attraction]{
        \formattedgraphics{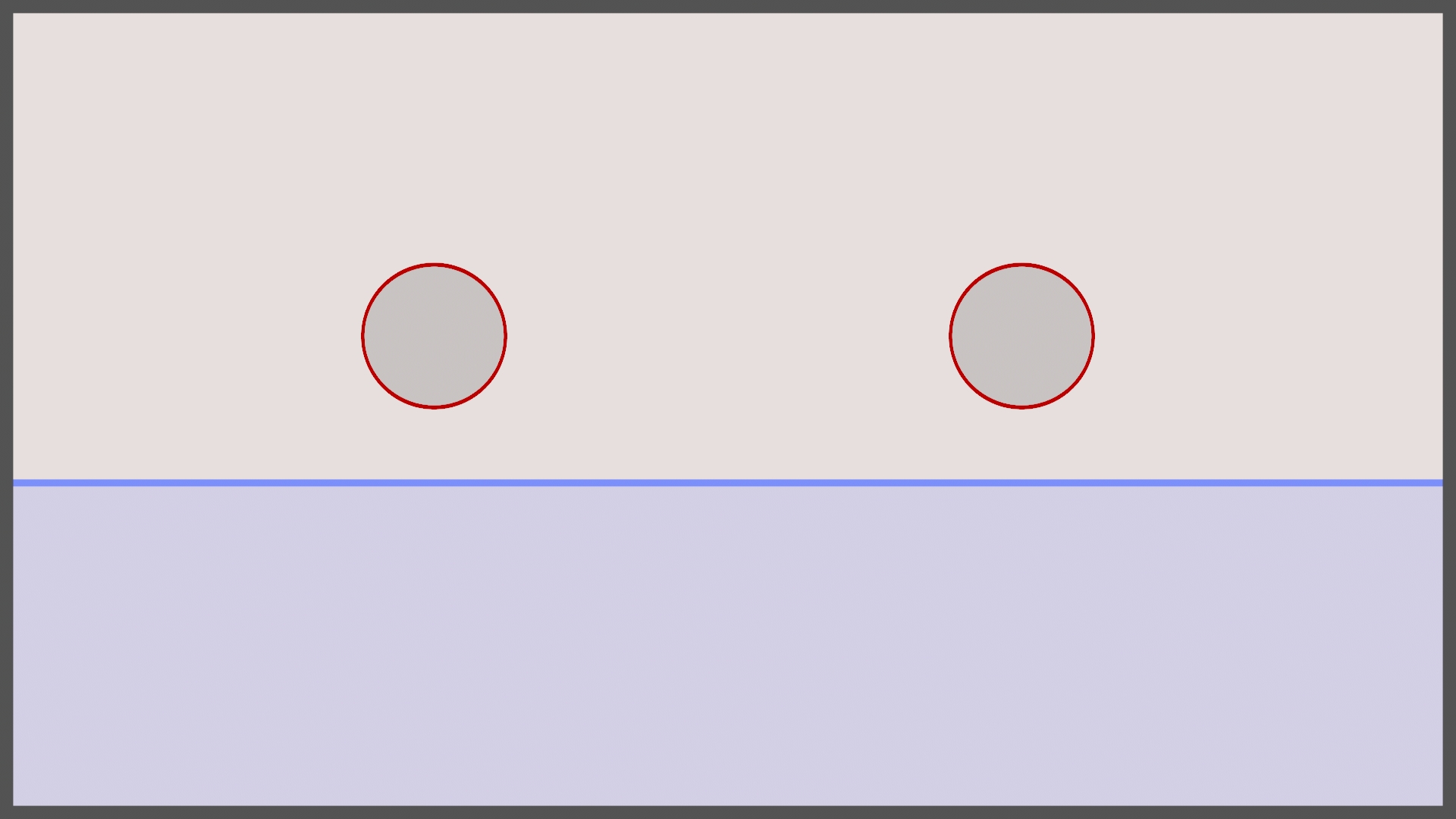}
        \formattedgraphics{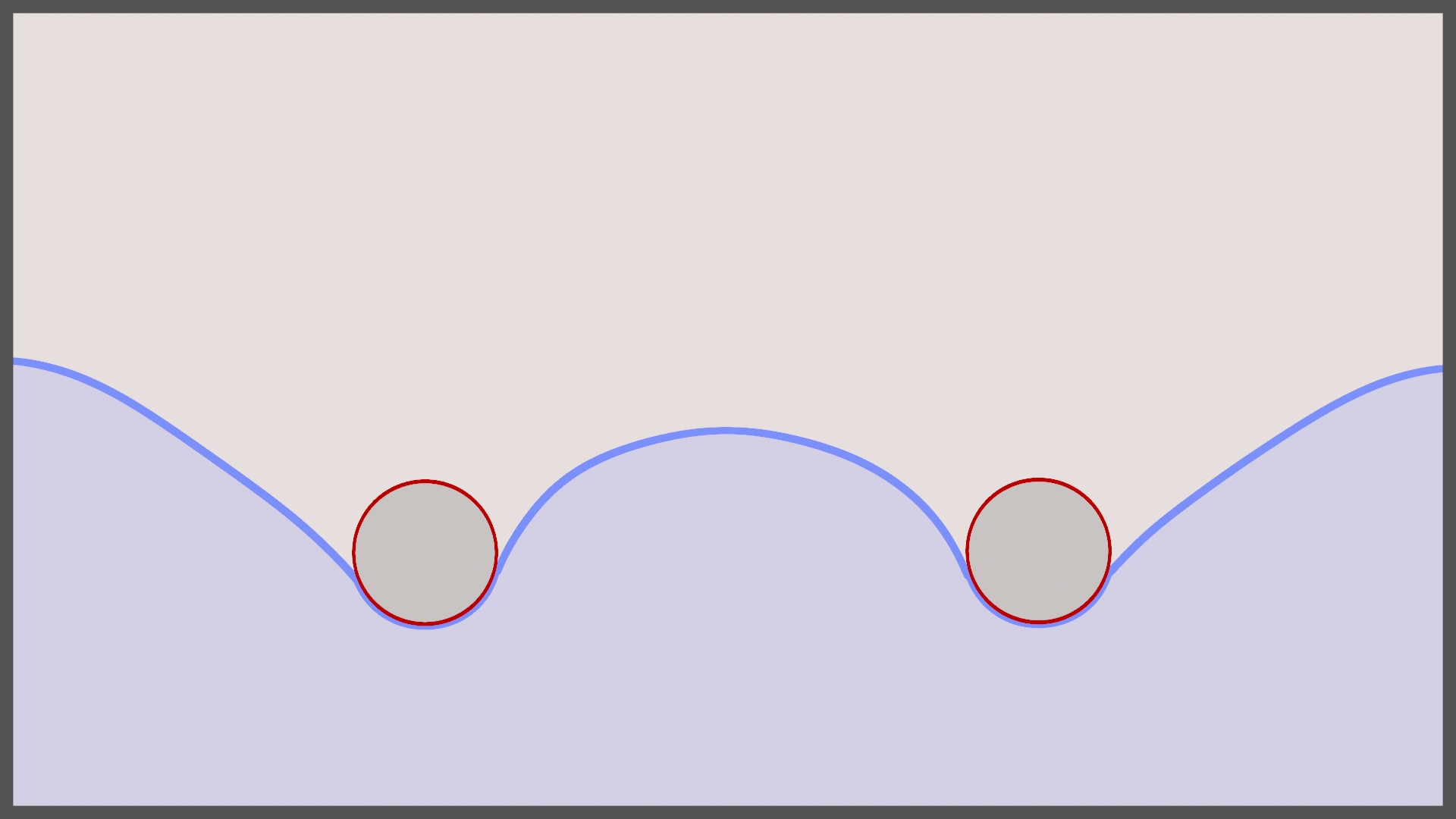}
        \formattedgraphics{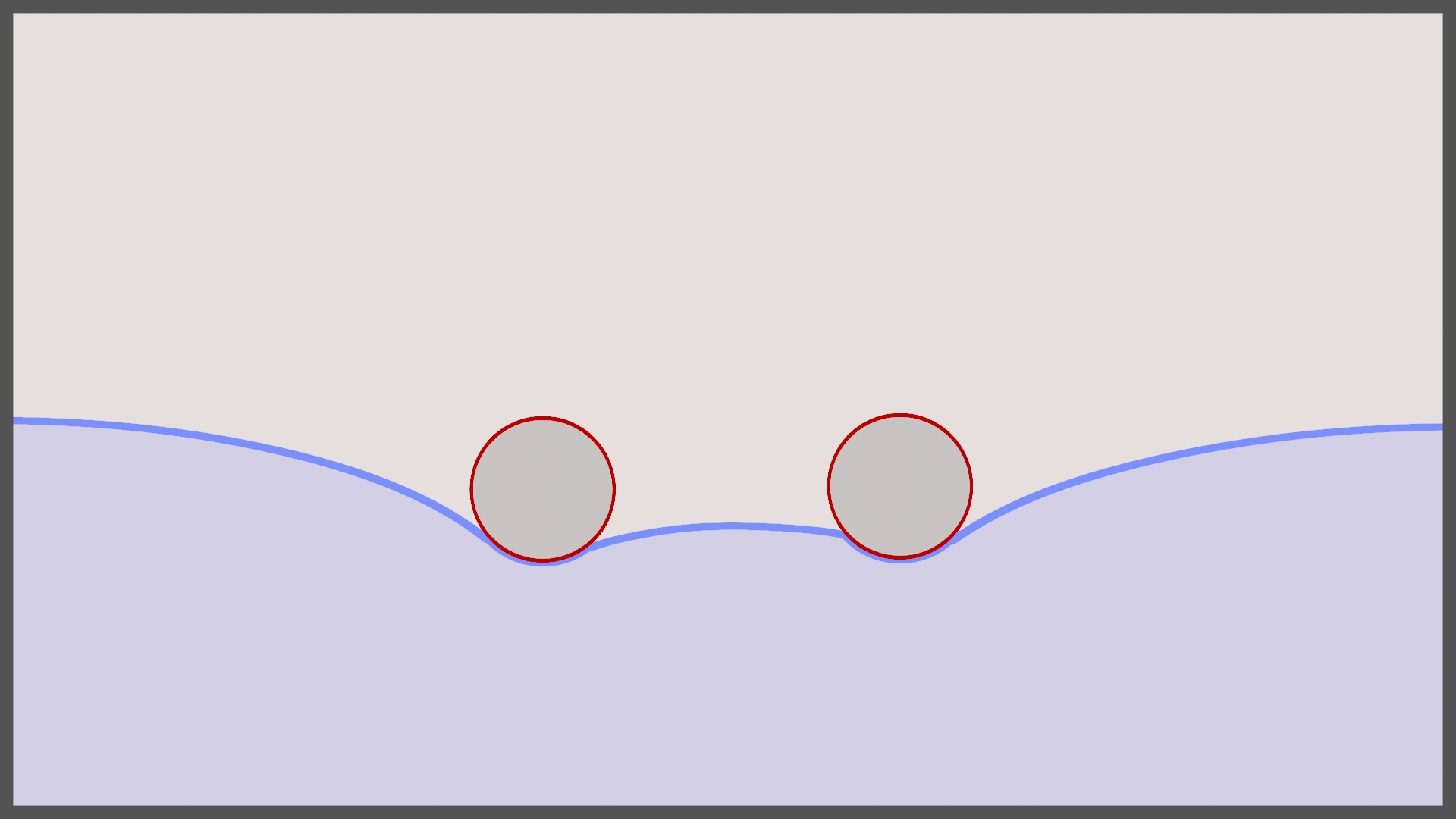}
        \formattedgraphics{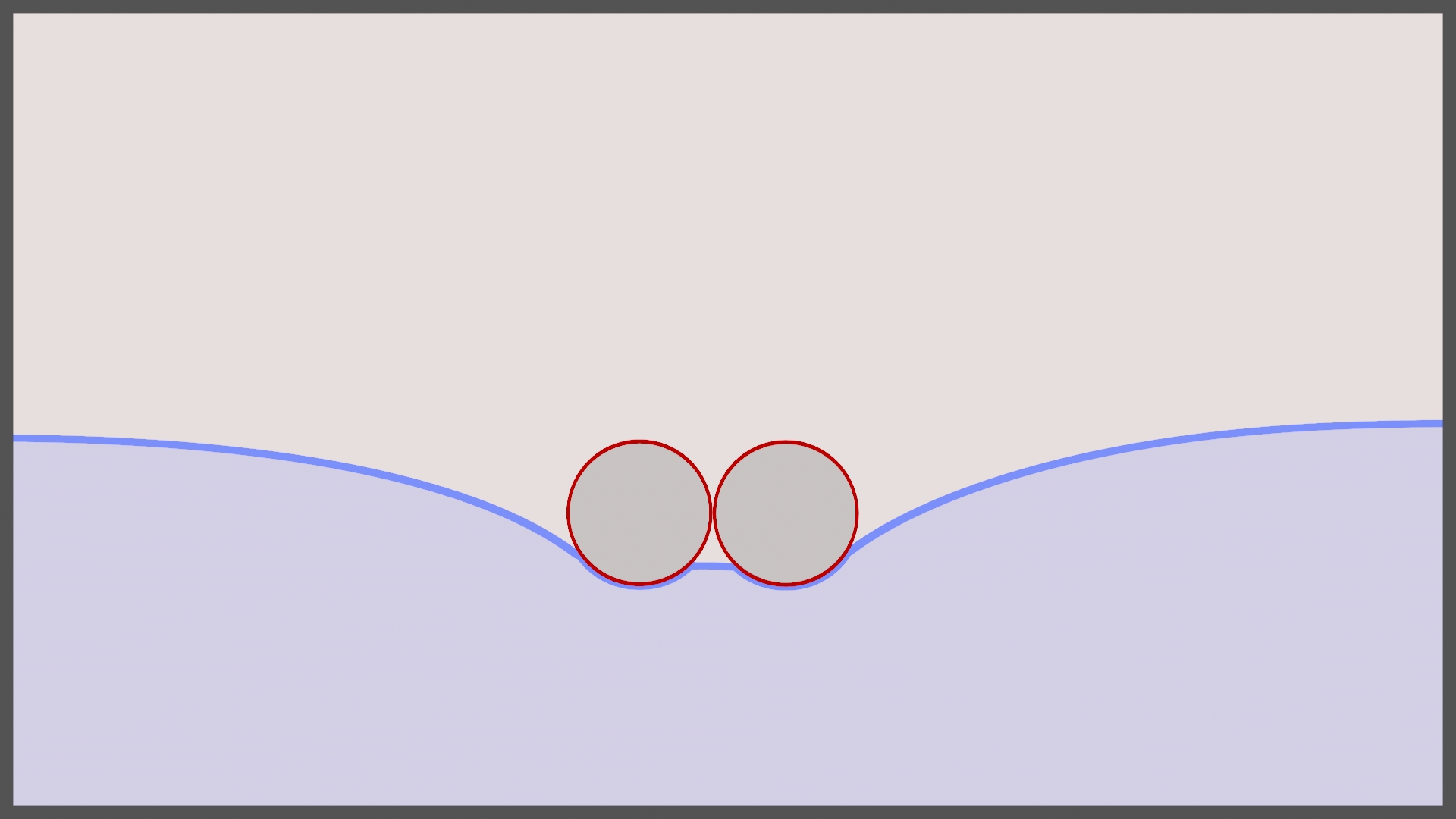}
    }
    \subfigure[Robot Swimming with Low Surface Tension]{
        \formattedgraphics{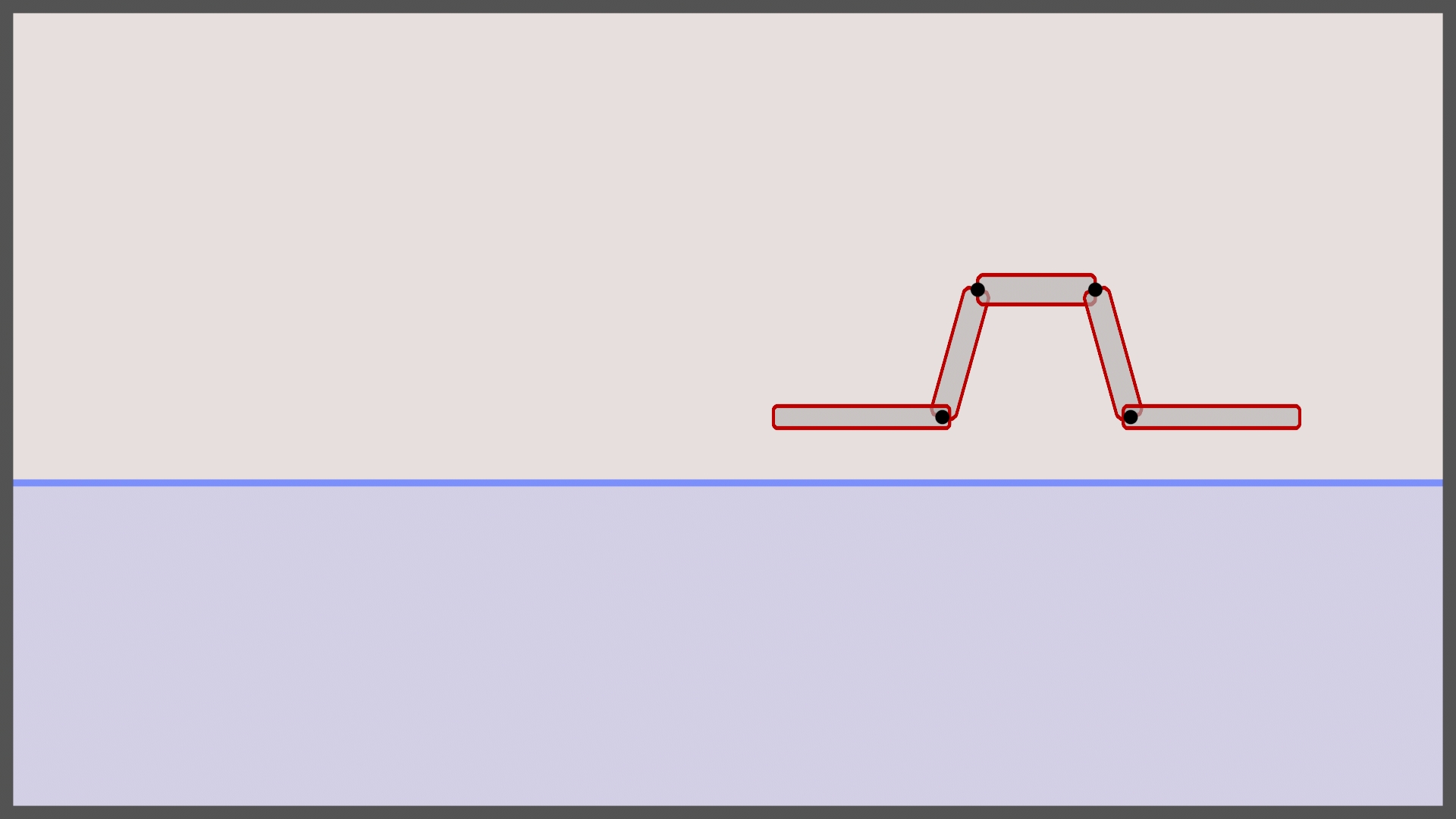}
        \formattedgraphics{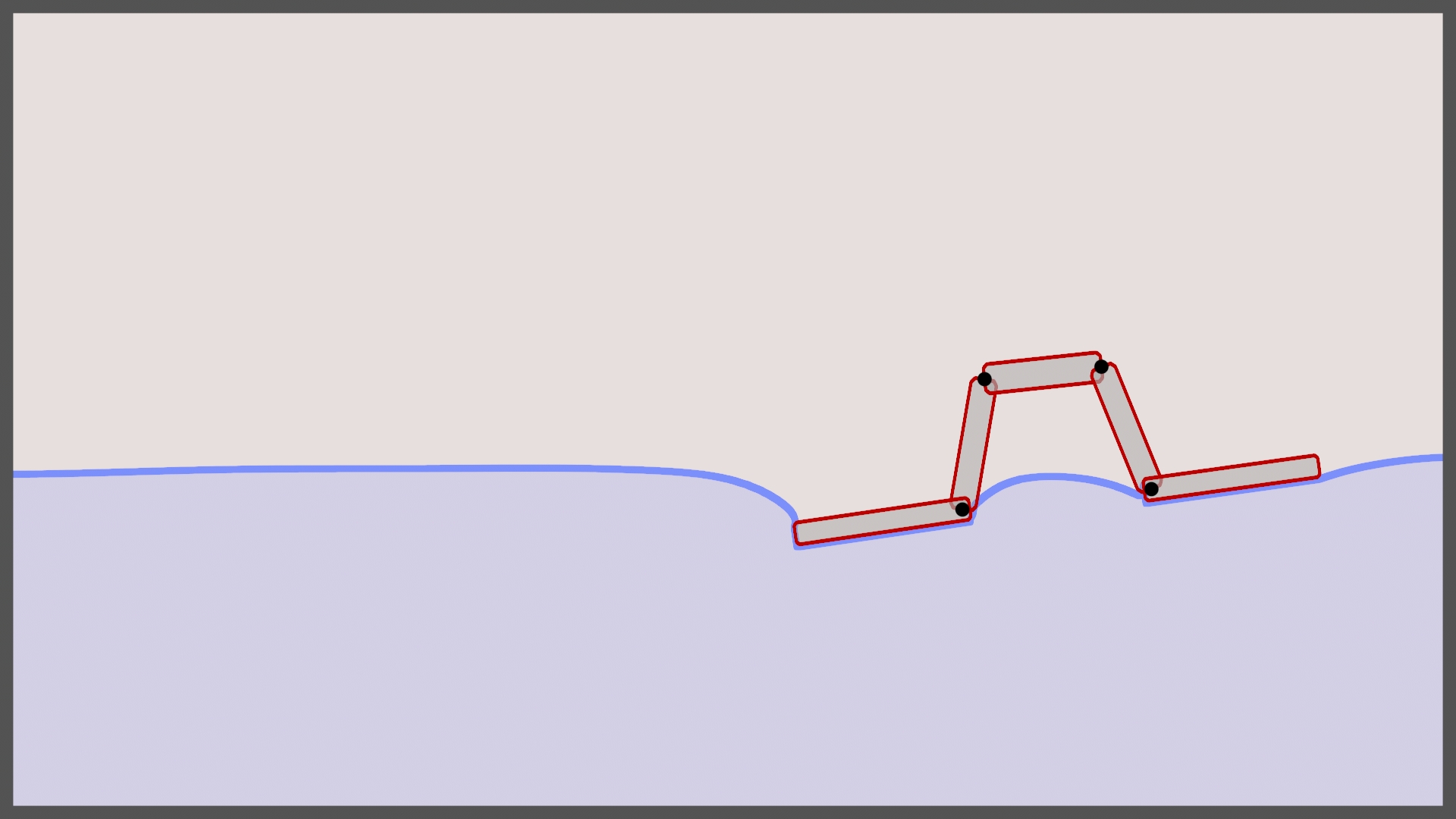}
        \formattedgraphics{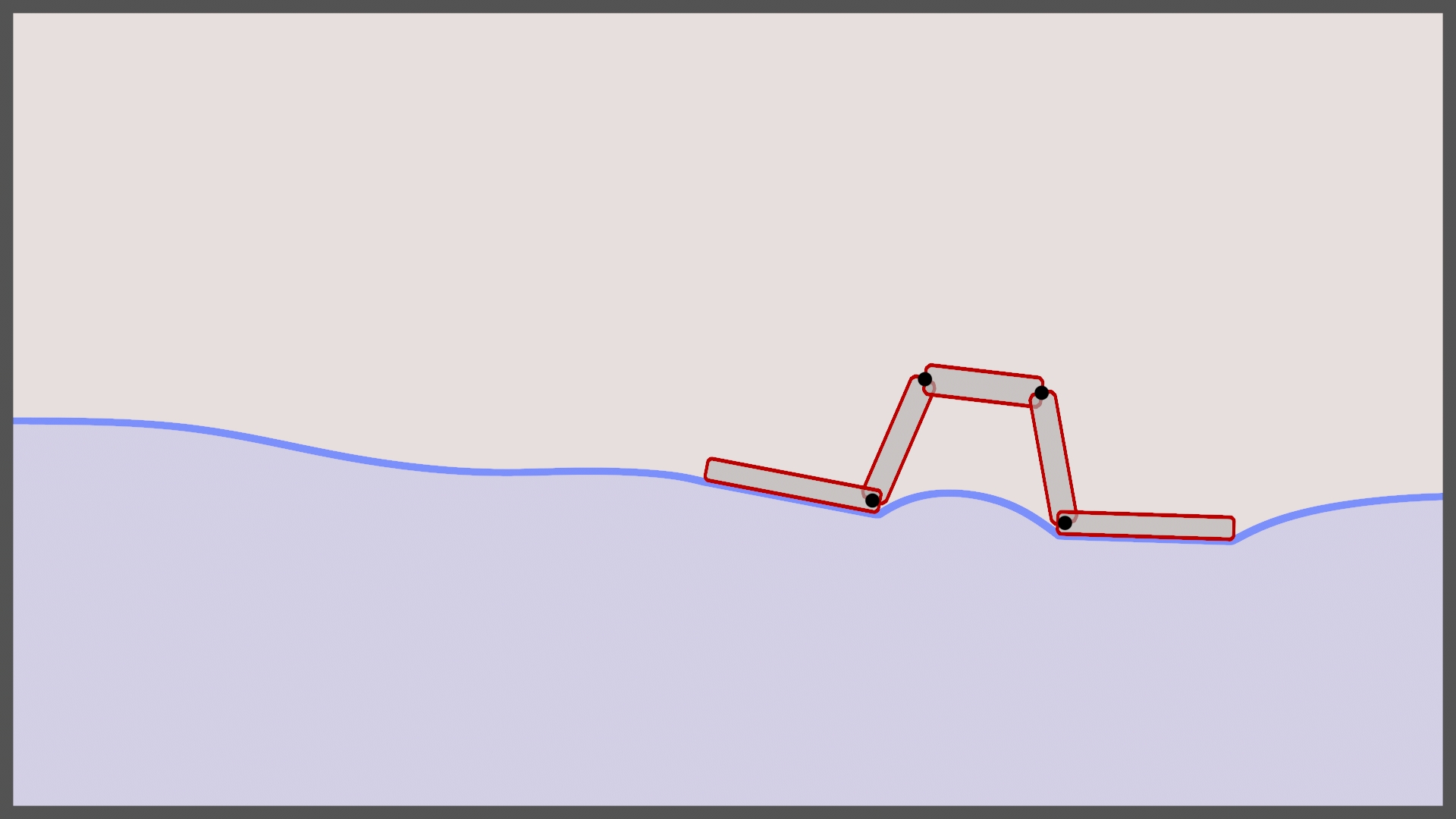}
        \formattedgraphics{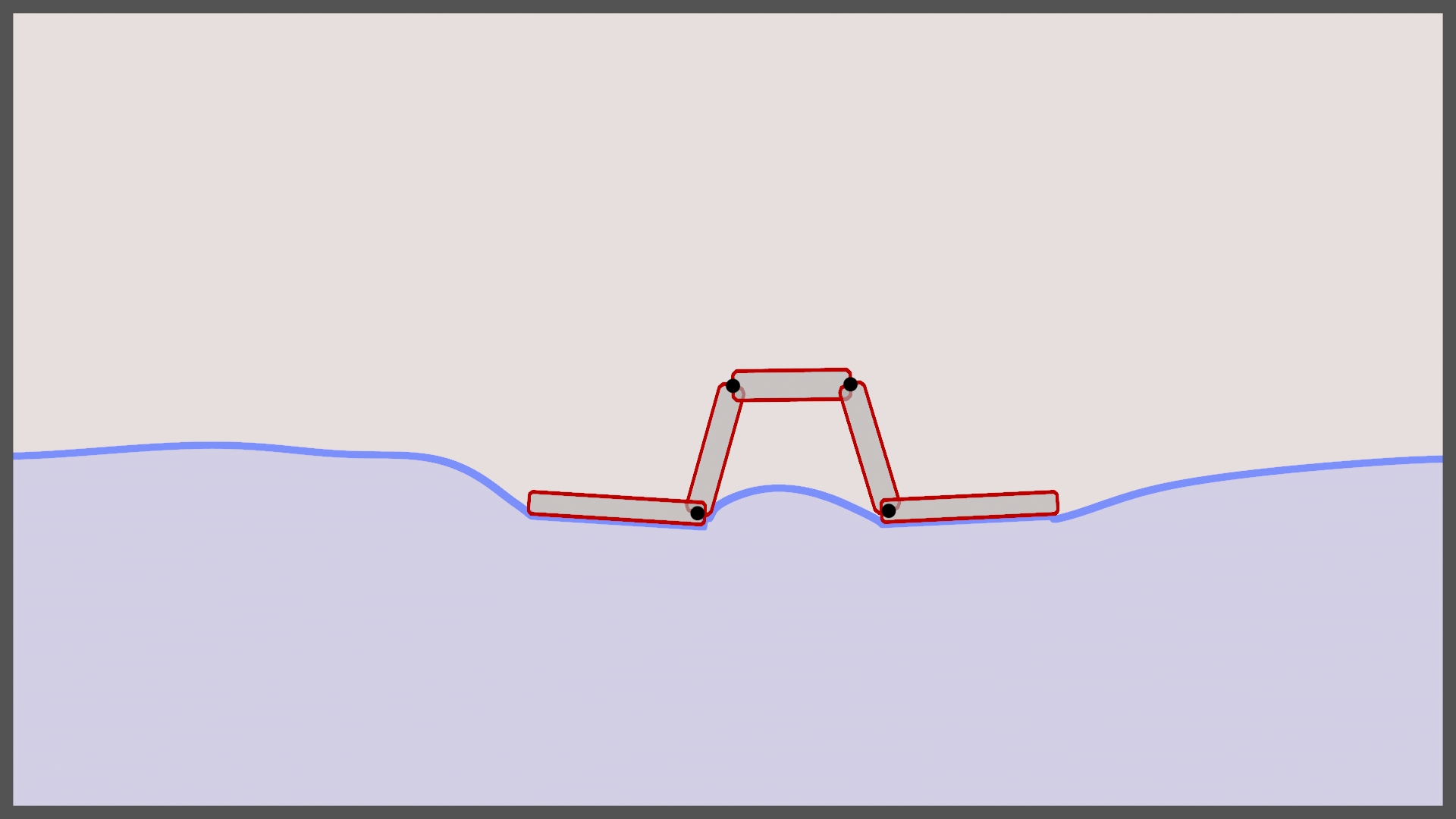}
    }
    \subfigure[Robot Swimming with High Surface Tension]{
        \formattedgraphics{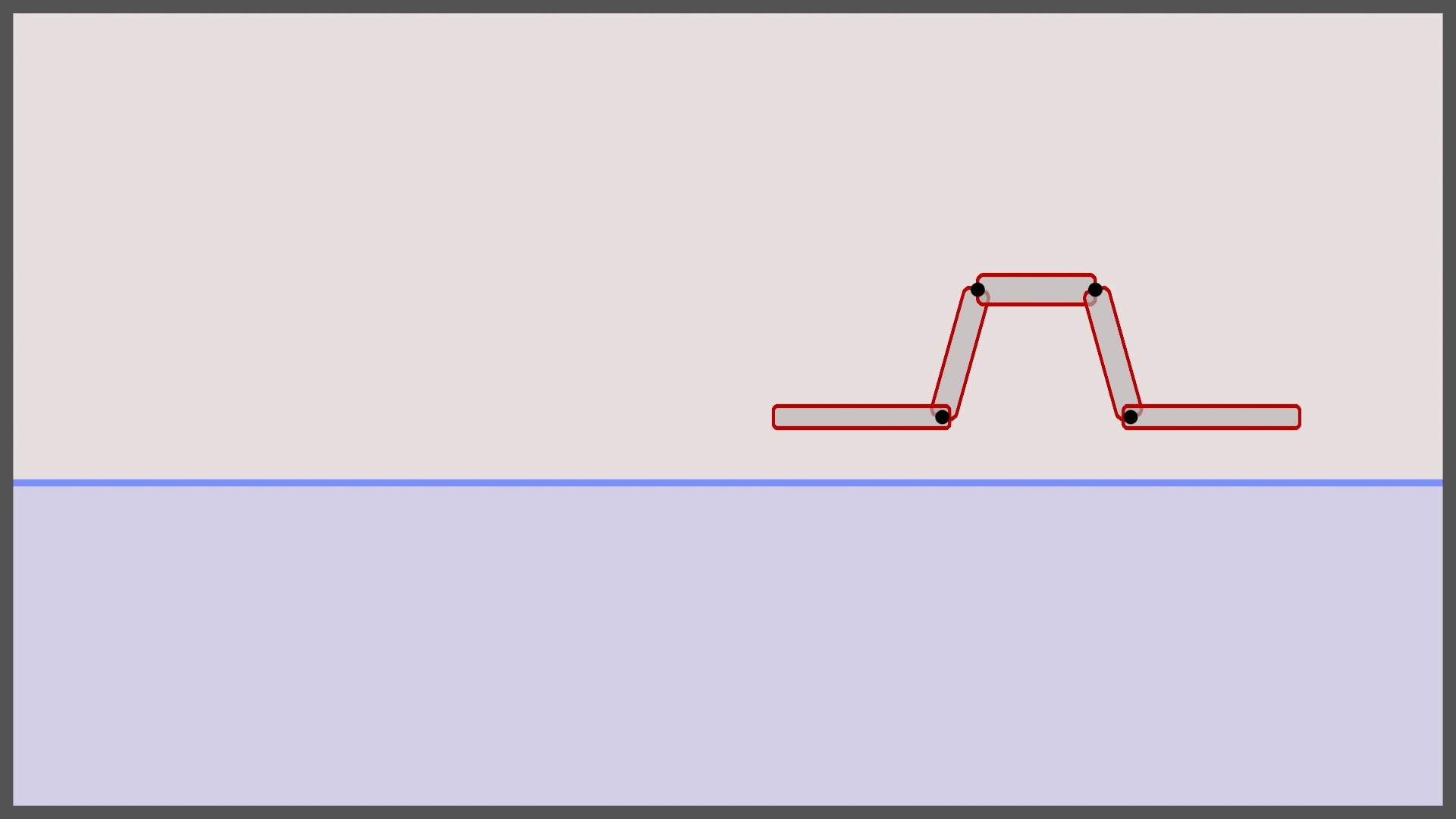}
        \formattedgraphics{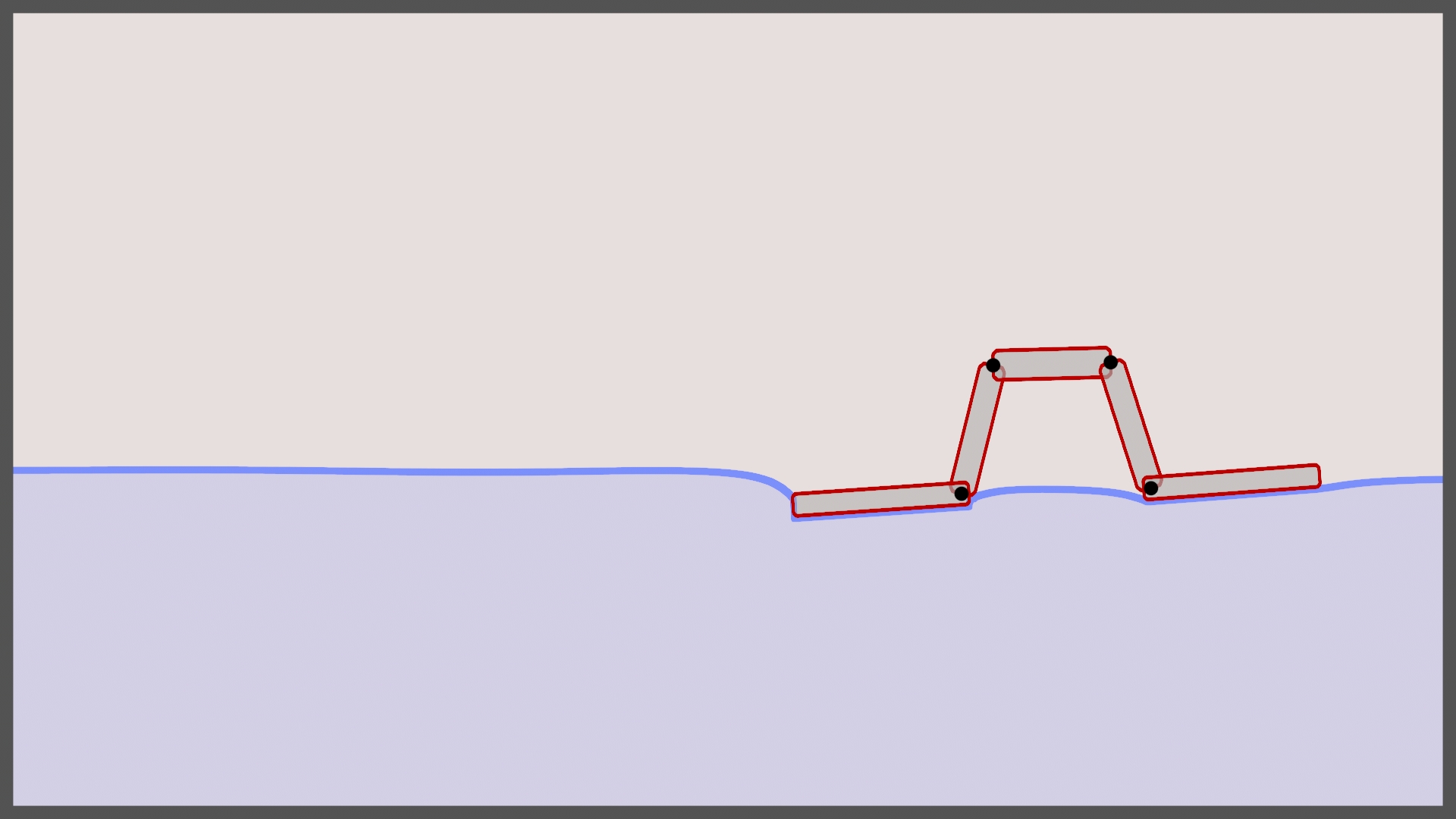}
        \formattedgraphics{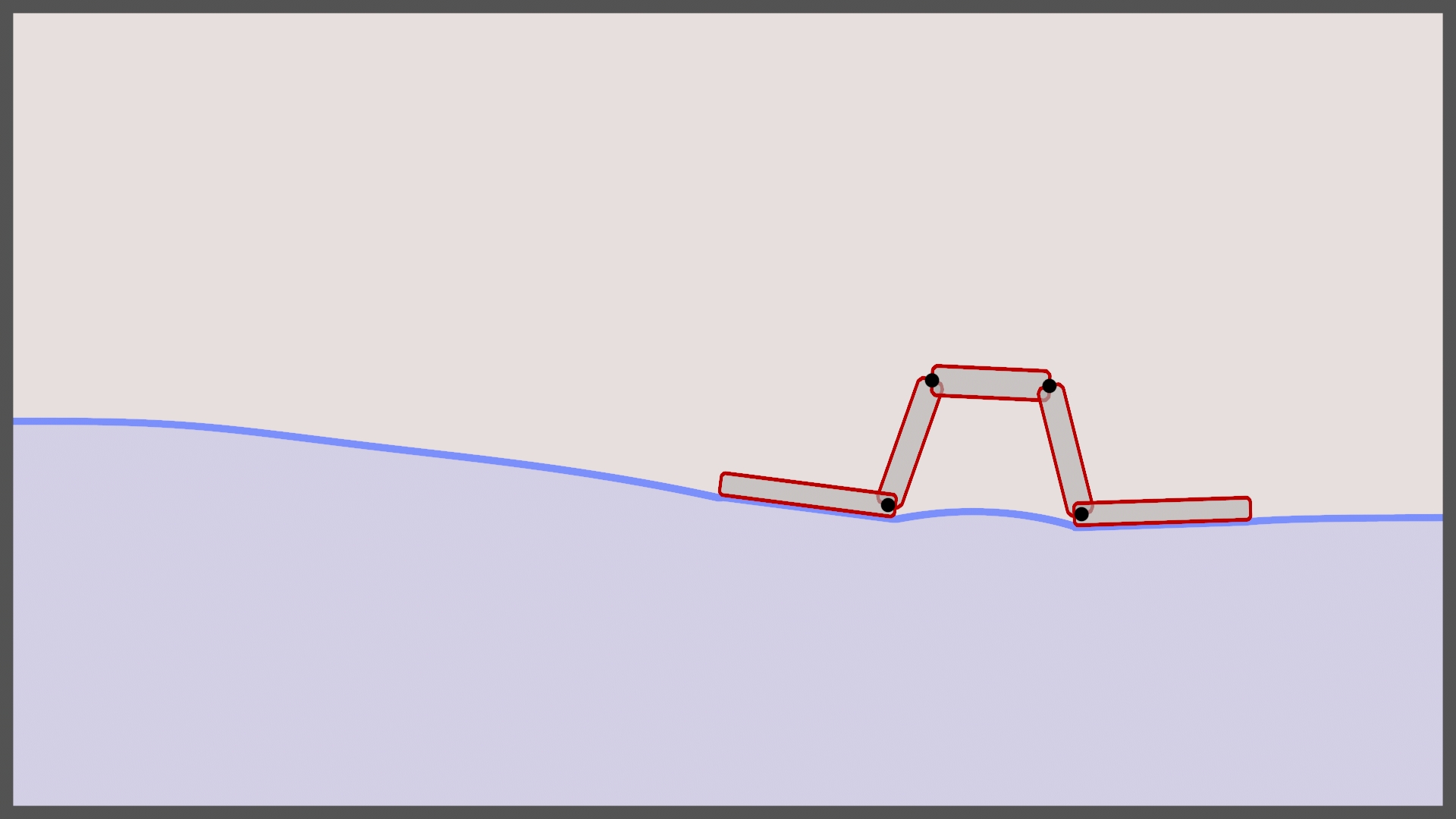}
        \formattedgraphics{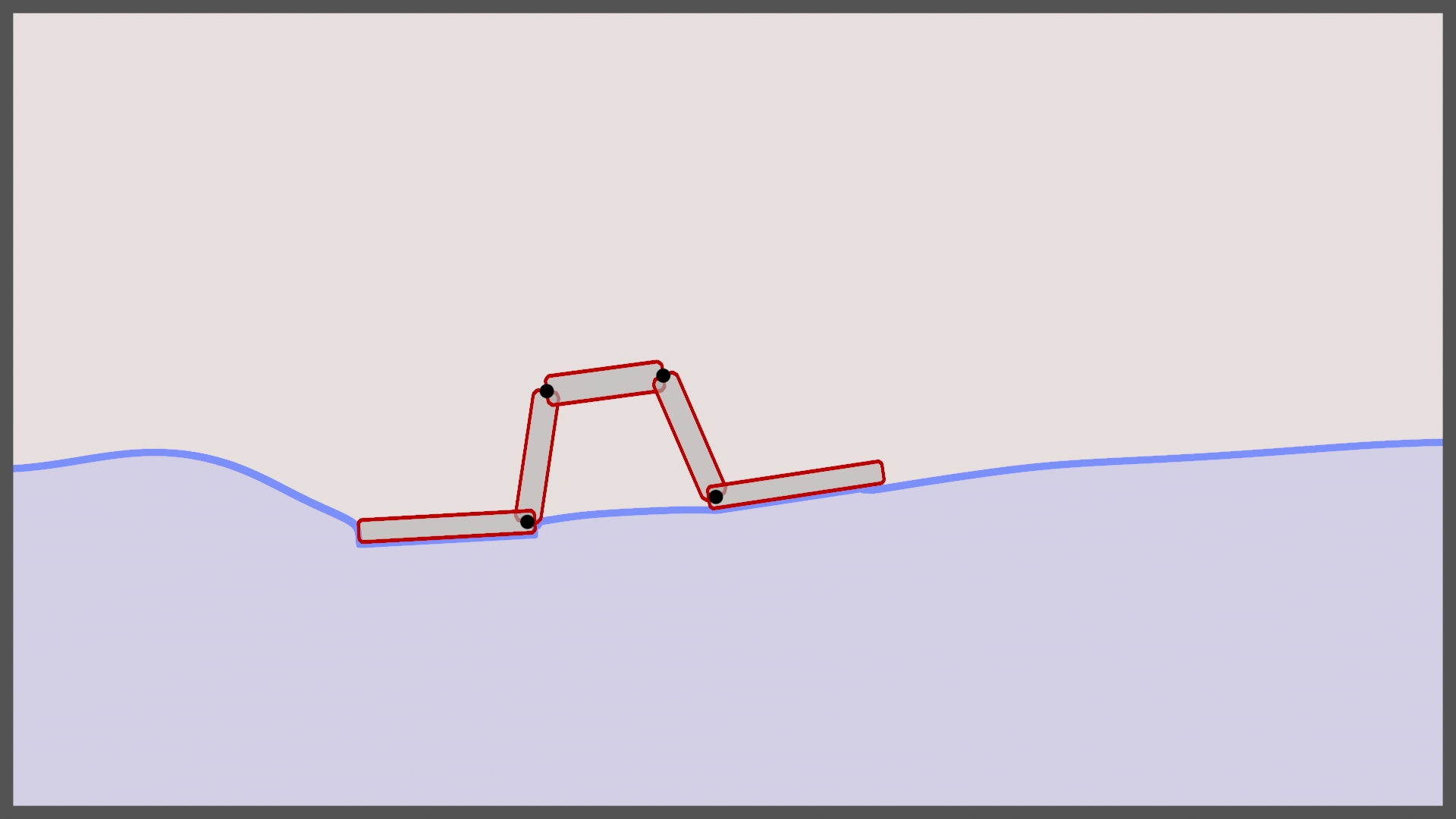}
    }
    \subfigure[Robot Jumping with Low Surface Tension]{
        \formattedgraphics{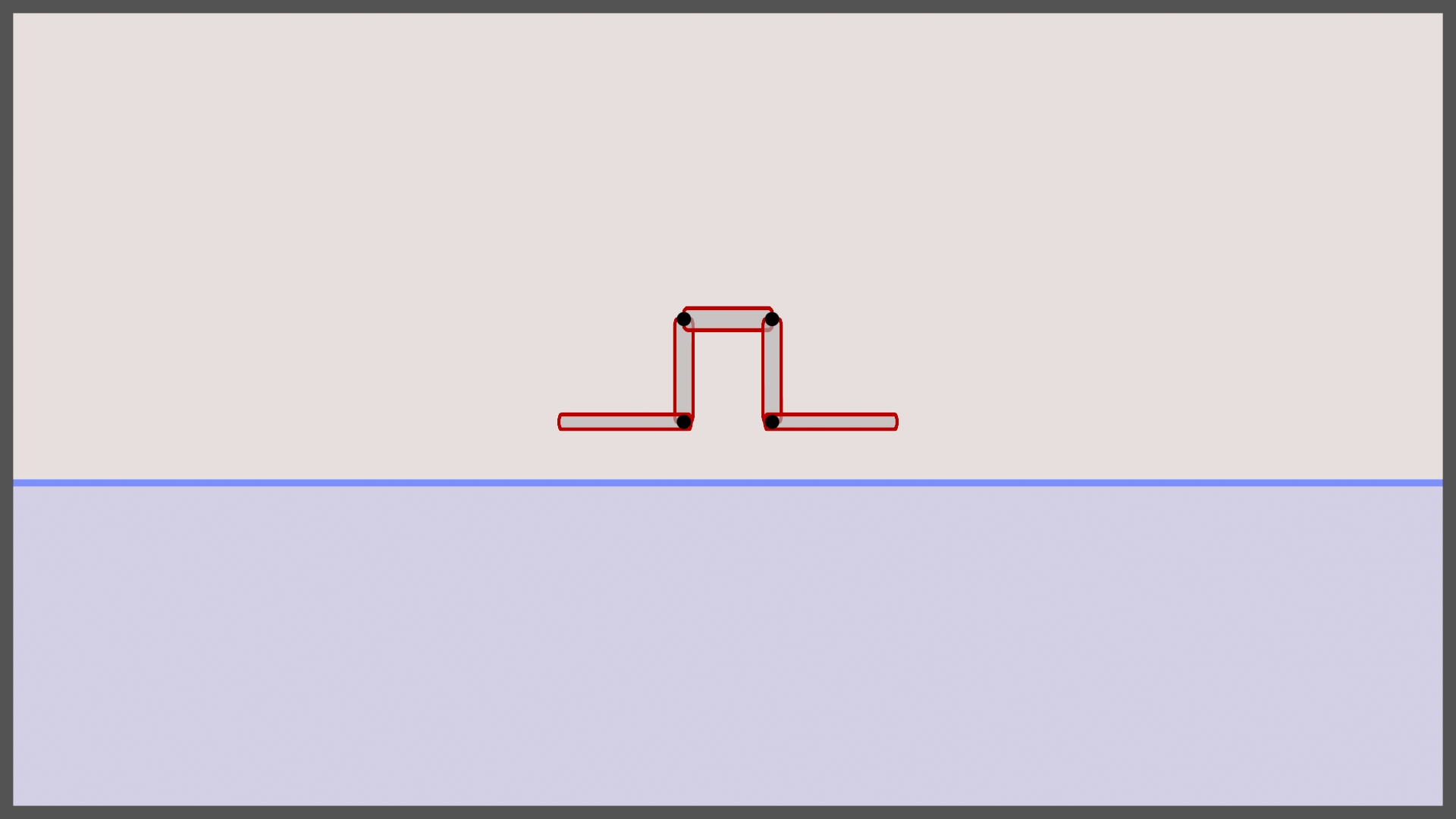}
        \formattedgraphics{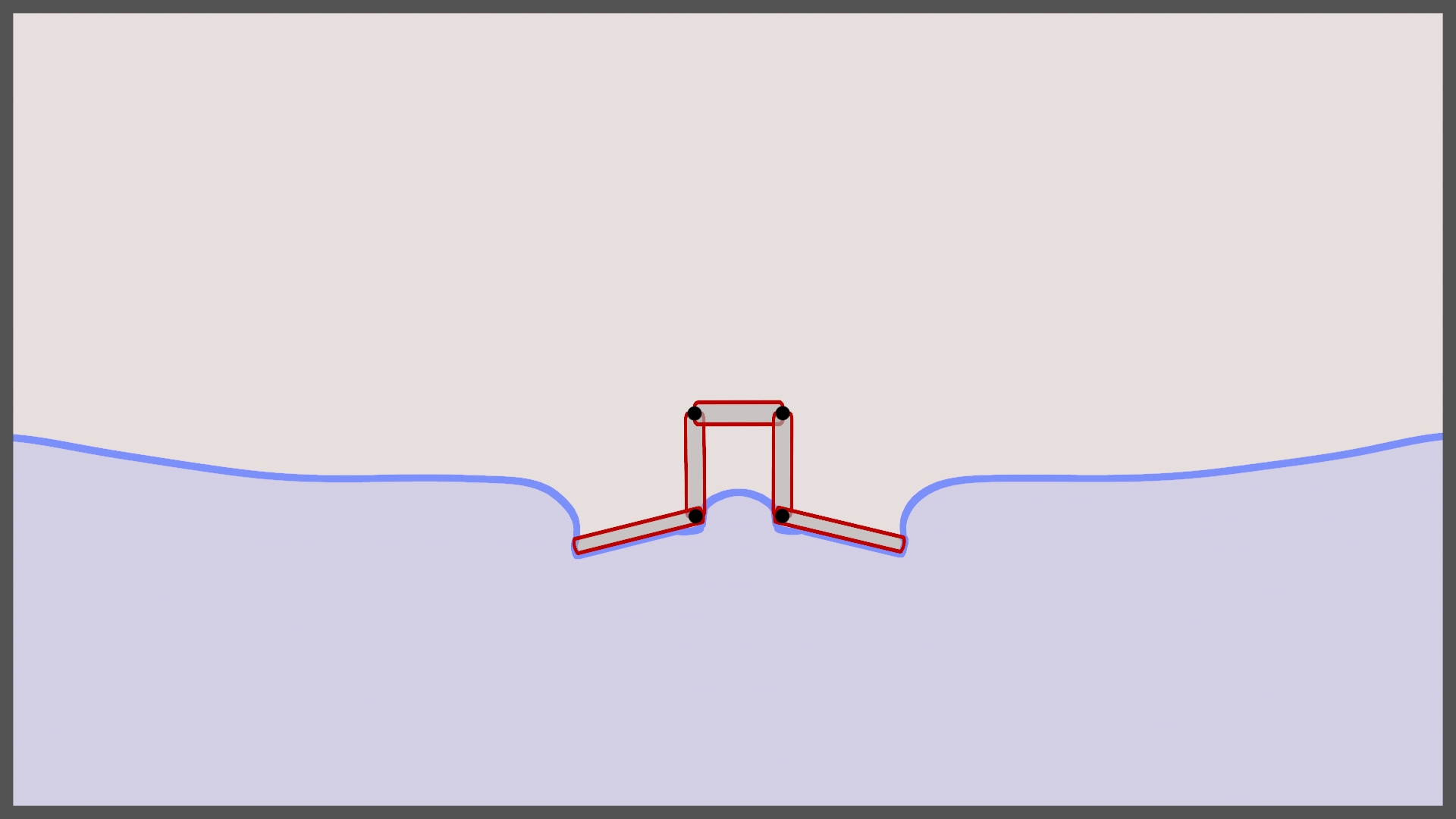}
        \formattedgraphics{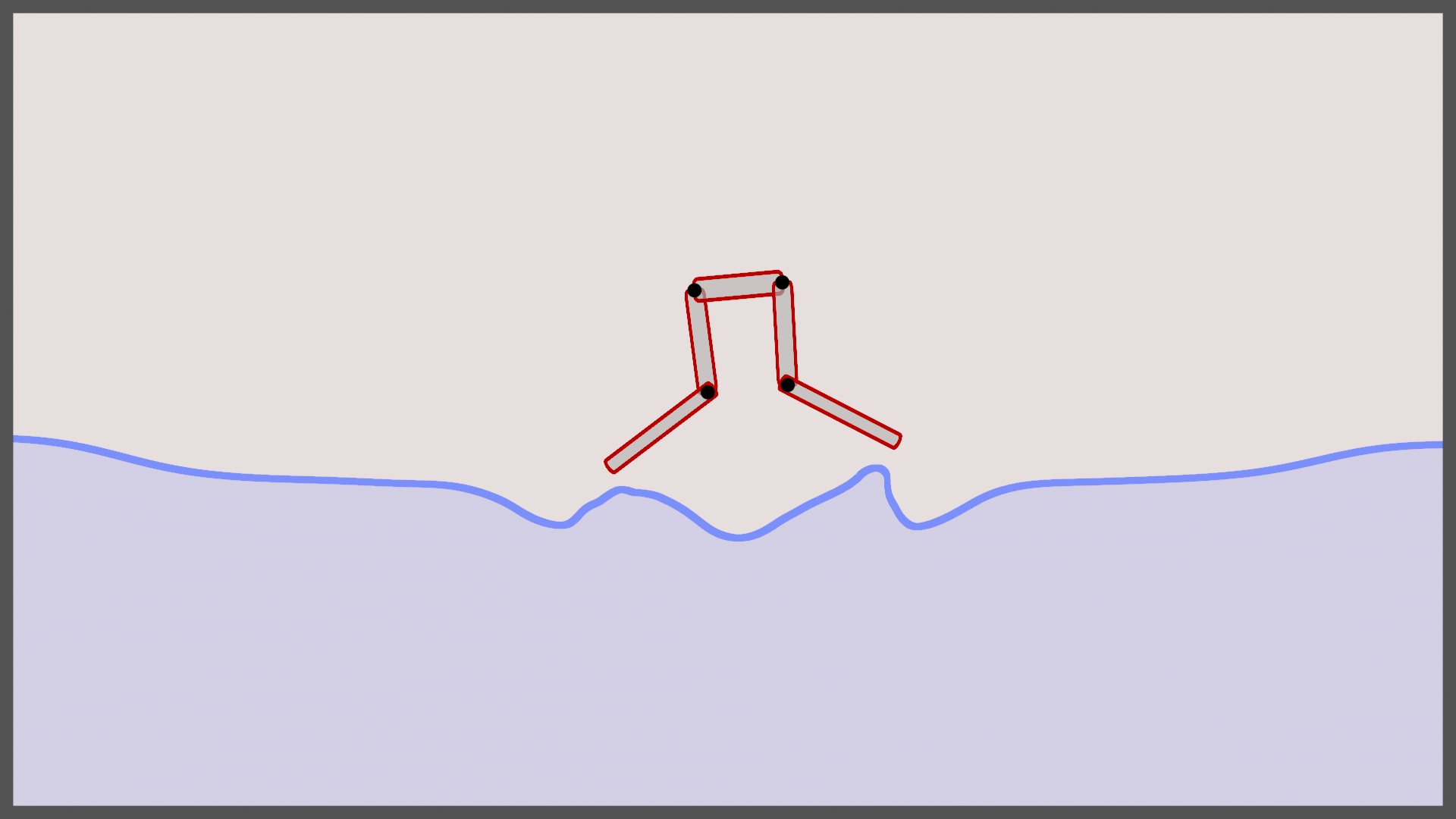}
        \formattedgraphics{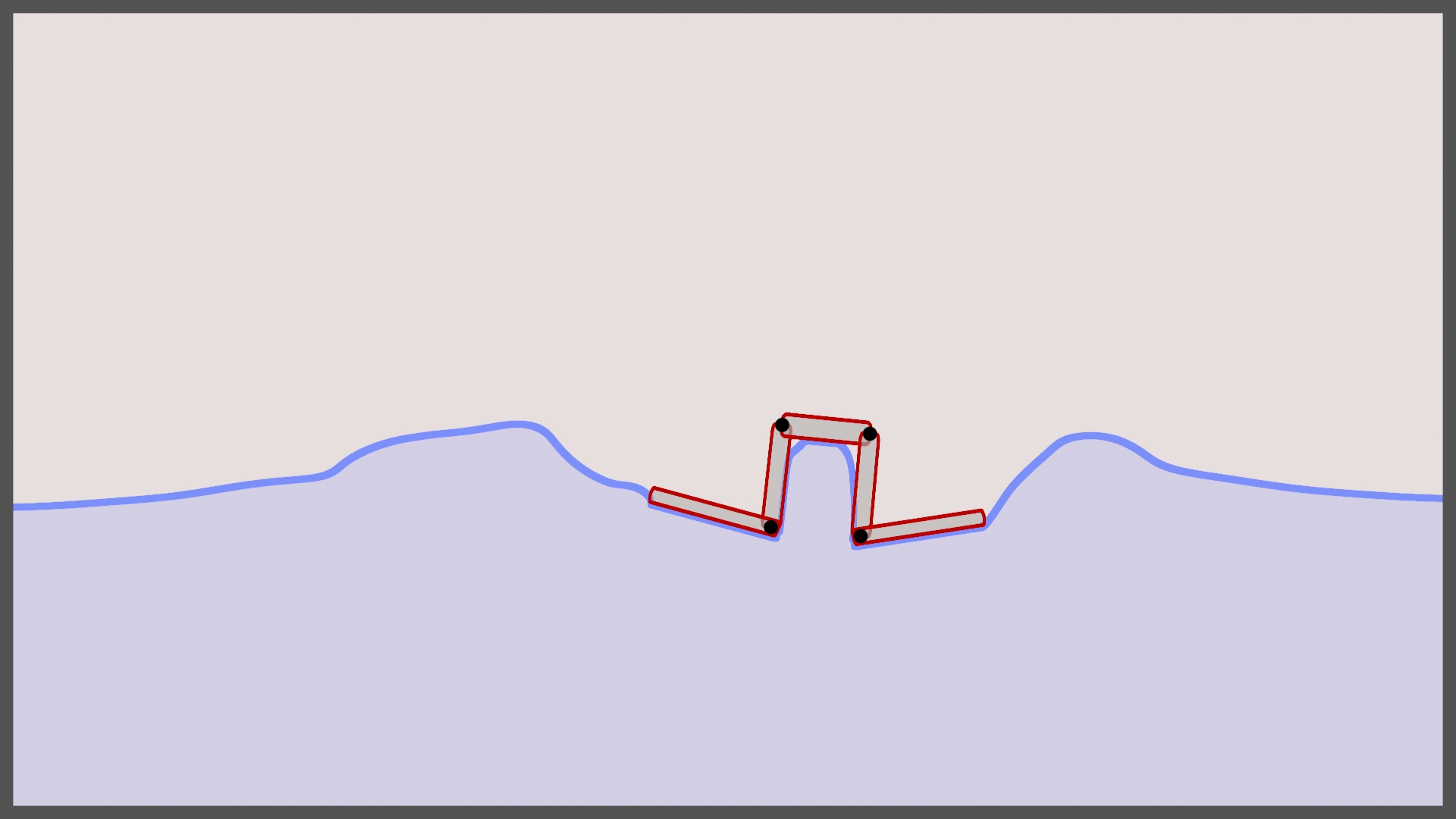}
    }
    \subfigure[Robot Jumping with High Surface Tension]{
        \formattedgraphics{Figures/2DJump/jump_hst_new_render0000.jpg}
        \formattedgraphics{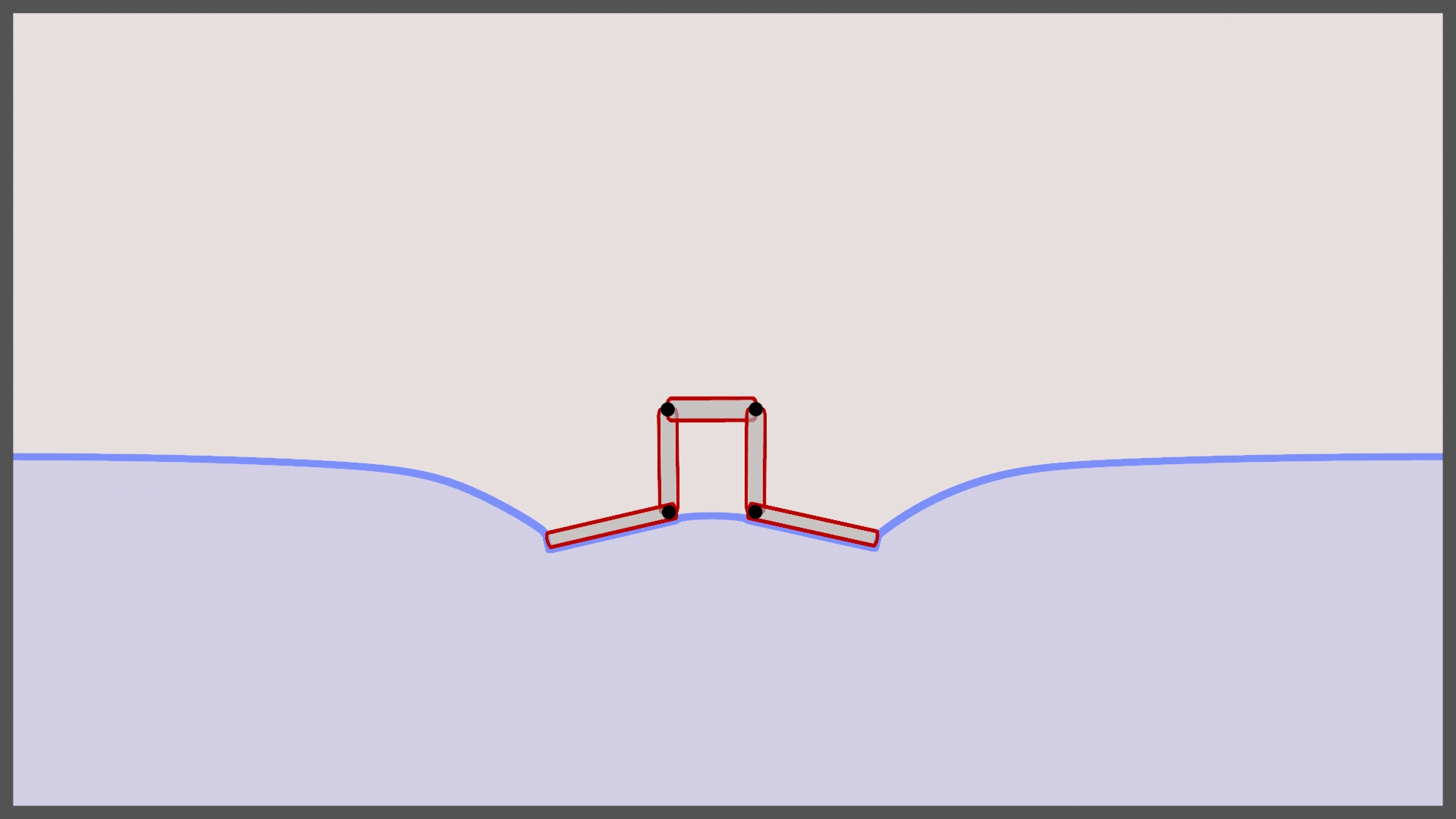}
        \formattedgraphics{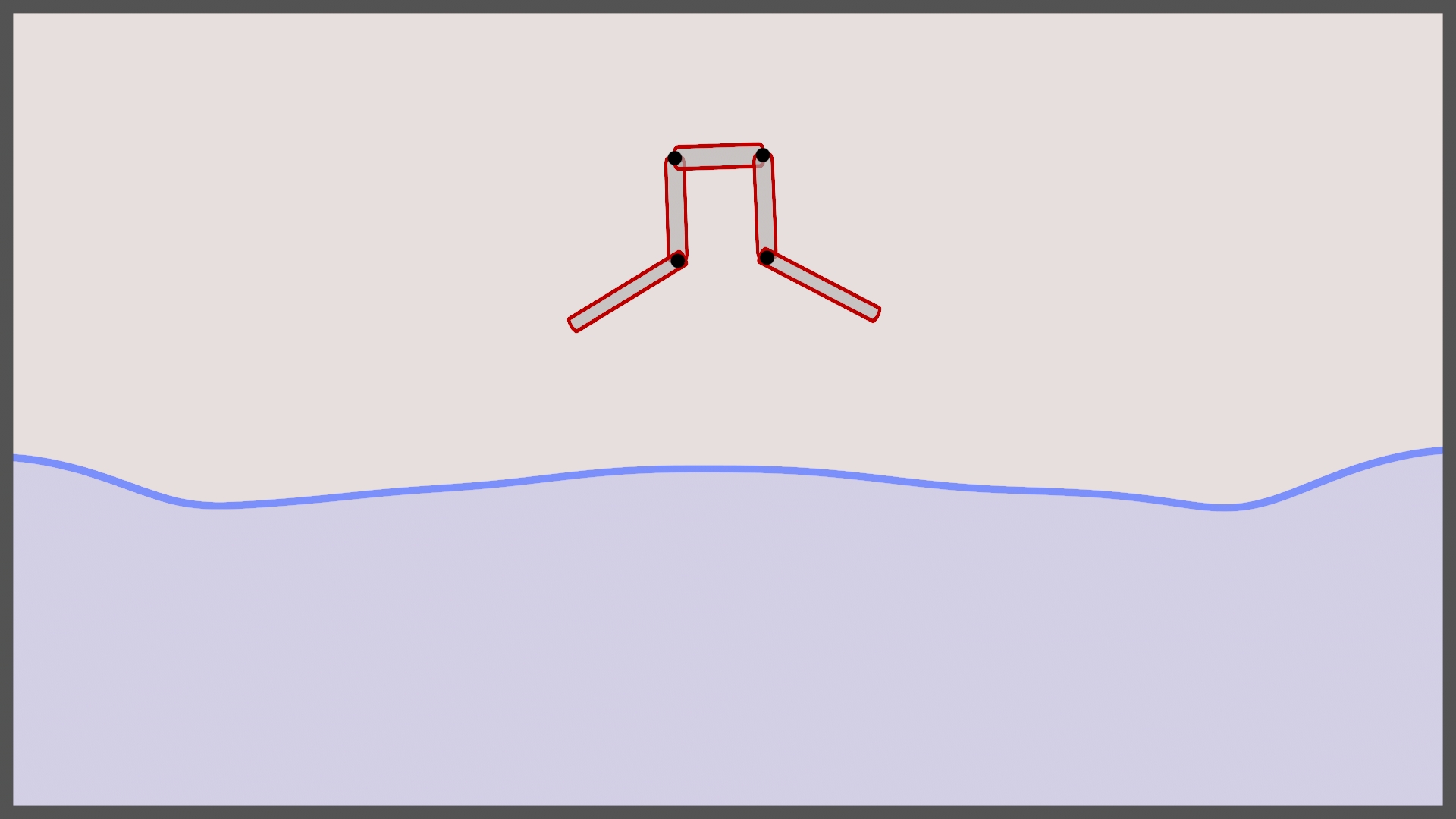}
        \formattedgraphics{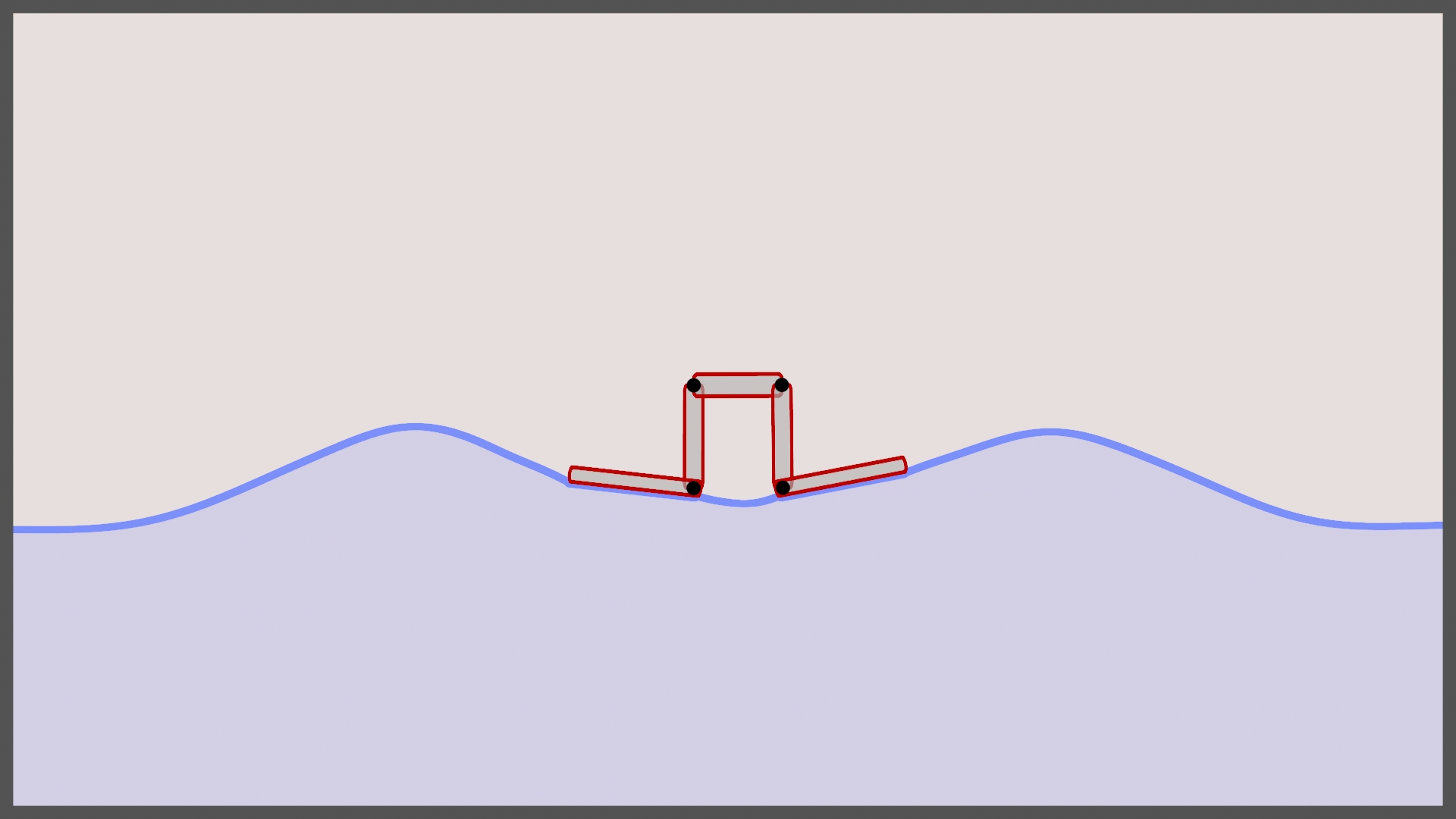}
    }
    \caption{Surface Tension. (a) "Cheerios effect", two hydrophobic objects lying on free surface tend to attract each other due to unbalance water surface. Comparison between (b) and (c) demonstrates that high surface tension fluid can help the robot slide more easily. Similar phenomenon occurs in (d) and (e), when the robot tries to jump up from the fluid surface. High surface tension fluid can produce more counter-impulse, thus helping the robot jump higher.}
    \label{fig:2d}
\end{figure*}

%% file: Figures/boat.tex
\begin{figure}
    \centering
    \includegraphics[width=0.98\linewidth]{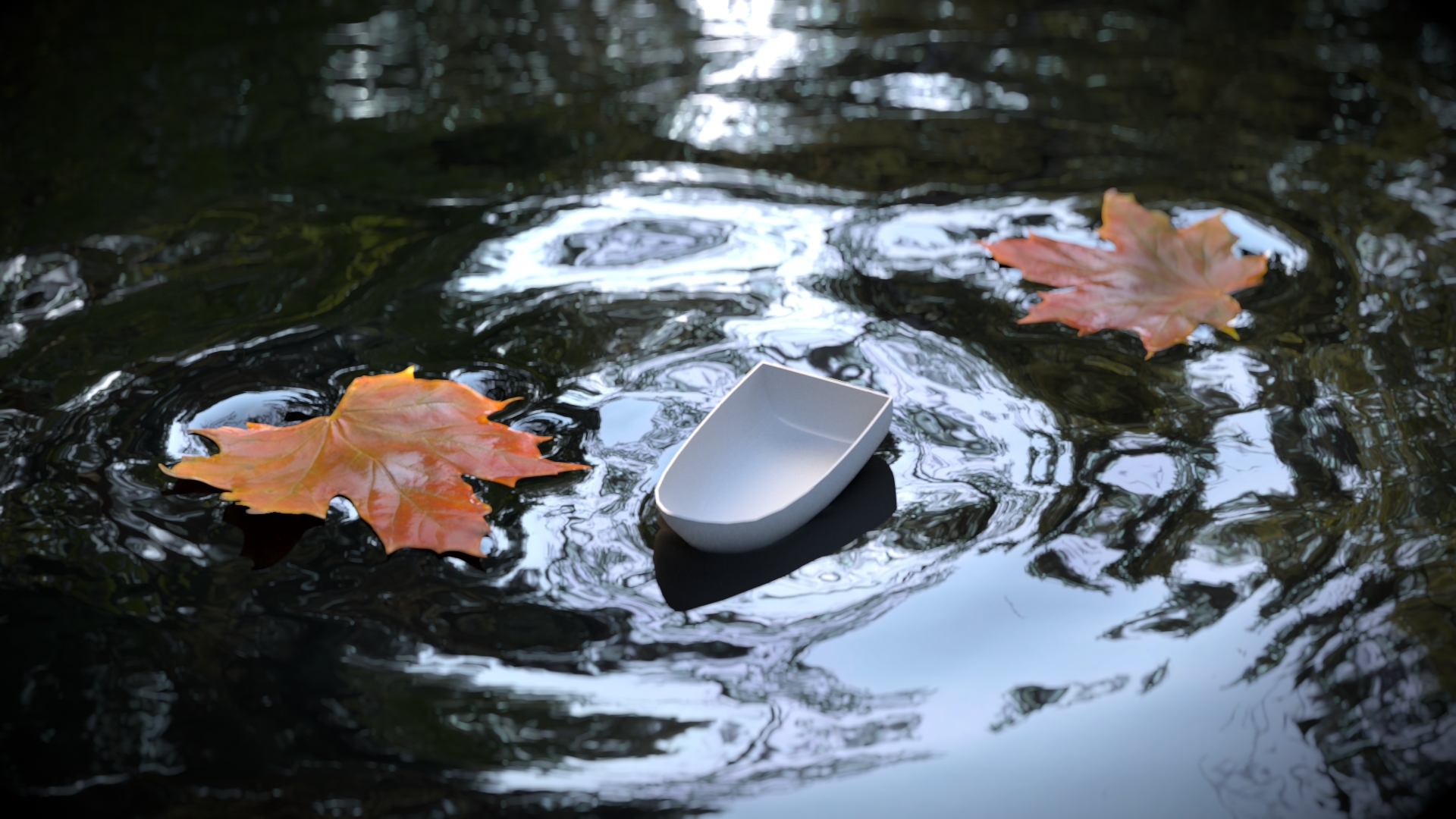}
    \caption{Boat and Leaves. The boat and leaves are modeled by surface meshes with zero thickness. Our algorithm can correctly resolve the collision between the membrane and thin shells.}
    \label{fig:boat}
\end{figure}

%% file: Figures/Sink.tex
\begin{figure*}[t]
	\newcommand{\formattedgraphics}[1]{\includegraphics[width=0.19\textwidth,trim= 0 0 0 0,clip]{#1}}
	\centering
    \formattedgraphics{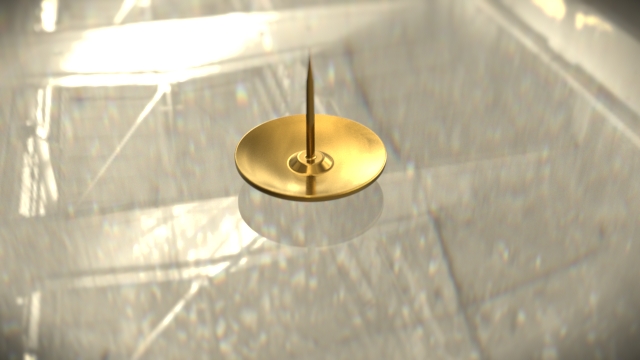}
    \formattedgraphics{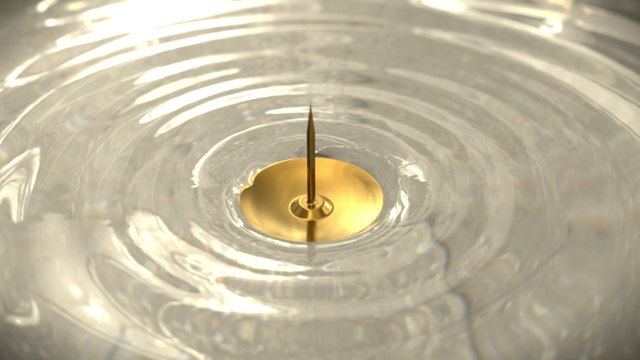}
    \formattedgraphics{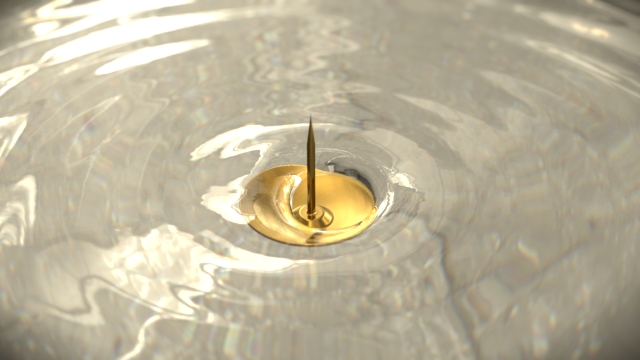}
    \formattedgraphics{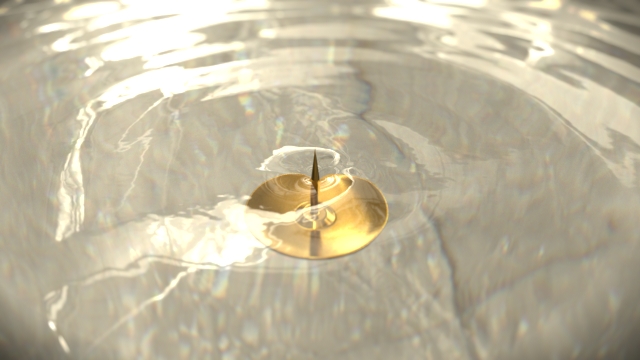}
    \formattedgraphics{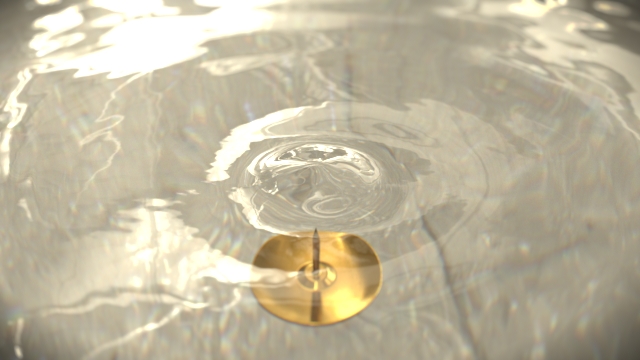}
    \caption{Breaking Surface and Sinking. Water gradually ``climbing'' onto the overweight pushpin until it completely submerges into the water.}
    \label{fig:sink}
\end{figure*}

%% file: Figures/cherry.tex
\begin{figure*}[t]
    \newcommand{\formattedgraphics}[1]{\includegraphics[width=0.16\textwidth,trim= 0 0 0 0,clip]{#1}}
    \centering
    \formattedgraphics{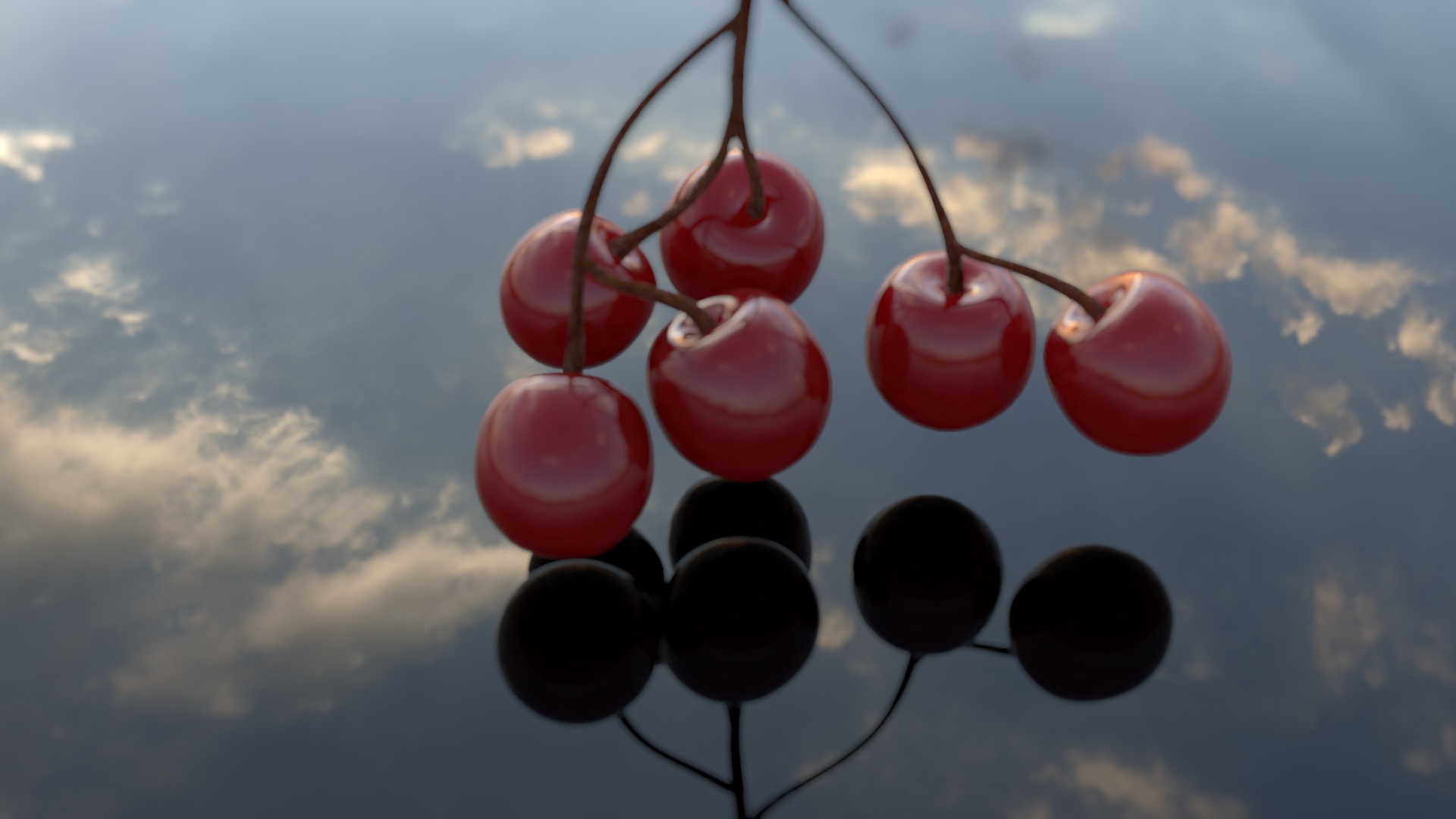}
    \formattedgraphics{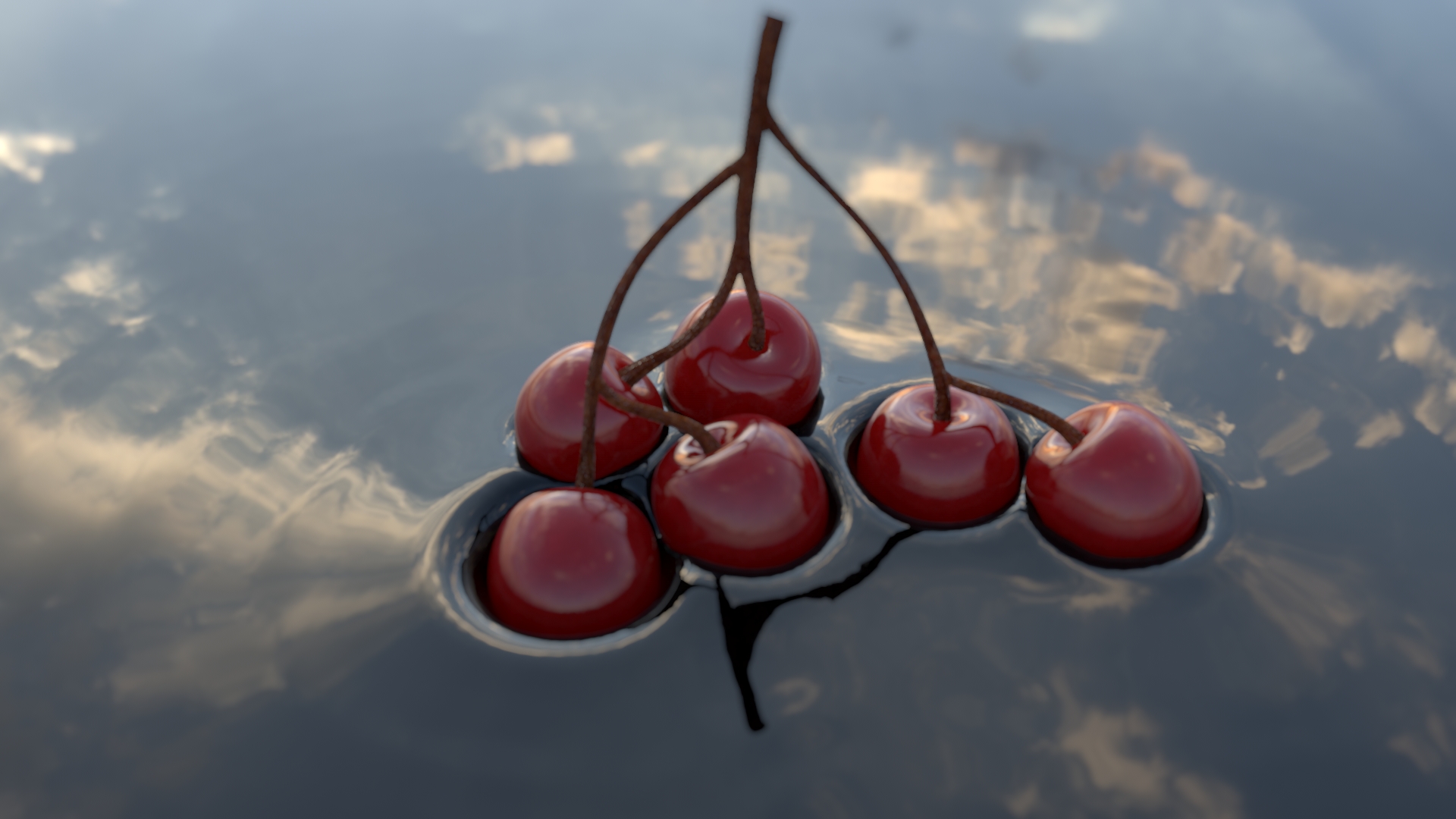}
    \formattedgraphics{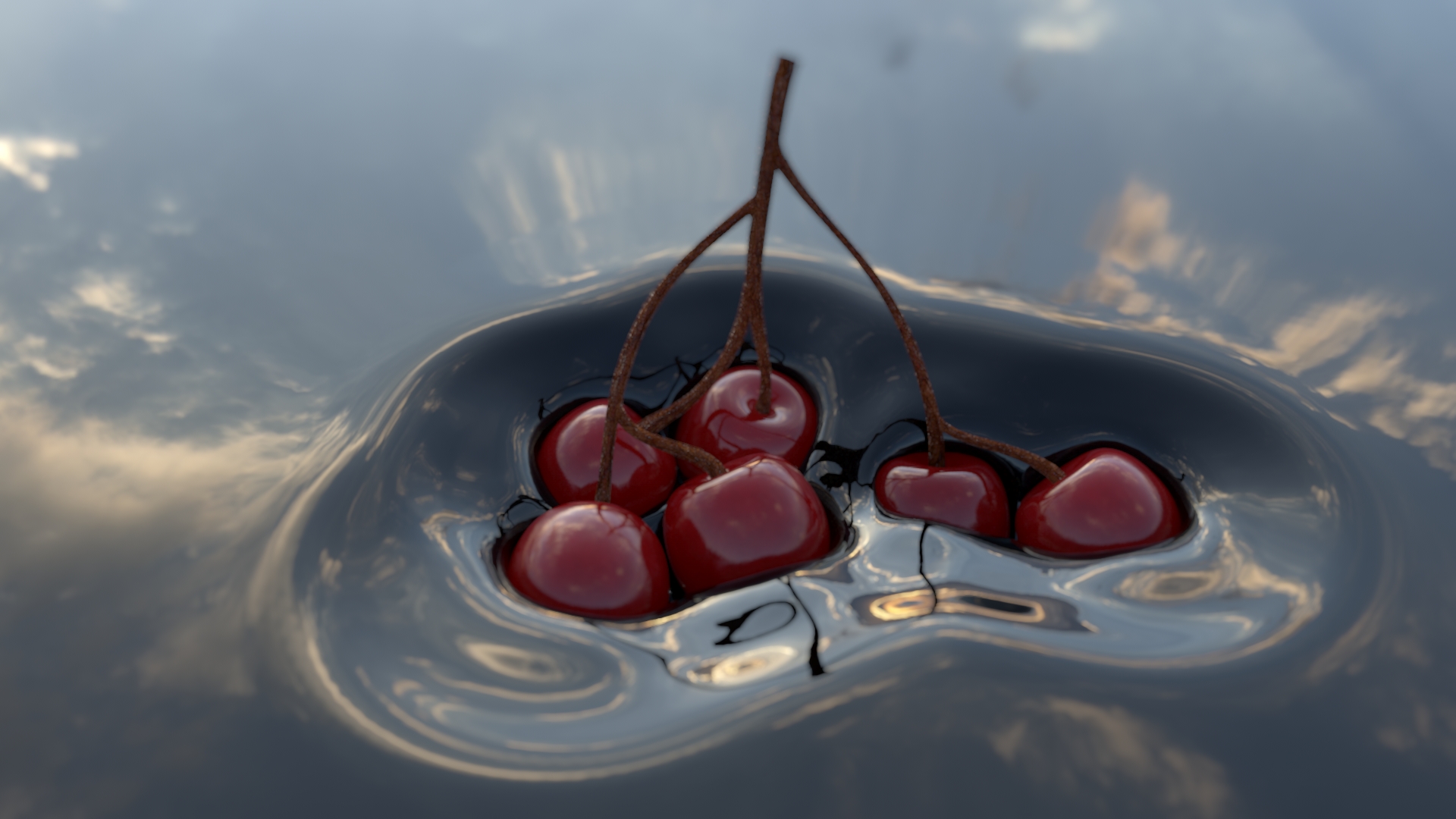}
    \formattedgraphics{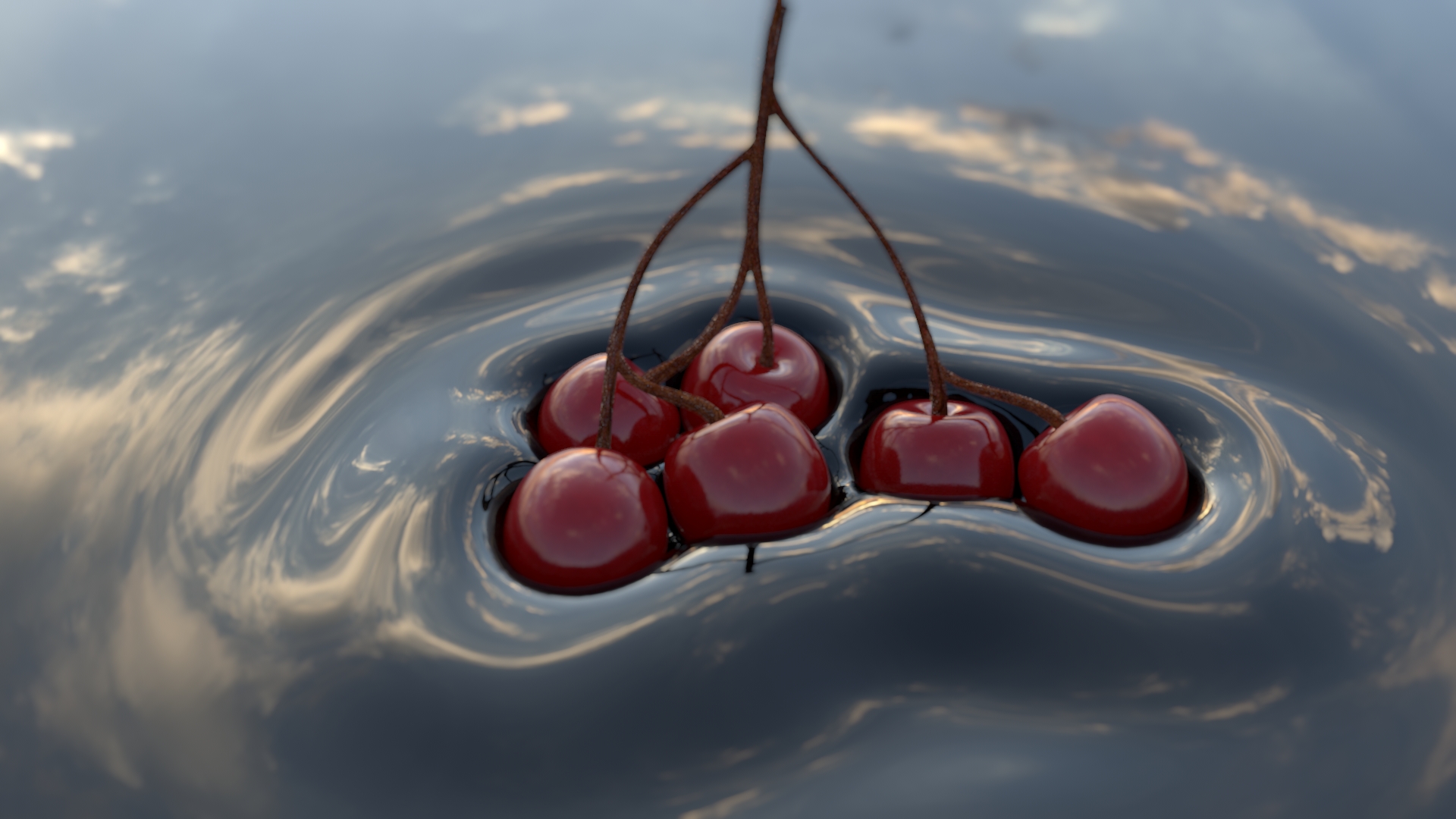}
    \formattedgraphics{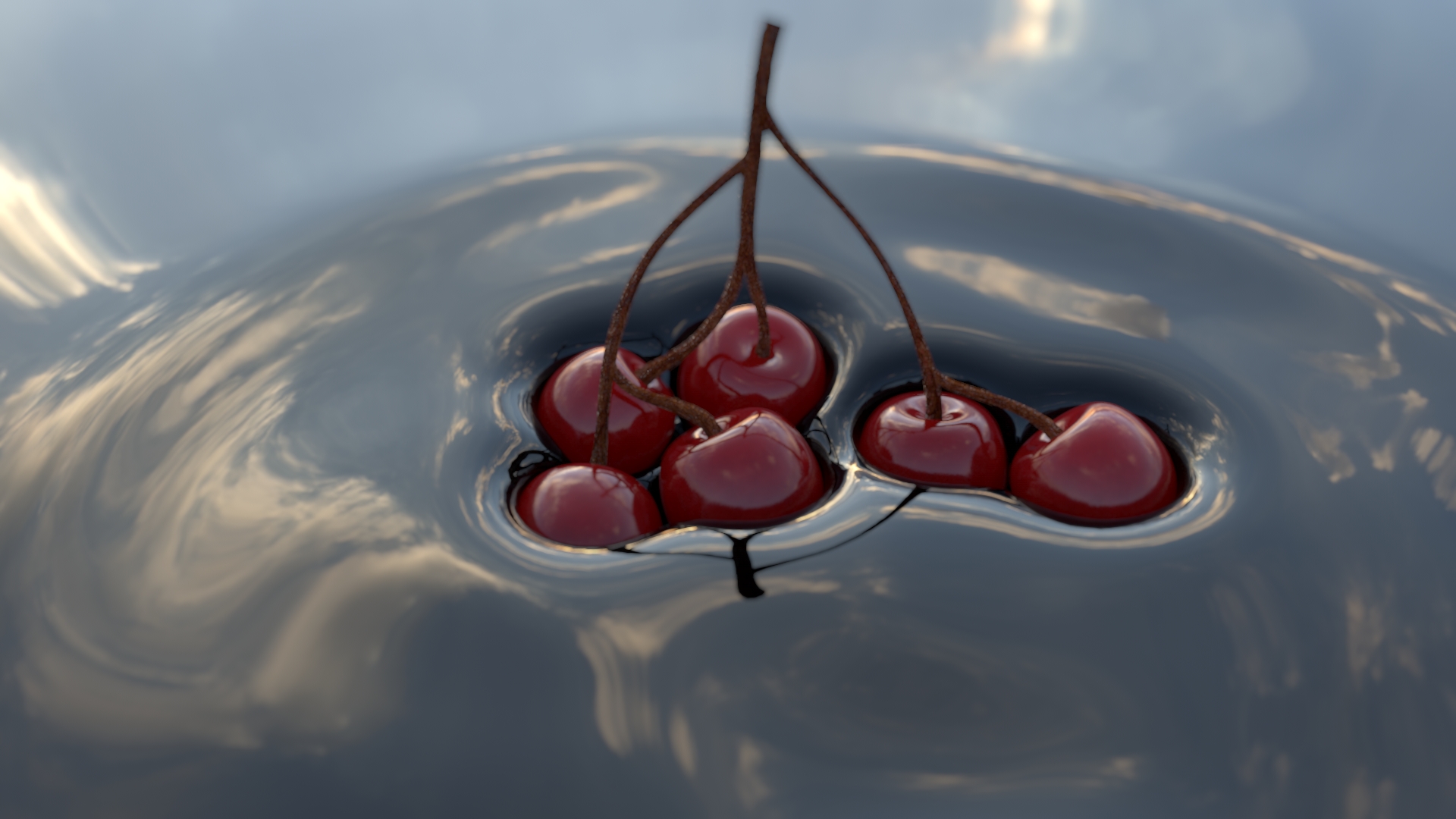}
    \formattedgraphics{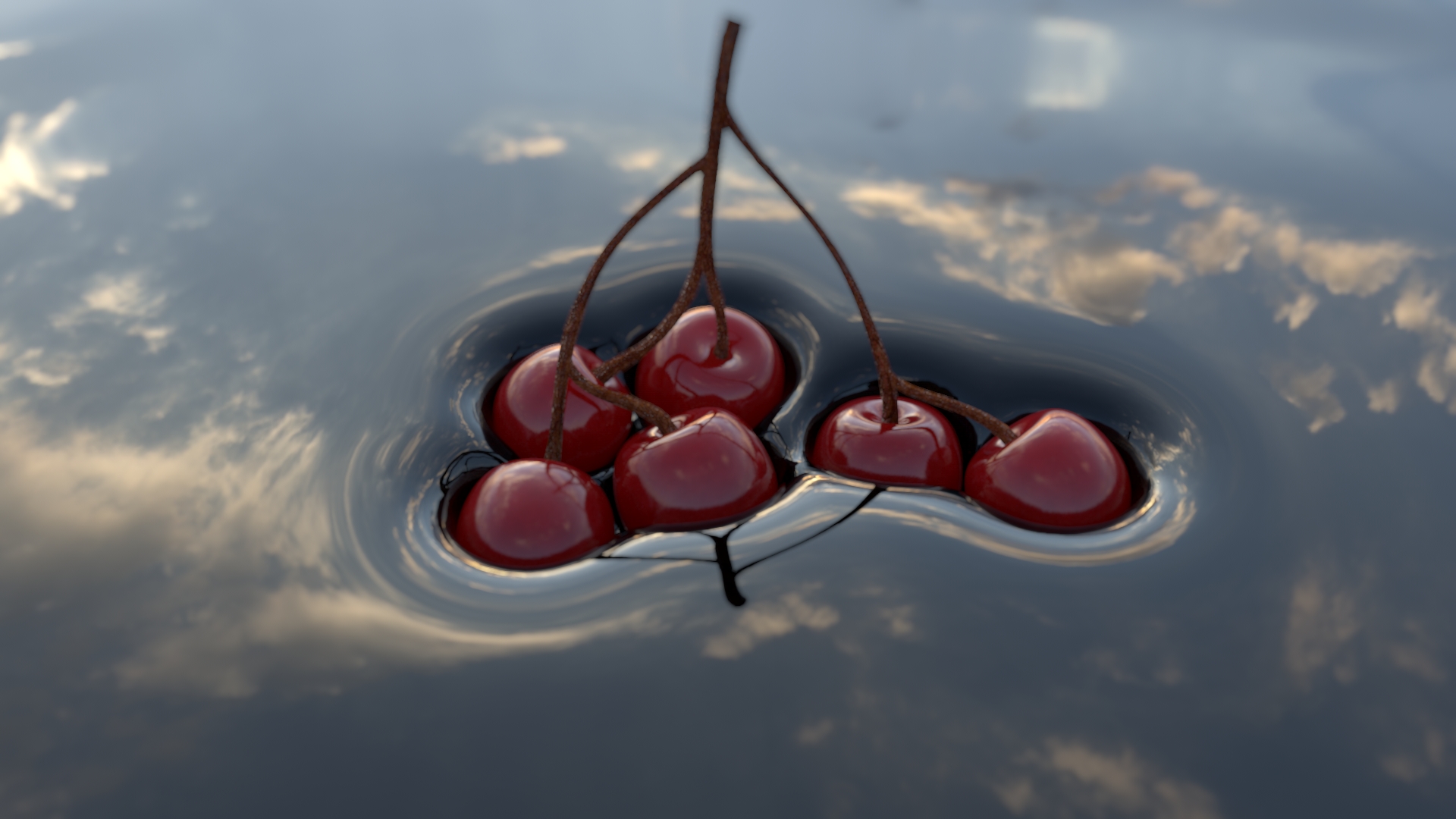}
    \\
    \formattedgraphics{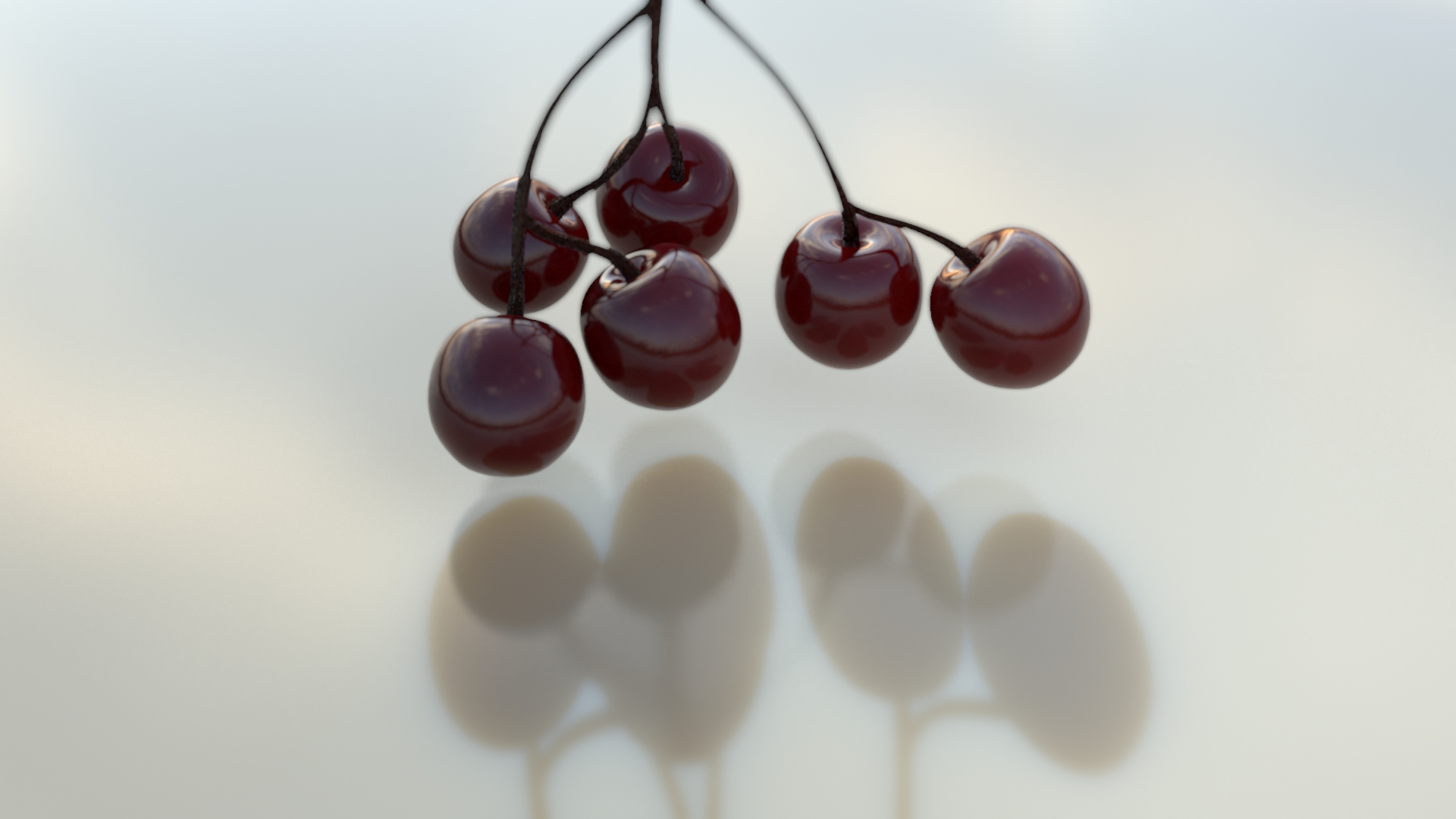}
    \formattedgraphics{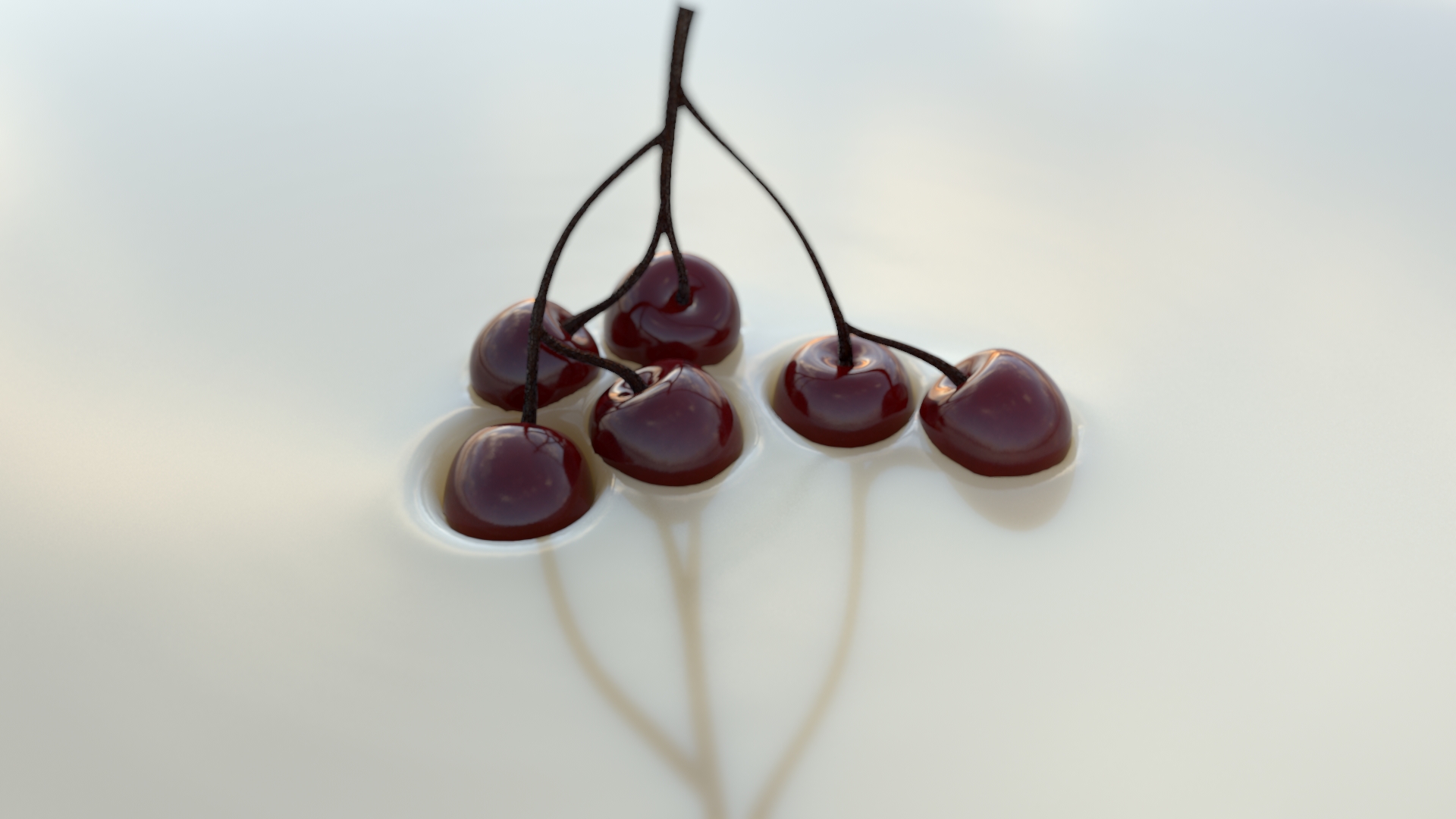}
    \formattedgraphics{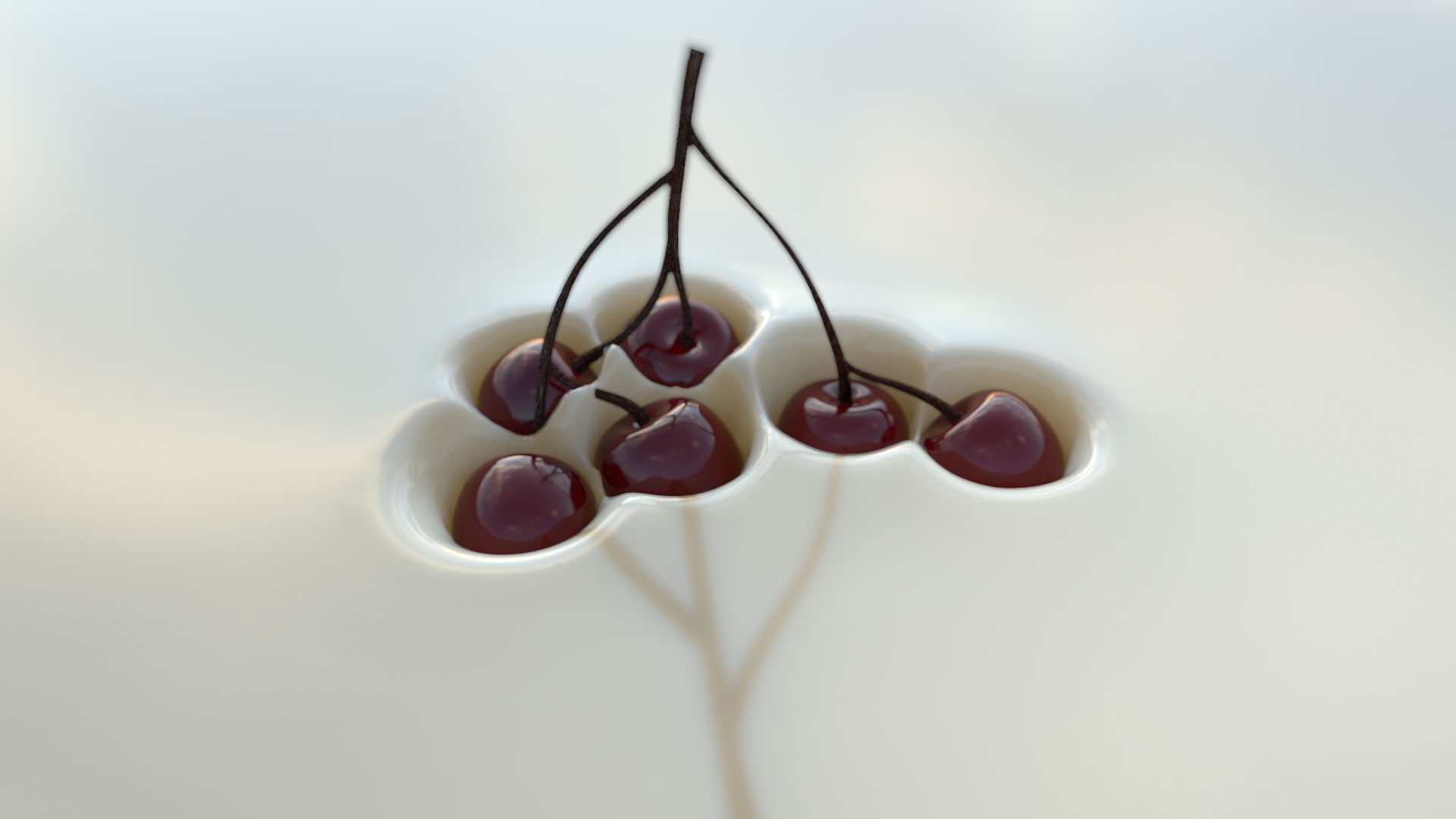}
    \formattedgraphics{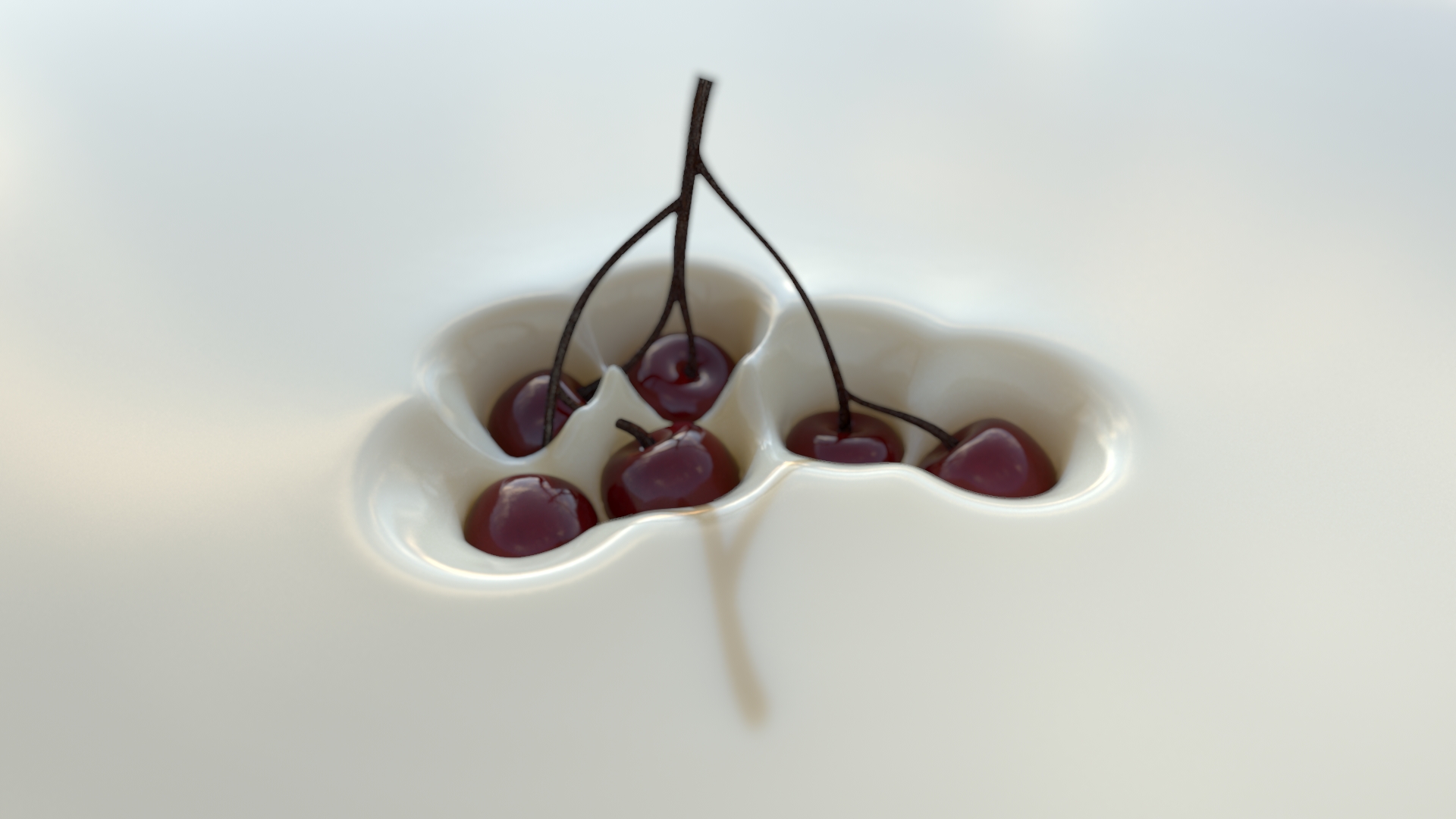}
    \formattedgraphics{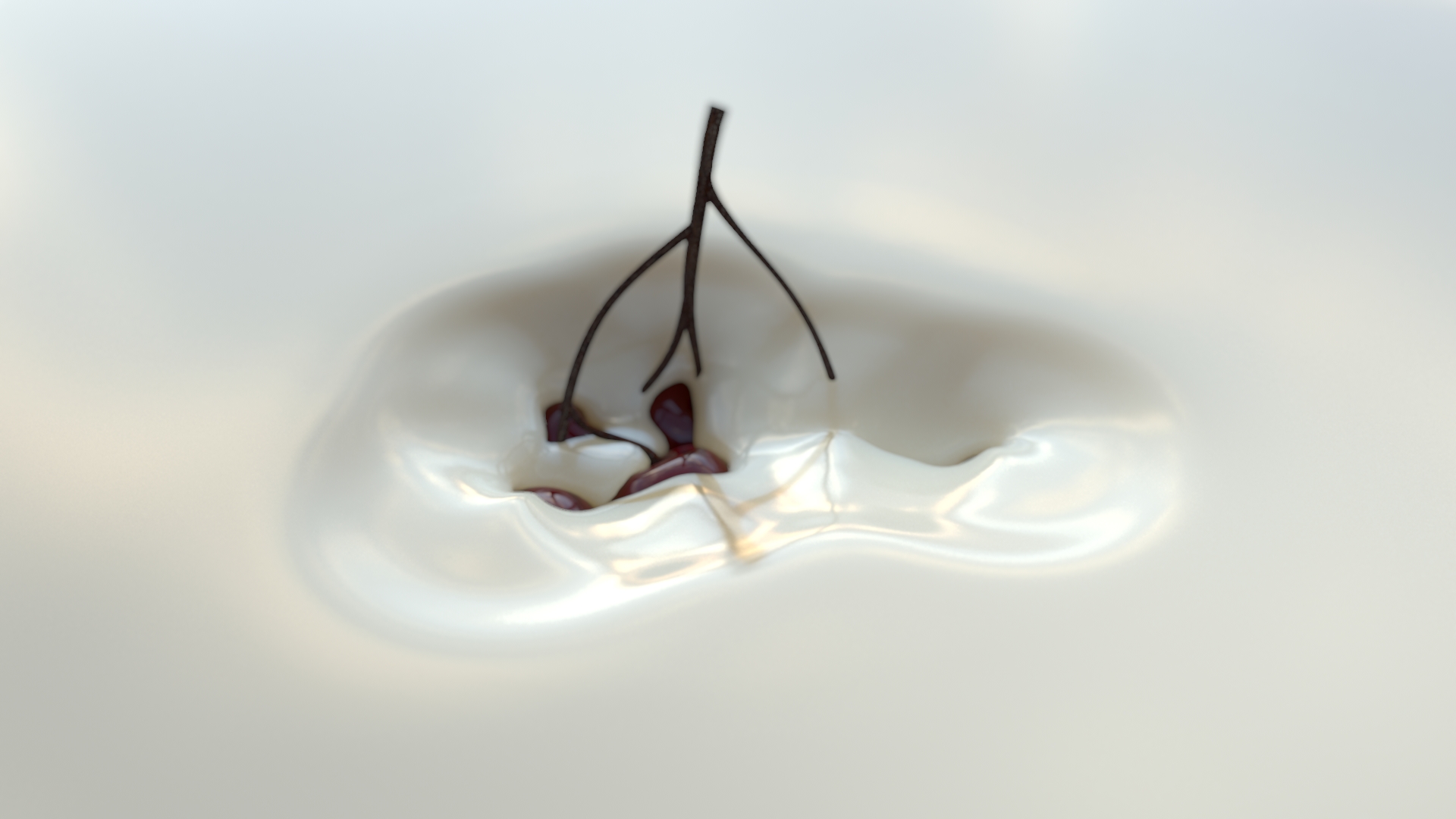}
    \formattedgraphics{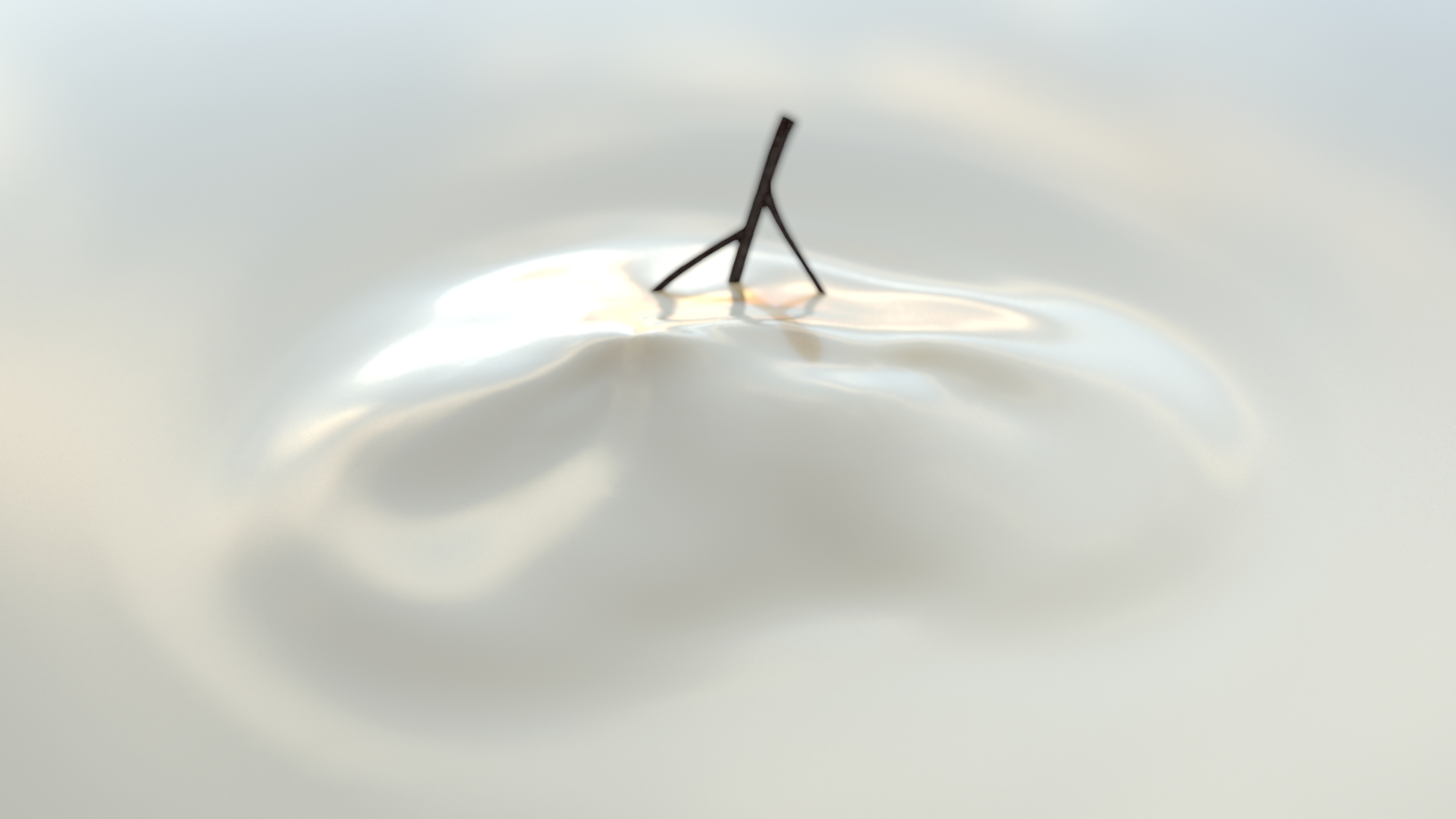}
    \caption{ Cherries Falling into Water and Milk. When cherries fall into a high surface tension fluid (water), they will drain away more fluid during the impact and thus slow down significantly. The cherries can be supported at the interface by buoyancy and capillary forces. Surface-active constituents in milk causes surface tension weakening effect, eventually causing the cherries to sink to the bottom.}
    \label{fig:cherry}
\end{figure*}

%% file: Figures/waterstrider.tex
\begin{figure*}[t]
    \newcommand{\formattedgraphics}[1]{\includegraphics[width=0.24\textwidth,trim= 0 0 0 0,clip]{#1}}
    \centering
    \formattedgraphics{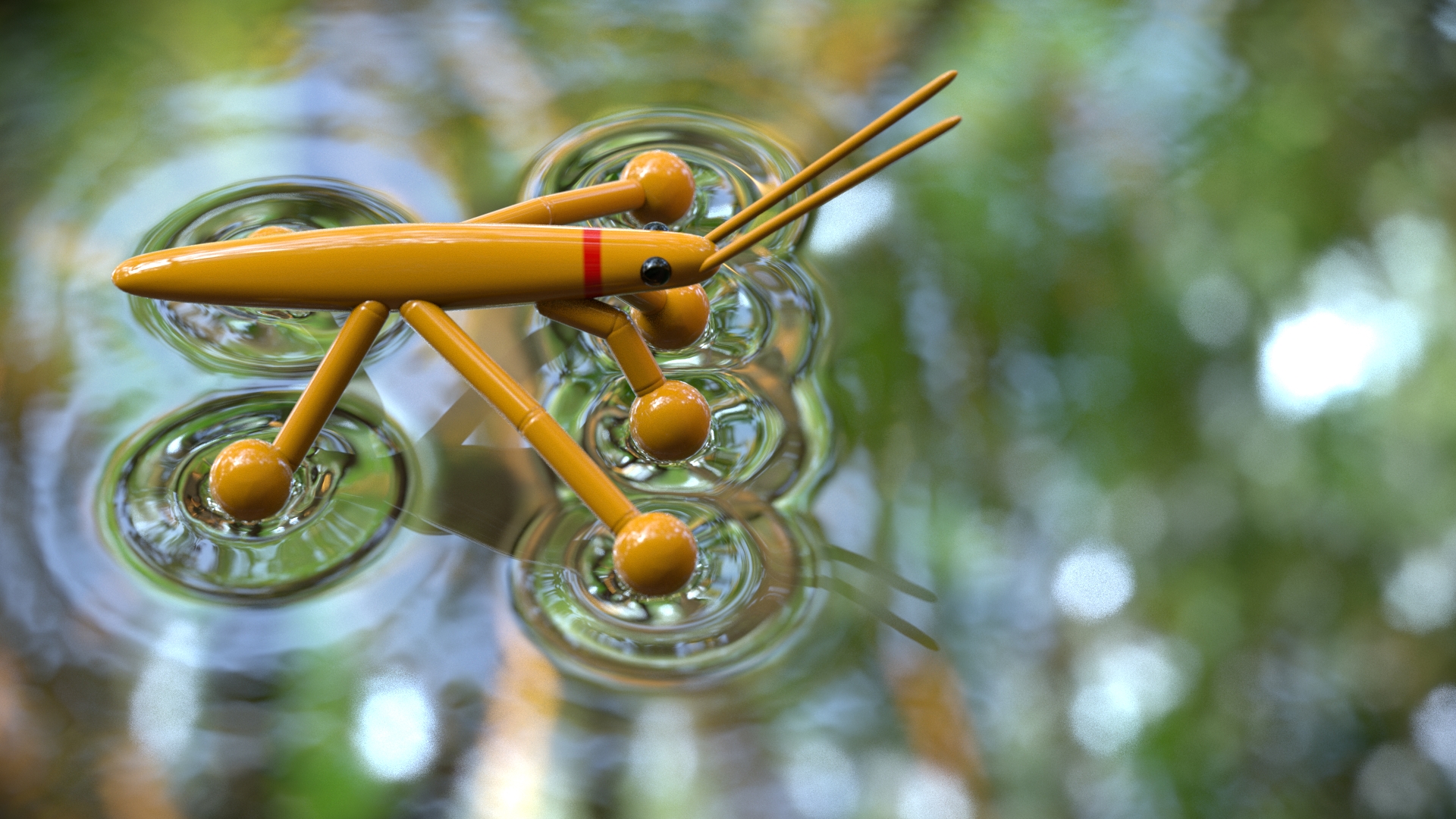}
    \formattedgraphics{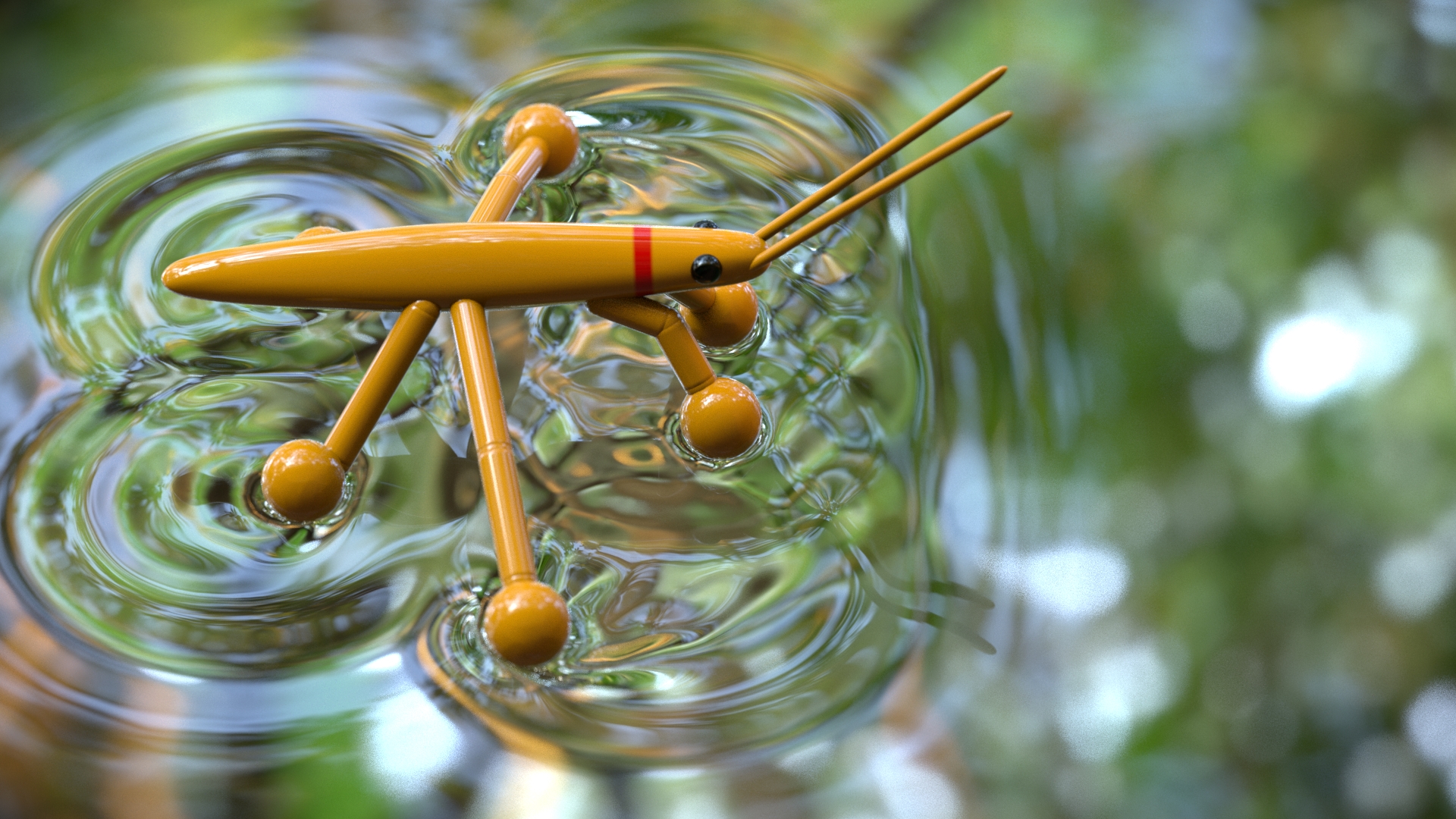}
    \formattedgraphics{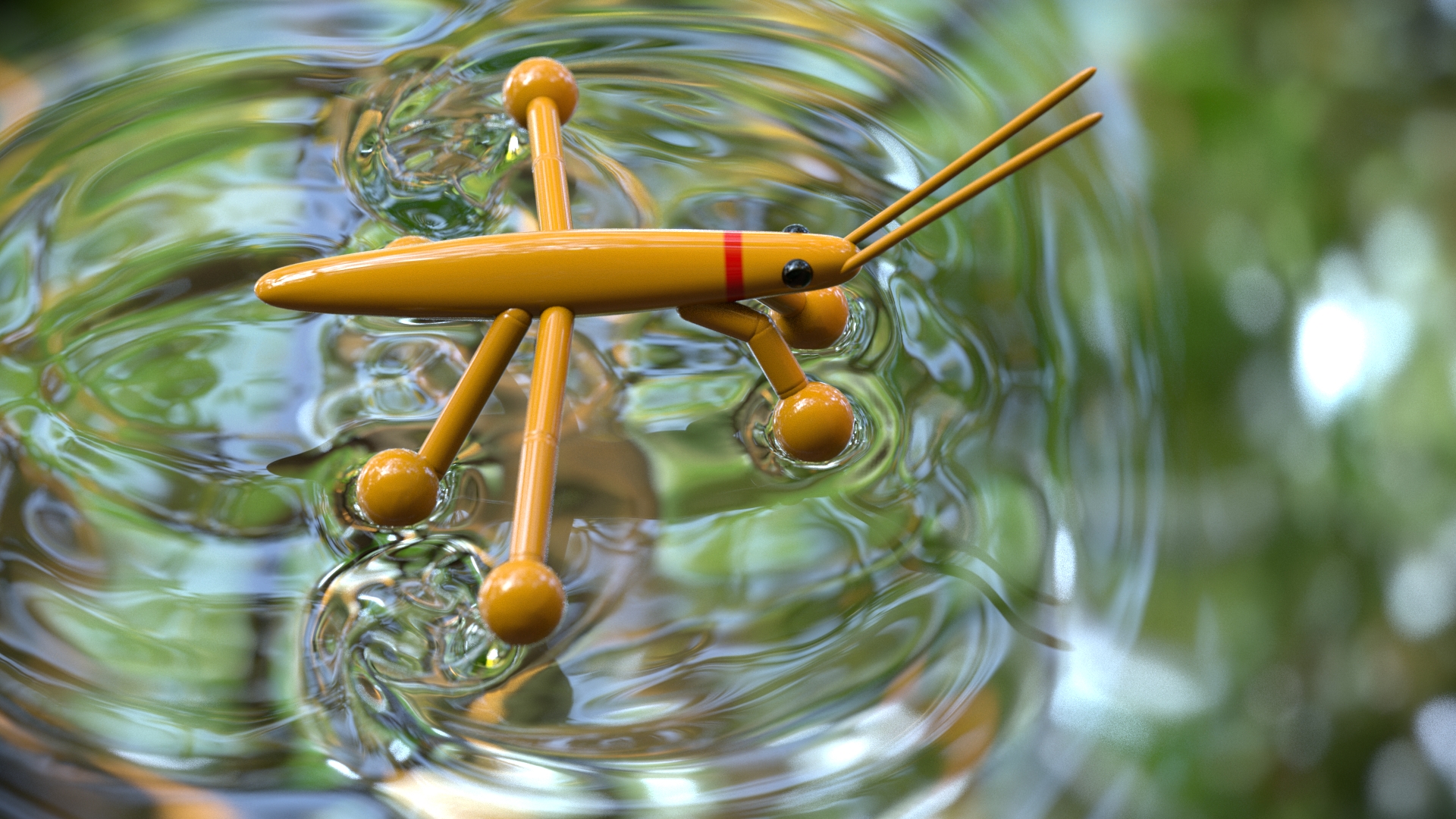}
    \formattedgraphics{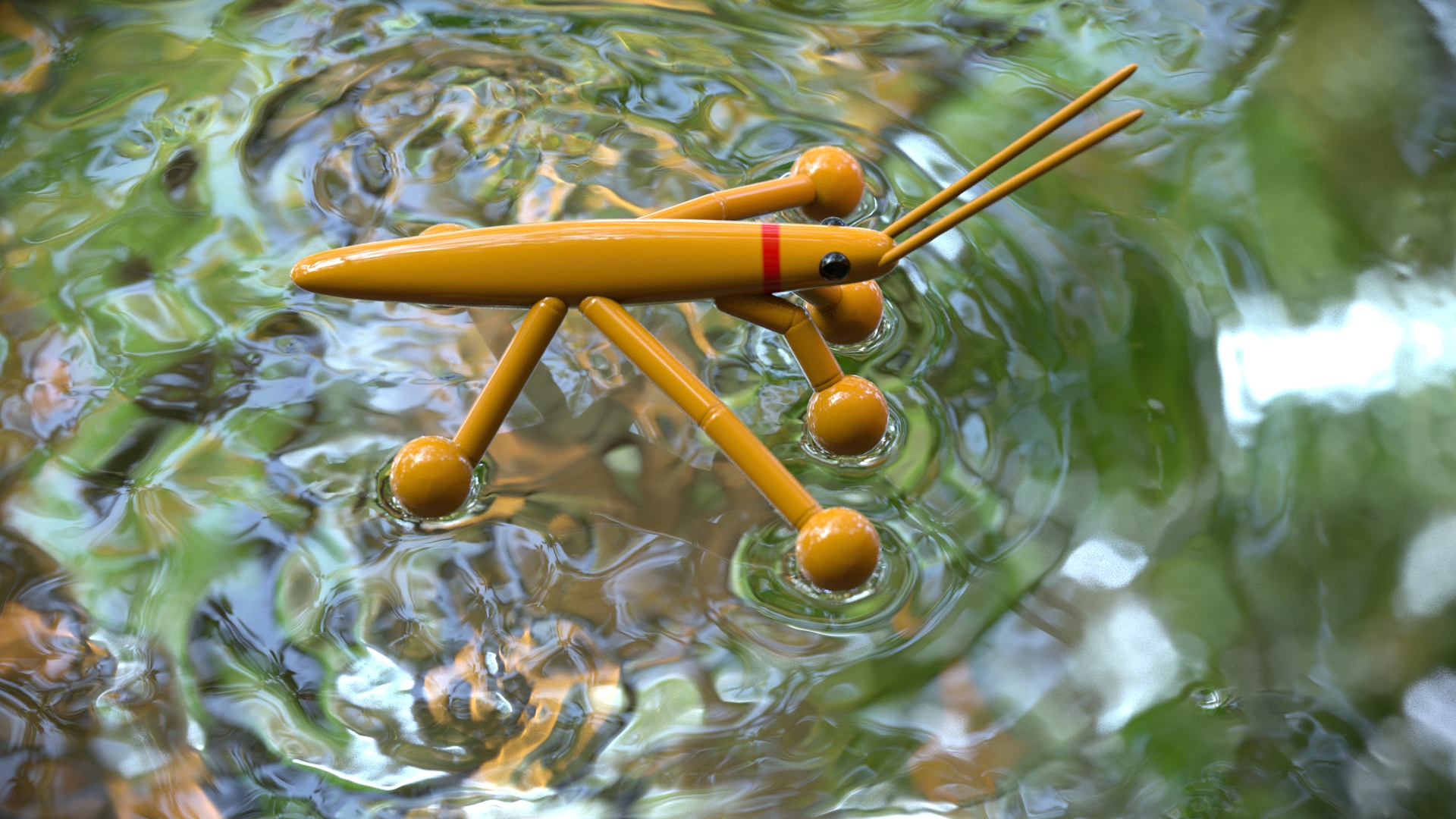}
    \caption{Water Strider Robot. Our water strider robot is modeled using an articulated rigid body. We kinematically control its two middle legs to push down, push back and lift up sequentially. The reaction force cause by surface tension makes our insect robot move forward.}
    \label{fig:waterstrider}
\end{figure*}

%% file: Sections/discussion_and_conclusion.tex
\section{Limitation and Future Work}
We propose a novel monolithic coupling framework to simulate the surface-tension-dominant contact between fluid and rigid bodies. By introducing a Lagrangian mesh to represent the liquid membrane with a small thickness, our model can handle the interactions between fluid and rigid bodies in a variety of realistic physical settings.

There are several limitations of our current method.
First, the three-way coupled system is not symmetric positive definite, which affects the solver's performance when solving large-scale problems. This problem can be addressed in the future by employing numerical techniques such as \cite{aanjaneya2018efficient} and \cite{8790953}.   

On the other hand, our explicit mesh representation, though exhibiting outstanding performance on handling contact, cannot handle the topological evolution of complex liquid surfaces efficiently. Our mesher supports some simple topological changes such as the ones caused by edge collapse (e.g., see the sinking pushpin example), but its ability is limited when handling more drastic mesh evolutions such as splash, droplet pinch-off, or water bulks merging. Particle-based surface or other connectivity-free Lagrangian methods are worth considering for future work.
Regarding the physical model, our current solver does not support hydrophilic materials which limits its scope of applications.

Overall, we believe that the three-way coupling numerical method we proposed in this paper is the first step toward building accurate numerical simulators to model complex solid-fluid interactions with strong surface tension. We hope these computational tools will be useful not only for producing new visual effects for computer graphics applications but also for facilitating scientific research to better understand these small-scale natural phenomena.  
